\documentclass[rmp,aps,nofootinbib,eqsecnum,showpacs,preprintnumbers]{revtex4}

\usepackage{epsfig}
\usepackage{bm}

\newcommand{\be}{\begin{equation}}
\newcommand{\dd}{\displaystyle}
\newcommand{\ee}{\end{equation}}
\newcommand{\bea}{\begin{eqnarray}}
\newcommand{\eea}{\end{eqnarray}}

\newcommand{\nn}{\nonumber}
\newcommand{\de}{\partial}
\newcommand{\dm}{{\delta\mu}}

\def\Zint{{Z \kern -.45 em Z}}
\def\complex{{\kern .1em {\raise .47ex \hbox
{$\scriptscriptstyle |$}} \kern -.4em {\rm C}}}
\def\real{{\vrule height 1.6ex width 0.05em depth 0ex
\kern -0.06em {\rm R}}}
\def\slash#1{#1\!\!\!\!/}
\def\inbar{\,\vrule height1.5ex width.4pt depth0pt}
\def\IR{\relax{\rm I\kern-.18em R}}
\def\IC{\relax\hbox{$\inbar\kern-.3em{\rm C}$}}

\begin{document}
\preprint{{\bf BARI-TH 461/03}}\preprint{{\bf CERN-TH/2003-082}}
\title{Inhomogeneous  Superconductivity
 in Condensed Matter and QCD}

\author{Roberto Casalbuoni\footnote{On leave from the
Department of Physics of the University of Florence, 50019,
Florence, Italy}} \email{casalbuoni@fi.infn.it}
\affiliation{TH-Division, CERN, CH-1211 Geneva 23, Switzerland}
\author{Giuseppe Nardulli}\email{giuseppe.nardulli@ba.infn.it}
\affiliation{Department of Physics, University of Bari, I-70124
Bari, Italy}\affiliation{INFN, Bari, Italy}

\date{\today}

\begin{abstract}Inhomogeneous superconductivity arises
when the  species participating in the pairing phenomenon have
different Fermi surfaces with a large enough separation. In these
conditions it could be more favorable for  each of the pairing
fermions to stay close to its Fermi surface and, differently from
the usual BCS state,  for the Cooper pair to have a non zero total
momentum. For this reason in this state the gap varies in space,
the ground state is inhomogeneous and a crystalline structure
might be formed. This situation was considered for the first time
by Fulde, Ferrell, Larkin and Ovchinnikov, and the corresponding
state is called LOFF.  The spontaneous breaking of the space
symmetries in the vacuum state is a characteristic feature of this
phase and  is associated to the presence of long wave-length
excitations of zero mass. The situation described here is of
interest both in solid state and in elementary particle physics,
in particular in Quantum Chromo-Dynamics at high density and small
temperature. In this review we present the theoretical approach to
the LOFF state and its phenomenological applications using the
language of the effective field theories.
\end{abstract}
\maketitle \tableofcontents
\section{Introduction}

   Superconductivity is one of the most fascinating chapters of
   modern physics. It has been a continuous source of inspiration for different
   realms of physics and has shown a tremendous
   capacity of cross-fertilization, to say nothing of its numerous  technological applications.
   This review is devoted to a less known chapter of its history, i.e.
   inhomogeneous superconductivity, which arises when the main property of the
   superconductor is not uniform in space. Before giving a more accurate definition of
   this phenomenon let us however briefly sketch the historical path leading to it.
   Two were the main steps in the discovery of superconductivity. The
   former
   was due to Kamerlingh Onnes \cite{Kamerlingh:1911ab} who discovered  that
   the electrical resistance of various metals, e. g. mercury, lead, tin and many others,
   disappeared when the
   temperature was lowered below some critical
   value $T_c$. The actual values of $T_c$ varied with the metal, but
   they were all of the order of a few K, or at most of the order of tenths of a K.
   Subsequently perfect diamagnetism in superconductors was discovered  \cite{meissner:1933ab}.
     This property not only implies that
   magnetic fields are excluded from superconductors, but also
   that any field originally present in the metal is expelled from it
  when lowering the temperature below its critical
   value. These two features were captured in the equations
   proposed by the brothers F. and H. London \cite{london:1935cd}
   who first realized the quantum character of the phenomenon.
   The decade starting in 1950 was the stage of two major theoretical
breakthroughs. First,
   Ginzburg and Landau (GL) created a theory describing the transition
   between the superconducting and the normal phases \cite{ginzburg:1950xz}.
    It can be noted that,
     when it appeared, the GL
   theory looked rather phenomenological and was not
   really appreciated in the western literature. Seven  years later
   Bardeen, Cooper and Schrieffer (BCS) created the microscopic theory
that bears their name
    \cite{Bardeen:1957kj}. Their theory was based on the fundamental
    theorem \cite{cooper:1956fz}, which states that,
     for a system of many electrons at small $T$,
    any weak
   attraction, no matter how small it is, can bind two electrons together,
   forming the so called Cooper pair. Subsequently in \cite{gorkov:1959hy}  it was realized that
   the GL theory was equivalent to the BCS theory around the critical
   point, and this result vindicated the GL theory as a masterpiece in physics.
   Furthermore Gor'kov proved that the fundamental quantities of the
two theories, i.e.  the BCS parameter gap $\Delta$ and the GL
wavefunction $\psi$, were related
   by a proportionality constant and $\psi$ can be thought of as the
Cooper pair  wavefunction in the center-of-mass frame. In
   a sense, the GL theory was the prototype of the modern effective
   theories; in spite of its limitation to the phase transition it
   has a larger field of application, as shown for example by its use in
    the inhomogeneous cases,
    when the gap is not uniform in space.
     Another remarkable advance in these years
     was the Abrikosov's theory of the type II
   superconductors \cite{abrikosov:1957jk}, a class of superconductors
   allowing a penetration of the magnetic field, within certain critical
   values.

   The inspiring power of superconductivity became soon evident in the
field of elementary   particle physics. Two pioneering papers
\cite{Nambu:1961tp,Nambu:1961fr} introduced the idea
   of generating elementary particle masses through the mechanism of
   dynamical symmetry breaking  suggested by superconductivity. This
   idea was so fruitful that it eventually was a crucial ingredient of  the
   Standard Model (SM) of the elementary particles, where the masses are
   generated by the formation of the Higgs condensate much in the same way as
   superconductivity originates from the presence of a gap. Furthermore,
   the Meissner effect, which is characterized by a penetration
   length, is the origin, in the elementary particle physics
   language, of the masses of the gauge vector bosons. These masses
   are nothing but the inverse of the penetration length.

   With the advent of QCD it was early realized that at high density,
   due to the asymptotic freedom
property   \cite{gross:1973ab,politzer:1973ab} and to the
existence of an attractive channel in the color interaction,
    diquark condensates might  be   formed
   \cite{collins:1975ab,Barrois:1977xd,Frautschi:1978rz,Bailin:1984bm}.
   Since these condensates break the color gauge symmetry, the
   subject took the name of color superconductivity. However,
   only in the last few years this has become a very active field of
   research; these developments are
   reviewed in   \cite{Rajagopal:2000wf,Hong:2000ck,Alford:2001dt,Hsu:2000sy,
   Nardulli:2002ma}. It should also be  noted that color
   superconductivity might have implications in
   astrophysics because  for some  compact stars, e.g. pulsars,
the baryon densities necessary for color superconductivity can
probably be  reached.

   Superconductivity in metals was the stage of another breakthrough in  the 1980s
   with the discovery of high $T_c$ superconductors. As we
   anticipated, however, the main subject of this review is a different and separate
development of
   superconductivity, which took place in 1964. It originates
 in high-field superconductors where a strong magnetic field,
   coupled to the spins of the conduction electrons, gives rise to a
   separation of the Fermi surfaces corresponding to electrons with
   opposite spins. If the separation is too high the pairing is
   destroyed and there is a transition (first-order at small
   temperature) from the superconducting state to the normal one.
   In two separate and contemporary papers, \cite{LO} and  \cite{FF}, it
   was   shown that a new state could be
   formed, close to the transition line. This state that hereafter
    will be called LOFF\footnote{In the literature the LOFF state is also
    known as the FFLO state.} has the feature
   of exhibiting an order parameter, or a gap, which is not a
   constant, but has a space variation whose typical wavelength is  of the
   order of the inverse of the difference in the Fermi energies of the  pairing electrons.
   The space modulation of the gap arises because the electron pair has
   non zero total momentum and it is a rather peculiar phenomenon that
leads to the
   possibility of a non uniform or anisotropic ground state, breaking
   translational and rotational symmetries. It has been also
   conjectured that the typical inhomogeneous ground state might have a
periodic
   or, in other words, a crystalline structure. For this reason other names of this
phenomenon are inhomogeneous or anisotropic or crystalline
superconductivity.

  Inhomogeneous  superconductivity in metals  has been the object of intense
   experimental investigations especially in the last decade; for
   reasons to be discussed below the experimental research has aimed
to rather unconventional superconductors, such as heavy fermion
superconductors, quasi-two dimensional layered organic
superconductors or high $T_c$ superconductors. While different
from the original LOFF proposal, these investigations still aim
 to a superconducting state characterized by non zero
total momentum of the Cooper pair and space modulation of its
wavefunction. At the moment they represent the main possibility
to discover the LOFF state in condensed matter physics.

   Quite recently it has been  also realized that at moderate density
    the mass difference
   between the strange and the up and down quarks at the weak
   equilibrium and/or color and electric neutrality  lead to a difference in the
   Fermi momenta, which renders in principle the  LOFF state
possible in color
   superconductivity   \cite{Alford:2000ze}. The same authors have
pointed out that this
   phenomenon might have some relevance in explaining the sudden
   variations of the rotation period of the pulsars (glitches).

The main aim of this review is to present ideas and methods of
the two main roads to inhomogeneous superconductivity, i.e. the
condensed matter and the QCD ways. Our approach will be mainly
theoretical and the discussion of phenomenological consequences
will be limited, first because we lack the necessary skills and
second because the theory of the LOFF superconductivity is up to
now much more advanced than experiment and its main
phenomenological implications belong to the future. For this
reason we will give large room to the theoretical foundations of
inhomogeneous superconductivity and will present only a summary
of experimental researches. Our scope is to show the similarities
of different physical situations and to present a formalism as
unified as possible. This not only to prove once again the
cross-fertilization power of superconductivity, but also to
expose experts in the two fields to results that may be easily
transferrable from one sector to the other. Moreover, by
presenting the LOFF phenomenon in a unified formalism, this
review can contribute, we hope, to establish a common language.
 To this end we discuss the
   LOFF state both in solid state and in QCD physics starting with
 Nambu Gor'kov (NG) equations. For the solid
   state part they will be derived by the effective theory of the
relevant degrees
 of freedom at the Fermi surface and  in the QCD sector by the so called
High Density
   Effective Theory (HDET) that, as we shall see, leads to equations
   of motion which coincide with the NG equations. In this way one is
   able to get in touch with a dictionary allowing to switch easily
   from one field to the other.

   The plan of this review is as follows. In Section
   \ref{section2} we start describing the general formalism, based
   on NG equations \cite{gorkov:1959hy,Nambu:1960cs}. As shown by
   \cite{Polchinski:1992ed} using the Renormalization Group approach,
   the excitations at the Fermi surface can be described by an effective
   field theory. Its    equations of motion are exactly the NG equations
   of ordinary (homogeneous) superconductivity.
   We will then apply this formalism to fermions
   with different Fermi surfaces. The difference can be due  to a magnetic
   field producing an energy splitting between spin up and spin down
   electrons, or, as in QCD, to a difference in the chemical potential
   originating from weak equilibrium, or color and electric neutrality, or mass difference between
   the pairing fermions. We will discuss the
   circumstances leading, in these cases, to inhomogeneous
   superconductivity. The Ginzburg Landau
   expansion can be used, as already mentioned,
   for the description of the inhomogeneous phase. It
    will be discussed in Section \ref{ginzburg_landau}, both at zero temperature and close to the tricritical
   point. The $T=0$ case is more interesting for QCD applications while
   the finite temperature case might be relevant in condensed
   matter.     In Section \ref{colorsuperconductivity} we will switch to QCD. We
   will first give a brief introduction to color
   superconductivity and then a description of the effective
   lagrangian for quarks at zero temperature near to the
   Fermi surface. We will also discuss more specifically the LOFF
   case for QCD with two  massless flavors.
   Since in the LOFF phase both translational and rotational
   symmetries are spontaneously broken, the Goldstone theorem
   requires the presence in the physical spectrum of
   long wave-length, gapless, excitations (phonons). In Section
   \ref{phonons} we discuss the phonon effective lagrangians  for two crystalline structures, i.e. the single plane wave and  the
   cubic structure. We will limit our presentation to the QCD case, though
   the presence of these excitations is obviously general.
   We will also discuss the gluon propagation inside these two
   crystalline media.
   In Section \ref{phenomenology} we will discuss the possible
   phenomenological applications of the LOFF phase. This discussion
   will go from  strongly type II superconductor to two-dimensional
   structures for condensed matter. For hadronic matter
   we will discuss  applications both in nuclear physics and in
   QCD, with particular emphasis on the physics of glitches in
   pulsars.

Let us conclude this introduction by  apologizing to the many
authors whose work is not reviewed here in depth. Space limits
forced us to sacrifice a more detailed exposition; the extensive
bibliography at the end should help to excuse, we hope, this
defect.

\section{The general setting}\label{section2}
In this Section we give a pedagogical introduction to
inhomogeneous superconductivity. We begin by reviewing
homogeneous superconductivity by a field theory with effective
Nambu-Gor'kov spin 1/2 fields describing quasi-particles. The
effective field theory considers only the relevant degrees of
freedom in the limit of small temperatures and high chemical
potential; they are the modes in a shell around the Fermi
surface. The dominant coupling in this limit is the four fermion
interaction as first introduced in the BCS model. The dominance
of this coupling can be also proved in a modern language by using
the renormalization group approach
\cite{Polchinski:1992ed,Shankar:1994pf,benfatto:1990ab}, which
shows that the BCS coupling is marginal and therefore, in absence
of relevant couplings, it can dominate over other irrelevant
couplings and produce the phenomenon of superconductivity.

After having derived the Nambu-Gor'kov equations and the gap
equation in Subsection \ref{NGor}, we discuss the case of
homogeneous superconductor in Section
\ref{homogeneous_superconductor} and analyze its phase diagram in
Section \ref{phasediagramhomogeneous}. We assume from the very
beginning that the two species participating in the Cooper
pairing have different chemical potentials, as this is the
necessary situation for the LOFF state. In Section \ref{section4}
we discuss the case of anisotropic superconductivity.
 In Section \ref{c5}
we will show that for appropriate values of the difference in
chemical potentials an anisotropic modulated gap $\Delta({\bf
r})\propto\exp(i{\bf q\cdot r})$ leads to a state that is
energetically favored in comparison to both the BCS and the
normal non superconducting states. This was the state first
discussed in \cite{FF}.

\subsection{Nambu-Gor'kov
equations\label{NGor}}
 To start with we consider, at $T=0$, a fermion liquid  formed by
two species, that we call $u$ and $d$, having different Fermi
energies. In the electron superconductivity, as in the original
LOFF papers \cite{LO,FF}, the species are the electron spin {up}
and {\rm down} states, but our formalism is general and will be
applied later to the case where the fermion forming the Cooper
pair are two quarks with different flavors.  In
 superconducting materials the difference of chemical potentials can be
 produced by the presence of paramagnetic impurities. All these cases
give rise to an effective exchange interaction that can described
by adding the following term to the hamiltonian\be
H_{exch}=-\delta\mu\,\psi^\dagger
 \sigma_3\psi\,.\label{exch}\ee
In the case of electron superconductivity $\delta\mu$ is
proportional to the magnetic field and  the effect of
(\ref{exch})
 is to change the chemical potentials of the two species:
\be \mu_u=\mu+\delta\mu,~~~~~ \mu_d=\mu-\delta\mu\
.\label{eq:2.2}\ee Adopting a BCS interaction, the
action can be written as follows \bea A&=&A_0+A_{BCS}\ , \\
A_0&=& \int dt\, \frac{d{\bf p}}{(2\pi)^3}\ \psi^\dagger({\bf
p})\left(i\partial_t-E({\bf p})+\mu+\delta\mu\sigma_3
\right)\psi({\bf p})\ , \label{a0}\\&&\cr A_{BCS}&=&
\frac{g}2\int dt \prod_{k=1}^4\frac{d{\bf p}_k}{(2\pi)^3}\left(
\psi^\dagger({\bf p}_1)\psi({\bf
p}_4)\right)\,\left(\psi^\dagger({\bf p}_2)\psi({\bf
p}_3)\right)\times(2\pi)^3 \, \delta({\bf p}_1+{\bf p}_2-{\bf
p}_3-{\bf p}_4)\,.\label{abcs}\eea Here and below, unless
explicitly stated, $\psi({\bf p})$  denotes the 3D Fourier
transform of the Pauli spinor $\psi({\bf r},t)$, i.e. $\psi({\bf
p})\equiv\psi_\sigma({\bf p},t)$. For non relativistic particles
the functional dependence of the energy would be $\dd E({\bf
p})={\bf p}^{\,2}/2m$, but we prefer to leave it in the more
general form (\ref{a0}).

The BCS interaction (\ref{abcs}) can be written as follows \be
A_{BCS}=A_{cond}+A_{int}\,,\ee with \bea A_{cond}&=& -\frac{g
}4\int dt\,  \prod_{k=1}^4\frac{d{\bf p}_k}{(2\pi)^3}
\Big[\,\tilde\Xi ({\bf p}_3,\,{\bf p}_4)\psi^\dagger({\bf p}_1)
C\psi^\dagger({\bf p}_2) \cr&-& \tilde\Xi^*({\bf p}_1,\,{\bf
p}_2) \psi({\bf p}_3) C\psi({\bf p}_4)\Big](2\pi)^3 \,
\delta({\bf p}_1+{\bf p}_2-{\bf p}_3-{\bf p}_4) \ ,\cr
 A_{int}&=&-\frac{g}4
\int dt\, \prod_{k=1}^4\frac{d{\bf p}_k}{(2\pi)^3}
 \left[\psi^\dagger({\bf p}_1)
C\psi^\dagger({\bf p}_2)+\tilde\Xi^* ({\bf p}_1,\,{\bf
p}_2)\right]\times\cr&\times&\left[ \psi({\bf p}_3) C\psi({\bf
p}_4)-\tilde\Xi ({\bf p}_3,\,{\bf p}_4)\right](2\pi)^3 \,
\delta({\bf p}_1+{\bf p}_2-{\bf p}_3-{\bf p}_4)\,,
\label{interaction}\eea where $C=i\sigma_2$ and\be \tilde\Xi({\bf
p},\,{\bf p}^{\prime})=<\psi({\bf p}) C\psi({\bf p}^{\prime})>\,
.\ee In the mean field approximation the interaction term can be
neglected while the gap term $A_{cond}$ is added to $A_0$. Note
that the spin 0 condensate $\tilde\Xi({\bf p},\,{\bf
p}^{\prime})$ is simply related to the condensate wave function
  \be \Xi({\bf r})= <\psi({\bf r},t) C\psi({\bf r},t)>\label{deltax}
\ee  by the formula \be \Xi({\bf r})=\int\frac{d{\bf
p}}{(2\pi)^3}\frac{d{\bf p}^{\prime}}{(2\pi)^3}\,e^{-i({\bf
p}+{\bf p}^{\prime})\cdot{\bf r}}\
 \tilde\Xi({\bf p},\,{\bf p}^{\prime})\ .\ee
  In general the condensate wavefunction  can depend on ${\bf r}$; only for homogeneous
materials it does not depend on the space coordinates; therefore
in this case $ \tilde\Xi({\bf p},\,{\bf p}^{\prime})$ is
proportional to $\delta({\bf p}+{\bf p}^\prime)$.

In order to write down the Nambu-Gor'kov (NG) equations we define
 the NG spinor \be \chi ({\bf p})=\frac{1}{\sqrt 2}
 \left(\matrix{\psi({\bf p}) \cr \psi^c(-{\bf p})}\right)
 \,,\label{ngspinor}\ee where we have introduced the
charge-conjugate field
 \be \psi^c=C\psi^\dagger\ .\ee
 We also define
\be\Delta({\bf p},-{\bf p}^{\prime})=\,\frac{g}2
\int\,\frac{d{\bf p}^{\prime\prime}}{(2\pi)^6}\tilde\Xi({\bf
p}^{\prime\prime}, {\bf p}+{\bf p}^{\prime} -{\bf
p}^{\prime\prime})\ .\ee

The free action can be therefore written as follows:
 \be A=\int dt\,
\frac{d{\bf p}}{(2\pi)^3}
 \frac{d{\bf p^\prime}}{(2\pi)^3}\,
\chi^\dagger({\bf p})\, S^{-1}({\bf p},\,{\bf p^\prime} )
\chi({\bf p^\prime})\\ ,\ee
 with
\bea S^{-1}({\bf p},\,{\bf p}^{\prime})=(2\pi)^3
\left(\matrix{(i\partial_t-\xi_{{\bf p}}+\delta\mu\sigma_3)
 \delta({\bf p}-{\bf p}^{\prime})& -
 \Delta({\bf p},{\bf p^\prime})\cr
 -\Delta^*({\bf p},{\bf p^\prime})&
(i\partial_t+\xi_{{\bf p}}+\delta\mu\sigma_3)\delta({\bf p}-{\bf
p}^{\prime})}\right)\,.\label{smenouno}\eea
 Here
 \be \xi_{{\bf p}}=E({\bf p})-\mu\approx {\bf v}_F
 \cdot({\bf p}-{\bf p}_F)\,,\label{vf}\ee
 where
 \be {\bf v}_F=\frac{\partial E({\bf p})}{\partial {\bf p}}
 \Big|_{{\bf p}\, =\,{\bf p}_F}\ee
 is the Fermi velocity. We have used the fact that we are considering
 only degrees of freedom near the Fermi surface, i.e. \be
 p_F-\delta<p <p_F+\delta\ ,\ee
 where
$\delta$ is the ultraviolet cutoff, of the order of the Debye
frequency. In particular in the non relativistic case
 \bea
\xi_{{\bf p}}=\frac{{{\bf
p}\,}^2}{2m}-\frac{p_F^{\,2}}{2m},~~~{\bf v}_F=\frac{ {\bf
p}_F}{m}\,. \label{diciannove}\eea $S^{-1}$ in (\ref{smenouno})
is the 3D Fourier transform of the inverse propagator. We can
make explicit the energy dependence  by Fourier transforming the
time variable as well. In this way we get for the inverse
propagator written as an operator: \be S^{-1}= \left(\matrix{
{\mathbf{(G_0^+)^{-1}}} & -{\mathbf\Delta}\cr -{\mathbf\Delta^*}
& -{\mathbf{(G_0^-)^{-1}}}}\right)\,, \ee and \bea [{\bf
G_0^+]^{-1}}&=& E-\xi_{{\bf P}}+\delta\mu\sigma_3
+\,i\,\epsilon\, {\rm sign\,} E\,,\cr&&\cr {\bf
[G_0^-]^{-1}}&=&-E-\xi_{{\bf P}}-\delta\mu\sigma_3
-\,i\,\epsilon\, {\rm sign\,} E \,,\eea with $\epsilon=0^+$ and
${\bf P}$ the momentum operator. The $i\epsilon$ prescription is
nothing but the usual one for the Feynman propagator, that is
forward propagation in time for the energy positive solutions and
backward propagation for the negative energy solutions. As for
the NG propagator $S$, one gets \be S=\left(\matrix{ {\bf G}&-
{\bf \tilde F} \cr- {\bf F}&{\bf \tilde G }} \right)\ . \ee $S$
has both spin, $\sigma,\,\sigma^\prime$, and $a,b$ NG indices,
i.e. $S^{ab}_{\sigma\sigma^\prime}$\footnote{We note that the
presence of the factor $1/{\sqrt 2}$ in (\ref{ngspinor}) implies
an extra factor of 2 in the
 propagator:
 $
S(x,x^\prime)=2\,<T\left\{\chi(x)\chi^\dagger(x^\prime)\right\}>$,
as it can be seen considering e.g.  the matrix element $S^{11}$:
$<T\left\{\psi(x)\psi^\dagger(x^\prime)\right\}>=
\left(i\partial_t-\xi_{-i\vec \nabla}-\delta\mu\sigma_3
\right)^{-1}\delta(x-x^\prime),$ with  $(x\equiv (t,{\bf r}))$.}.
The NG equations in compact form are
 \be S^{-1}S=1\ ,\ee or, explicitly,
 \bea [{\bf G_0^+}]^{-1}{\bf G}+{\bf \Delta F}&=&{\bf 1}\ ,\cr
  -[{\bf G_0^-}]^{-1} {\bf F}
 +\bf{\Delta^* G}&=&{\bf 0}\label{NG}\ .\eea
 Note that we will use
 \be
 <{\bf r}\,|{\bf \Delta}|{\bf r}^{\,\prime}>=
\frac{g}2\,\Xi({\bf r})\,\delta({\bf r}-{\bf
r}^{\,\prime})=\Delta({\bf r})\,\delta({\bf r}-{\bf
r}^{\,\prime})\ ,\label{dra}\ee or \be <{\bf p}\,|{\bf
\Delta}|{\bf p}^{\,\prime}>= \Delta
 ({\bf p},{\bf p}^{\,\prime})\ee
 depending on our choice of the coordinate or momenta representation.
 The
formal solution of the system (\ref{NG}) is \bea {\bf F}&=&{\bf
G_0^-\Delta^*G}\ ,\cr
 {\bf G}&=&{\bf G_0^+}-{\bf G_0^+\Delta F}\ ,\eea so that ${\bf F}$
satisfies the equation \be {\bf
F=G_0^-\Delta^*\left(G_0^+-G_0^+\Delta F\right)}\ \label{18}\ee
and is therefore given by \be {\bf F=\frac{1}{ \Delta^*
[G_0^+]^{-1} [\Delta^{*}]^{-1} [G_0^-]^{-1} \,+\, \Delta^*
\Delta}\,\Delta^*} \label{18bis}\ .\ee

In the configuration  space the NG Eqs. (\ref{NG}) are as follows
\bea \left(E-E(-i\bm{\nabla})+\mu +\delta\mu\sigma_3
\right)G({\bf r},{\bf r}^{\,\prime},E)+\,\Delta({\bf r}) F({\bf
r},{\bf r}^{\,\prime},E) &=&\delta({\bf r}-{\bf r}^{\,\prime})\
,\cr \left(-E-E(-i\bm{\nabla})+\mu -\delta\mu\sigma_3
\right)F({\bf r},{\bf r}^{\,\prime},E)- \Delta^*({\bf r}) G ({\bf
r},{\bf r}^{\,\prime},E)&=&0\,.\eea

The gap equation at $T=0$ is the following consistency condition
\be \Delta^*({\bf r})=-i\ \frac{g}2\int\frac{dE}{2\pi}\, {\text
Tr} F({\bf r},{\bf r},E)\,,\label{consistency}\ee where $F$ is
given by (\ref{18bis}). To derive the gap equation  we observe
that \bea \Delta^*({\bf r})&=&\,\frac{g}2\, \Xi^*({\bf
r})=\,\frac{g}2\,\int\frac{d{\bf p}_1}{(2\pi)^3} \frac{d{\bf
p}_2}{(2\pi)^3}\ e^{i({\bf p}_1+{\bf p}_2)\cdot{\bf r}}\
\tilde\Xi^*({\bf p}_1,\,{\bf p}_2)\cr &=&-\,\frac{g
}2\int\frac{dE}{2\pi}\frac{d{\bf p}_1}{(2\pi)^3}\frac{d{\bf
p}_2}{(2\pi)^3}\ e^{i({\bf p}_1+{\bf p}_2)\cdot{\bf r}}\
<\psi^\dagger({\bf p}_1,E)\psi^c({\bf p}_2,E)>\cr
&=&+\,i\,\frac{g}2\sum_\sigma\int\frac{dE}{2\pi}\frac{d{\bf
p}_1}{(2\pi)^3}\frac{d{\bf p}_2}{(2\pi)^3}e^{i({\bf p}_1-{\bf
p}_2)\cdot {\bf r}} S^{21}_{\sigma\sigma} ({\bf p}_2,{\bf
p}_1)\cr &=&+\,i\,\frac{g
}2\,\sum_\sigma\int\frac{dE}{2\pi}S^{21}_{\sigma\sigma}({\bf
r},{\bf r})\ , \eea which gives (\ref{consistency}).

 At finite temperature, introducing the Matsubara frequencies
 $\omega_n=(2n+1)\pi T$, the gap equation reads \be \Delta^*({\bf
 r})\,=\,\frac{g}2\, T\,\sum_{n=-\infty}^{+\infty} {\text
 Tr}F({\bf r},{\bf r},E)\Big|_{E=i\omega_n}\label{gap2}\ .\ee
\subsection{Homogeneous superconductors
\label{homogeneous_superconductor}}

It is useful to specialize these relations to the case of
homogeneous materials. In this case
we have \bea \Xi({\bf r})&=&{\rm const.}\equiv\frac{2\Delta}{g}\ ,\label{xir}\\
\tilde\Xi({\bf p}_1,\,{\bf
p}_2)&=&\frac{2\Delta}{g}\frac{\pi^2}{p_F^2\delta}\,(2\pi)^3\delta({\bf
p}_1+{\bf p}_2)\ \label{xip} .\eea
 Therefore one gets \be
\Delta({\bf p}_1,{\bf p}_2)= \Delta\, \delta({\bf p}_1-{\bf p}_2)
\label{deltaP}\ee and from (\ref{dra}) and (\ref{xir}) \be
\Delta({\bf r})=\Delta^*({\bf r})=\Delta\ .\label{homogeneous}\ee
Therefore $F({\bf r},{\bf r},E)$ is independent of ${\bf r}$ and,
from Eq. (\ref{18bis}), one gets \be TrF({\bf r},{\bf
r},E)=-2\,\Delta\int\frac{d^3p}{(2\pi)^3}\frac{1}
{(E-\delta\mu)^2-\xi_{{\bf p}}^2-\Delta^2}\label{trf} \ee which
gives the gap equation at $T= 0$:
 \be \Delta=i\, g\Delta\int\frac{dE}{2\pi}\frac{d^3p}
 {(2\pi)^3}\frac{1}{(E-\delta\mu)^2-\xi_{{\bf p}}^2-\Delta^2}\,,
\label{gapt0}\ee and at $T\neq 0$: \be \Delta\,=\, \,g
T\,\sum_{n=-\infty}^{+\infty}\int\frac{d^3p}
{(2\pi)^3}\frac{\Delta}{(\omega_n+i\delta\mu)^2+ \epsilon({\bf
p},\Delta)^2}\,,\label{gaptne0}\ee with \be \epsilon({\bf
p},\Delta)=\sqrt{\Delta^2+\xi^2_{{\bf p}}}\, . \ee We now use the
identity \be \frac 1 2\left[1-n_u-n_d \right] =\epsilon({\bf
p},\Delta) T
\sum_{n=-\infty}^{+\infty}\frac{1}{(\omega_n+i\delta\mu)^2+\epsilon^2({\bf
p},\Delta)}\,, \label{28}\ee where \be n_u({\bf p})=\frac{1}{\dd
e^{(\epsilon+\delta\mu)/T}+1}\ ,~~~~~~~~~~~ n_d({\bf
p}=\frac{1}{\dd e^{(\epsilon-\delta\mu)/T}+1} \, .\ee The gap
equation can be therefore written as
\be\Delta=\frac{g\,\Delta}2\,\int \frac{d^3p}{(2\pi)^3}\,
\frac{1}{\epsilon({\bf p},\Delta)}\,\left(1-n_u({\bf p})-n_d({\bf
p})\right)\,.\label{gap1} \ee In the Landau theory of the Fermi
liquid $n_u,\,n_d$ are interpreted as
 the equilibrium distributions for the quasiparticles of type $u,d$.
It can be noted  that the last two terms act as blocking factors,
reducing the phase space, and producing eventually $\Delta\to 0$
when  $T$ reaches a critical value $T_c$ (see below).

Before considering the solutions of the gap equations in the
general case let us first consider the case $\delta\mu=0$; the
corresponding gap is denoted $\Delta_0$. At $T=0$ there is no
reduction of the phase space and  the gap equation becomes \be
1=\frac{g}2\,\int \frac{d^3p}{(2\pi)^3}\, \frac{1}{\epsilon({\bf
p},\Delta_0)}\label{gap1bis}\ , \ee whose solution  is (we have
assumed $d^3p=p_F^2 dp d\Omega$)\be
\Delta_0=\frac{\delta}{\sinh\frac 2{g\rho}}\label{gap0}\ .\ee
Here \be \rho=\frac{p_F^2}{\pi^2 v_F}\ee is the density of states
and we have used  $\xi_{{\bf p}}\approx v_F(p-p_F)$, see Eqs.
(\ref{vf})-(\ref{diciannove}). In the weak coupling limit
(\ref{gap0}) gives \be \Delta_0=2\delta\, e^{-2/\rho
g}\label{bcsgap}\ .\ee

Let us now consider the case $\delta\mu\neq 0$.
 By (\ref{gap1}) the gap equation is written as
 \be-1+\frac{g}2\int\frac{d^3p}{(2\pi)^3}\frac{1}{\epsilon}
 =\frac{g}2\int\frac{d^3p}{(2\pi)^3}\,\frac{n_u+n_d}{\epsilon}\ .\ee
Using the gap equation for the BCS superconductor, the l.h.s can
be written, in the weak coupling limit, as \be {\text
{l.h.s}}=\frac{g\rho}{2}\ln\frac{\Delta_0}{\Delta}\ ,\ee where we
got rid of the cutoff $\delta$ by using $\Delta_0$, the  gap at
$\delta\mu=0$ and $T=0$. Let us now evaluate the r.h.s. at $T=0$.
 We get \be
\text {r.h.s.}\Big|_{T=0}\,=\,\frac{g\rho}{2}
\int_{0}^{\delta}\frac{d\xi_{{\bf p}}}{\epsilon}
\left[\theta(-\epsilon-\delta\mu)+\theta(-\epsilon+\delta\mu
)\right]\,. \label{Gap}\ee The gap equation at $T=0$ can
therefore be written as follows:
\be\ln\frac{\Delta_0}{\Delta}=\theta(\delta\mu-\Delta) {\rm
arcsinh}
\frac{\sqrt{\delta\mu^2-\Delta^2}}{\Delta}\,,\label{gap5}\ee i.e.
\be
\ln\frac{\Delta_0}{\delta\mu+\sqrt{\delta\mu^2-\Delta^2}}=0\,.\label{54}\ee
One can immediately see that there are no solutions for
$\delta\mu>\Delta_0$. For $\delta\mu\le \Delta_0$ one has two
solutions.
\bea a)~~~\,\,\Delta&=&\Delta_0\,,\label{d2b}\\
b)~~~ \Delta^2&=&2\,\delta\mu\Delta_0-\Delta_0^2\,.\label{d1b}
\eea The first arises since for $\Delta=\Delta_0$ the l.h.s. of
the Eq. (\ref{gap5}) is zero. But  since we may have solutions
only for $\delta\mu\le\Delta_0$ the $\theta$-function in Eq.
(\ref{gap5}) makes zero also the r.h.s.. The existence of this
solution can also be seen  from Eq. (\ref{gapt0}). In fact in
this equation one can shift the integration variable as follows:
$E\to E+\delta\mu$, getting the result that, in the
superconductive phase, {\it the gap $\Delta$  is independent of}
$\delta\mu$, i.e. $\Delta=\Delta_0$.

To compute the free energy we make use of the theorem saying that
for small variations of an external parameter of the system all
the thermodynamical quantities vary in the same way
\cite{landau1}. We apply this to the grand potential to get \be
\frac{\de\Omega}{\de g}=\Big\langle\frac{\de H}{\de
g}\Big\rangle\,.\ee From the expression of the interaction
hamiltonian (see Eq. (\ref{abcs})) we get immediately  (cfr.
\cite{abrikosov}, cap. 7):  \be \Omega=-\int \frac{dg}{g^2} \int
d\vec x\, |\Delta(\vec x)|^2\ \label{eq:59}. \ee For homogeneous
media
 this gives
\be \frac{ \Omega}V\,=-\,\int \frac{dg}{g^2}\, |\Delta|^2\ .\ee
Using the result (\ref{bcsgap}) one can trade the integration over
the coupling constant $g$ for an integration over $\Delta_0$, the
BCS gap at $\delta\mu=0$, because ${d\Delta_0}/{\Delta_0}=
{2dg}/{\rho g^2}$. Therefore the difference in free energy between
the superconductor and the normal state is (we will use
indifferently the symbol $\Omega$ for the grand potential and its
density $\Omega/V$) \be
\Omega_\Delta-\Omega_0=\,-\,\frac{\rho}{2}\int_{\Delta_f}^{\Delta_0}
\Delta^2\ \frac{d\Delta_0}{\Delta_0}\,. \label{d0bis}\ee Here
$\Delta_f$ is the value of $\Delta_0$ corresponding to $\Delta=0$.
$\Delta_f=0$ in the case a) of Eq. (\ref{d2b}) and
 $\Delta_f=2\delta\mu$  in the case b) of Eq. (\ref{d1b}); in the latter case
 one sees immediately that $\Omega_\Delta-\Omega_0>0$ because from Eq.
 (\ref{d1b}) it follows that $\Delta_0<2\delta\mu$.  The free energies
for $\delta\mu\neq 0$ corresponding to the cases a), b) above can
be computed substituting (\ref{d2b}) and (\ref{d1b}) in
(\ref{d0bis}). Before doing that let us derive the density of free
energy at $T=0$ and $\delta\mu\neq 0$ in the normal non
superconducting state. Let us start from the very definition of
the grand potential for free spin  1/2 particles\be \Omega_0(0,T)
\,=\,-2{VT}\int \frac{d^3p}{(2\pi)^3}\,{\rm
ln}\left(1+e^{(\mu-\epsilon({\bf p}))/T}\right)\,.\label{eq:61}\ee
Integrating by parts this expression we get, for $T\to 0$, \be
\Omega_0(0) \,=\,-\frac{V}{12\pi^3}\int d\Omega_{{\bf p}}\, p^3\,
d\epsilon\,\theta(\mu-\epsilon)\,.\ee From this expression we can
easily evaluate the grand-potential for two fermions with
different chemical potentials expanding at the first non-trivial
order in $\delta\mu/\mu$. The result is
 \be
\Omega_0 (\delta\mu)\,=\,\Omega_0 (0)-\frac{\delta\mu^2}{2}\rho
\label{om1}\,.\ee Therefore
 from  (\ref{d2b}), (\ref{d1b})  and (\ref{d0bis}) in the cases a), b)
 one has
\bea a)~~~\Omega_\Delta(\delta\mu)&=&\Omega_0(\delta\mu)
-\frac \rho 4 \,(-2\,\delta\mu^2+\Delta_0^2)\label{oma1}\ ,\\
b)~~~ \Omega_\Delta(\delta\mu)&=&\Omega_0(\delta\mu) -\frac \rho 4
\,(-4\,\delta\mu^2+4\delta\mu\Delta_0-\Delta_0^2)\ .\label{omb1}
\eea Comparing  (\ref{oma1}) and (\ref{omb1}) we see that the
solution $a)$ has lower $\Omega$. Therefore, for
$\delta\mu<\Delta_0/\sqrt 2$ the BCS superconductive state is
stable \cite{clogston}. At $\delta\mu=\Delta_0/\sqrt 2$ it becomes
metastable,
 as the normal state has a lower free energy. This transition would be first
 order since the gap does not depend on $\delta\mu$.

\begin{center}
\begin{figure}[ht]
\epsfxsize=9truecm \centerline{\epsffile{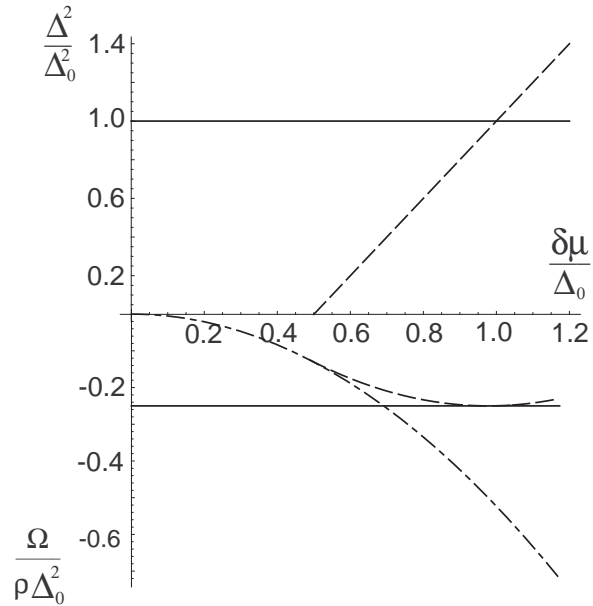}} \noindent
\caption{\it Gap and grand potential as functions of $\delta\mu$
for the two solutions a) and b) discussed in the text, see
Eqs.(\ref{d2b}), (\ref{d1b}) and (\ref{oma1}), (\ref{omb1}).
Upper solid (resp. dashed) line: Gap for solution a) (resp.
solution b)). In the lower part we plot the grand potential for
the solution a) (solid line) and solution b) (dashed line); we
also plot the grand potential for the normal gapless state with
$\delta\mu\neq 0$ (dashed-dotted line). All the grand potentials
are referred to the value $\Omega_0(0)$ (normal state with
$\delta\mu=0$). \label{fig0}}
\end{figure}
\end{center}

The grand potentials for the two cases a) and b) and for the
gapless phase, Eq. (\ref{om1}),  are depicted in Fig. \ref{fig0},
together with the corresponding gaps.

A different proof is obtained integrating the gap equation written
in the form
 \begin{equation}\frac{\partial\Omega}{\partial\Delta}=0\end{equation}
The normalization can be obtained considering the homogeneous case
with $ \delta\mu=0$, when, in the weak coupling limit, from Eqs.
(\ref{eq:59}) and (\ref{bcsgap}) one gets\be \Omega=-\frac{\rho}
4\Delta_0^2\,,\ee see below Eq. (\ref{eq:70b}). In this way one
obtains again the results (\ref{oma1}) and (\ref{omb1}).

This analysis shows that at $\delta\mu=\delta\mu_1=\Delta_0/{\sqrt
2}$ one would pass abruptly from the superconducting ($\Delta\neq
0$) to the normal ($\Delta=0$) phase. However, as we shall discuss
below,
 the real ground state  for $\delta\mu>\delta\mu_1$
 turns out to be an inhomogeneous one, where the assumption (\ref{homogeneous})
of a uniform gap is not justified.

\subsubsection{Phase diagram of homogeneous superconductors
\label{phasediagramhomogeneous}}

We will now study the phase diagram of the homogeneous
superconductor for small values of the gap parameter, which allows
to perform a Ginzburg-Landau expansion of gap equation and grand
potential. In order to perform a complete study we need to expand
the grand-potential  up to the 6$^{\text{th}}$ order in the gap.
As a matter of fact in the plane $(\delta\mu, T)$ there is a first
order transition at $(\delta\mu_1,0)$ and a second order one at
$(0,T_c)$ (the usual BCS second order transition). Therefore we
expect that a second order and a first order lines start from
these points and meet at a tricritical point, which by definition
is the meeting point of a second order and a first order
transition line. A tricritical point is characterized by the
simultaneous vanishing of the $\Delta^2$ and  $\Delta^4$
coefficients in the grand-potential expansion, which is why one
needs to introduce in the grand potential the 6$^{\text{th}}$
order term. For stability reasons the corresponding coefficient
should be positive; if not, one should include also the $\Delta^8$
term.

We consider the  grand potential, as measured from the normal
state, near a second order phase transition \be\Omega=\frac 1 2
\alpha\Delta^2+\frac 1 4\beta\Delta^4+\frac 1
6\gamma\Delta^6\,.\label{eq:potential1}\ee Minimization gives the
gap equation: \be
\alpha\Delta+\beta\Delta^3+\gamma\Delta^5=0\,.\ee Expanding Eq.
(\ref{gaptne0}) up to the $5^{th}$ order in $\Delta$ and
comparing with the previous equation one determines the
coefficients $\alpha$, $\beta$ and $\gamma$ up to a normalization
constant. One gets \be \Delta=2\,g\,\rho\,T
\,Re\,\sum_{n=0}^{\infty}\int_0^\delta
d\xi\left[\frac{\Delta}{(\bar\omega_n^2+\xi^2)}-
\frac{\Delta^3}{(\bar\omega_n^2+\xi^2)^2}+
\frac{\Delta^5}{(\bar\omega_n^2+\xi^2)^3} +\cdots\,\right]\,,
\label{eq:gap_expan}\ee
with\be\bar\omega_n=\omega_n+i\delta\mu=(2n+1)\pi
T+i\delta\mu\,.\label{eq:70}\ee The grand potential can be
obtained, up to a normalization factor, integrating in $\Delta$
the gap equation. The normalization can be obtained by the simple
BCS case, considering the grand potential as obtained, in the
weak coupling limit, from Eqs. (\ref{eq:59}) and (\ref{bcsgap})
\be \Omega=-\frac{\rho} 4\Delta_0^2\,.\label{eq:70b}\ee The same
result can be obtained multiplying the gap equation
(\ref{gap1bis}) by $\Delta_0$ and integrating the result provided
that we multiply it by the factor $2/g$, which fixes the
normalization. Therefore \bea \alpha&=&\frac 2 g\left(1-2\,
g\,\rho\,T\,
Re\sum_{n=0}^\infty\int_0^\delta\frac{d\xi}{(\bar\omega_n^2+\xi^2)}\right)\,,
\label{eq:alpha}\\
\beta&=&4\rho\,T\,Re\sum_{n=0}^\infty
\int_0^\infty\frac{d\xi}{(\bar\omega_n^2+\xi^2)^2}\,,\label{eq:beta}\\
\gamma&=&-4\rho\,T\,Re\sum_{n=0}^\infty
\int_0^\infty\frac{d\xi}{(\bar\omega_n^2+\xi^2)^3}\,.\label{eq:gamma}
\eea In the coefficients $\beta$ and $\gamma$ we have extended
the integration in $\xi$ up to infinity since both the sum and
the integral are convergent. To evaluate $\alpha$ is less
trivial. One can proceed in two different ways.  One can sum over
the Matsubara frequencies and then integrate over $\xi$ or one
can perform the operations in the inverse order. Let us begin
with the former method. We get \be \alpha=\frac 2
g\left[1-\frac{g\,\rho}4\int_0^\delta\frac{d\xi}{\xi}
\left(\tanh\left(\frac{\xi-\mu}{2T}\right)+\tanh\left(\frac{\xi+\mu}{2T}\right)
\right)\right]\,.\ee Performing an integration by part we can
extract the logarithmic divergence in $\delta$. This can be
eliminated using the result (\ref{gap0}) valid for
$\delta\mu=T=0$ in the weak coupling limit \be
1=\frac{g\,\rho}2\log\frac{2\delta}{\Delta_0}\,.\ee  We find \be
\alpha=\rho\left[\log\frac{2T}{\Delta_0}+\frac 14 \int_0^\infty
dx {\rm\, ln } x\, \left(\frac  1{\cosh^2\frac{x+y} 2} +\frac
1{\cosh^2\frac{x-y}2} \right)\right]\,,\label{eq:alpha2}\ee where
\be y=\frac{\delta\mu}{T}\,.\ee Defining \be\log\frac{\Delta_0}{2
T_c(y)}=\frac 14 \int_0^\infty dx {\rm\, ln } x\, \left(\frac
1{\cosh^2\frac{x+y} 2} +\frac 1{\cosh^2\frac{x-y}2}
\right)\,,\label{eq:78}\ee we get \be \alpha(v,t)=\rho\log\frac
t{t_c(v/t)}\,,\label{eq:79}\ee where \be
v=\frac{\delta\mu}{\Delta_0},~~~ t=\frac T{\Delta_0},~~~ t_c
=\frac{T_c}{\Delta_0}\,.\ee Therefore the equation \be
t=t_c(v/t)\ee defines the line of the second order phase
transition. Performing the calculation in the reverse order
brings to a more manageable result for $t_c(y)$
\cite{buzdin:1997ab}. In Eq. (\ref{eq:alpha}) we first integrate
over $\xi$ obtaining a divergent series which can be regulated
cutting the sum at a maximal value of $n$ determined by
\be\omega_N=\delta \Rightarrow N\approx \frac{\delta}{2\pi
T}\,.\ee We obtain \be \alpha=\frac 2
g\left(1-\pi\,g\,\rho\,T\,Re\sum_{n=0}^N\frac{1}{\bar\omega_n}\right)\,.\ee
The sum can be performed in terms of the Euler's function
$\psi(z)$:\be Re\sum_{n=0}^N\frac{1}{\bar\omega_n}=\frac{1}{2\pi
T}\, Re\left[\psi\left(\frac 3 2+i\frac
y{2\pi}+N\right)-\psi\left(\frac 1 2+i\frac
y{2\pi}\right)\right]\approx \frac 1 {2\pi T}\left(
\log\frac{\delta}{2\pi T}-Re\,\psi\left(\frac 1 2+i\frac
y{2\pi}\right)\right)\,.\ee Eliminating the cutoff as we did
before we get \be \alpha(v,t)=\rho\left(\log(4\pi
t)+Re\,\psi\left(\frac 1 2+i\frac v{2\pi
t}\right)\right)\,.\label{eq:alpha_f}\ee By comparing with Eq.
(\ref{eq:alpha2}) we get the following identity
 \be Re\,\psi\left(\frac 1 2+i\frac y{2\pi}\right)
 =-\log(2\pi)+
\frac 14 \int_0^\infty dx {\rm\, ln } x\, \left(\frac
1{\cosh^2\frac{x+y} 2} +\frac 1{\cosh^2\frac{x-y}2} \right)\,.\ee
The equation (\ref{eq:78}) can be re-written as \be
\log\frac{\Delta_0}{4\pi T_c(y)}=Re\psi\left(\frac 1
2+i\frac{y}{2\pi}\right)\,.\ee In particular at $\delta\mu=0$,
using ($C$ the Euler-Mascheroni constant)\be\psi\left(\frac 1
2\right)=-\log(4\gamma), ~~~~\gamma=e^C,~~~~C=0.5777\dots\,,\ee
we find from Eq.
(\ref{eq:alpha_f})\be\alpha(0,T/\Delta_0)=\log\frac{\pi
T}{\gamma\Delta_0}\,,\ee reproducing the critical temperature for
the BCS case \be T_c=\frac\gamma\pi\Delta_0\approx
0.56693\,\Delta_0\,.\label{eq:90}\ee The other terms in the
expansion of the gap equation are easily evaluated integrating
over $\xi$ and  summing over the Matsubara frequencies. We get
\bea \beta&=&\pi\,\rho\, T\,Re\sum_{n=0}^\infty\frac
1{\bar\omega_n^3}=-\frac\rho{16\,\pi^2\, T^2}
Re\,\psi^{(2)}\left(\frac 1 2+i\frac{\delta\mu}{2\pi
T}\right)\,,\label{eq:beta2}\\
\gamma&=&-\frac 3 4\pi\,\rho\, T\,Re\sum_{n=0}^\infty\frac
1{\bar\omega_n^5}=\frac 3 4\frac\rho{768\,\pi^4 \,T^4}
Re\,\psi^{(4)}\left(\frac 1 2+i\frac{\delta\mu}{2\pi T}\right)\,,
\label{eq:gamma2}\eea where \be\psi^{(n)}(z)=\frac
{d^n}{dz^n}\psi(z)\,.\ee

\begin{figure}[ht]
\centerline{\epsfxsize=11cm\epsfbox{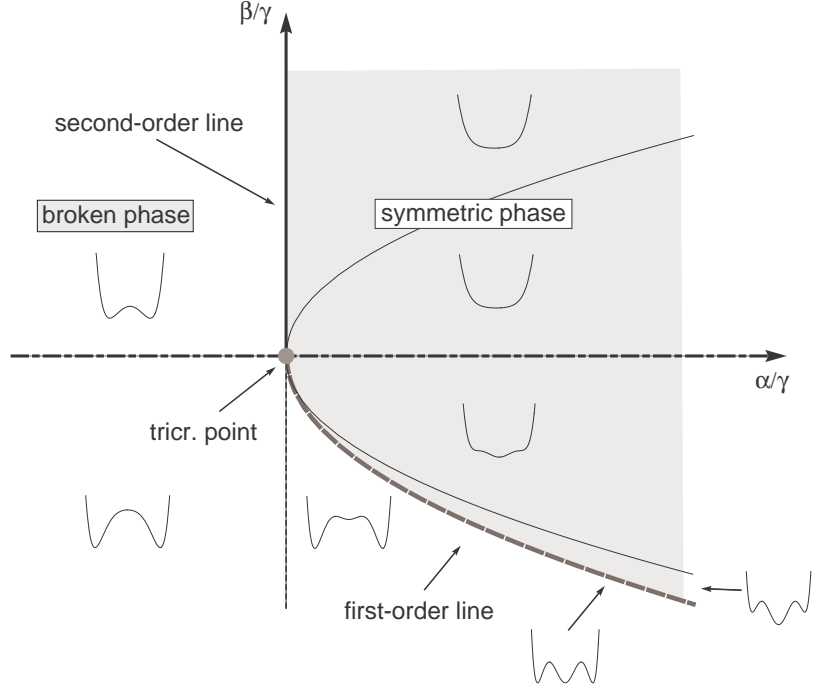}} \caption{{\it
The graph shows the first order and the second order transition
lines for the potential of Eq. (\ref{eq:potential1}). We show the
tricritical point and the regions corresponding to the symmetric
and the broken phase. Also shown is the behavior of the grand
potential in the various regions. The thin solid line is the
locus of the points $\beta^2-4\alpha\gamma=0$. In the interior
region we have $\beta^2-4\alpha\gamma<0$.\label{fig1a}}}
\end{figure}

Let us now briefly review some results on the grand potential in
the GL expansion (\ref{eq:potential1}). We will assume $\gamma>0$
in order to ensure the stability of the potential. The
minimization leads to the solutions \be
\Delta=0\,,\label{eq:94}\ee\be\Delta^2=\Delta_\pm^2=\frac
1{2\gamma}\left(-\beta\pm\sqrt{\beta^2-4\alpha\gamma}
\right)\,.\label{eq:95}\ee The discussion of the minima of
$\Omega$ depends on the signs of the parameters $\alpha$ and
$\beta$. The results are the following:
\begin{enumerate}
\item \fbox{${\alpha>0,~~\beta>0}$}\\\\In this case there is a
single minimum given by (\ref{eq:94}) and the phase is
\underline{symmetric}. \item \fbox{${\alpha>0,~~\beta<0}$}\\\\
Here there are three minima, one is given by (\ref{eq:94}) and
the other two are degenerate minima
at\be\Delta=\pm\Delta_+\,.\label{eq:96}\ee The line along which
the three minima become equal is given
by:\be\Omega(0)=\Omega(\pm\Delta_+)~~\longrightarrow~~
\beta=-4\,\sqrt{\frac{\alpha\gamma}3}\,.\label{eq:101}\ee Along
this line there is a first order transition with a discontinuity
in the gap given by \be\Delta_+^2=-\frac{4\alpha}\beta=-\frac 3
4\frac\beta\gamma\,.\label{eq:102}\ee To the right of the first
order line we have $\Omega(0)<\Omega(\pm\Delta_+)$. It follows
that to the right of this line there is the symmetric phase,
whereas the broken phase is in the left part (see Fig.
\ref{fig1a}). \item \fbox{${\alpha<0,~~\beta>0}$}\\\\In this case
Eq. (\ref{eq:94}) gives a maximum, and there are  two degenerate
minima given by Eq. (\ref{eq:96}).
 Since for $\alpha>0$ the two minima
disappear, it follows that there is a  second order phase
transition along the line $\alpha=0$. This can also be seen by
noticing that going from the broken phase to the symmetric one we
have \be\lim_{\alpha\,\to\, 0} \Delta_+^2=0\,.\ee \item
\fbox{${\alpha<0,~~\beta<0}$}\\\\ The minima and the maximum are
as in the previous case.
\end{enumerate}
Notice also that the solutions $\Delta_\pm$ do not exist in the
region $\beta^2<4\alpha\gamma$.  The situation is summarized in
Fig. \ref{fig1a}. Here we show the behavior of the grand
potential in the different sectors of the plane
$(\alpha/\gamma,\beta/\gamma)$, together with the transition
lines. Notice that in the quadrant $(\alpha>0,\beta<0)$ there are
metastable phases corresponding to non absolute minima. In the
sector included between the line $\beta=-2\sqrt{\alpha/\gamma}$
and the first order transition line the metastable phase is the
broken one, whereas in the region between the first order  and
the $\alpha=0$ lines the metastable phase is the symmetric one.

\begin{figure}[ht]
\centerline{\epsfxsize=12cm\epsfbox{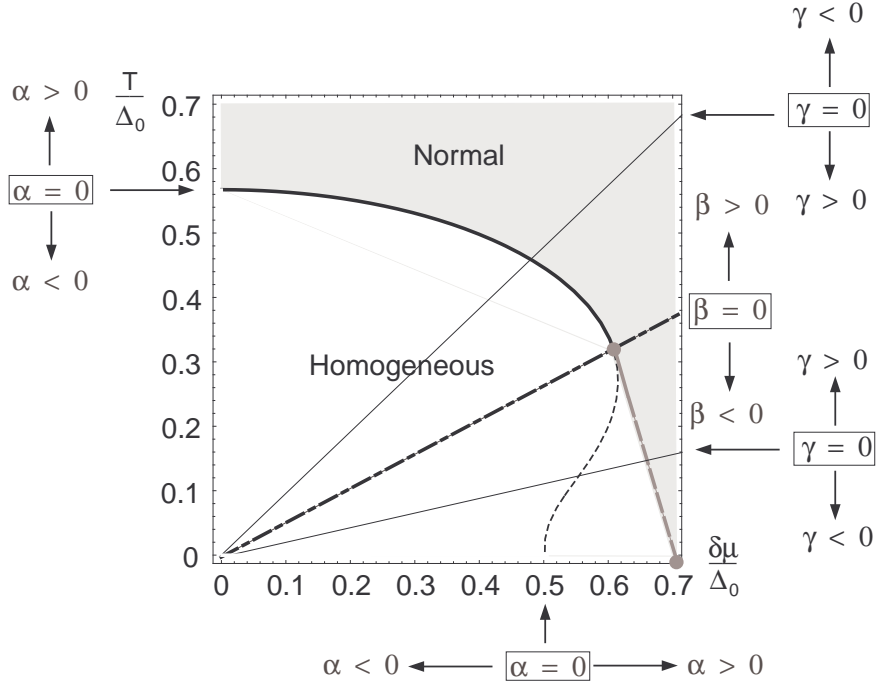}} \caption{{\it
The curve shows the points solutions of the equation $\Delta=0$
in the plane $(v,t)=(\delta\mu/\Delta_0,T/\Delta_0)$. The
tricritical point at $(\delta\mu,T)\approx(0.62,0.28)\,\Delta_0$
is also shown. The upper part of the curve (solid line) separates
the homogeneous phase from the normal one. Along the dashed line
$\Delta=0$ but this is not the absolute minimum of the grand
potential. \label{fig1}}}
\end{figure}

Using Eqs. (\ref{eq:alpha_f}), (\ref{eq:beta2}) and
(\ref{eq:gamma2}) which give the parameters $\alpha$, $\beta$ and
$\gamma$ in terms of the variables $v=\delta\mu/\Delta_0$ and
$t=T/\Delta_0$, we can map the plane $\alpha$ and $\beta$ into
the plane $(\delta\mu/\Delta_0,T/\Delta_0)$. The result is shown
in Fig. \ref{fig1}. From this mapping we can draw several
conclusions. First of all the region where the previous
discussion in terms of the parameters $\alpha$, $\beta$ and
$\gamma$ applies is the inner region of the triangular part
delimited by the lines $\gamma=0$. In fact, as  already stressed,
our expansion does not hold  outside this region. This statement
can be made quantitative by noticing that along the first order
transition line the gap increases when going away from the
tricritical point
as\be\Delta_+^2=-\,\frac{4\alpha}\beta=\sqrt{\frac{3\alpha}\gamma}\,.\ee

Notice  that the lines $\beta(v,t)=0$ and $\gamma(v,t)=0$ are
straight lines, since these zeroes are determined  by the
functions $\psi^{(2)}$ and $\psi^{(4)}$ which depend only on the
ratio $v/t$. Calculating the first order line around the
tricritical point one gets the result  plotted as a solid line in
Fig. \ref{fig1}. Since we know that
$\delta\mu=\delta\mu_1=\Delta_0/\sqrt{2}$ is a first order
transition point, the first order line must end there. In Fig.
\ref{fig1} we have simply connected the line with the point with
grey dashed line. To get this line a numerical evaluation at all
orders in $\Delta$ would be required. This is feasible but we
will skip it since the results will not be necessary in the
following, see \cite{sarma:1963ab}. The location of the
tricritical point is determined by the intersection of the lines
$\alpha=0$ and $\beta=0$. One finds
\cite{combescot:2002ab,buzdin:1997ab}\be
\frac{\delta\mu}{\Delta_0}\Big|_{\rm tric}=0.60822,~~~~
\frac{T}{\Delta_0}\Big|_{\rm tric}=0.31833\,.\ee
 We  also note that the line $\alpha=0$ should cross the temperature axis at the
 BCS point. In this way one reobtains the result in Eq. (\ref{eq:90})
 for the BCS critical temperature, and also the value for the
 tricritical temperature
\be \frac{T_{\rm tric}}{T_{\rm BCS}}=0.56149\,.\ee

The results given in this Section are valid as long as other
possible condensates are neglected. In fact, we will see that
close to the first order transition of the homogeneous phase the
LOFF phase with inhomogeneous  gap can be formed.
\subsection{Gap equation for anisotropic superconductor: One
plane wave (FF state)\label{section4}}

Let us now consider again the condensate wave function $\Xi({\bf
r})$ of Eq. (\ref{deltax}):
  \be \Xi({\bf r})= \langle vac|\psi({\bf r},t) C\psi({\bf r},t)|vac\rangle
  \nonumber\ .\ee Here $|vac\rangle$ is the
  ground state. We develop it as follows\be
  |vac\rangle=\sum_{N=0}^\infty c_N|N\rangle\,,\ee
  where $N$ is even, the state $|N\rangle$  contains N/2 quark
  pairs of momenta
  \be{\bf p}_1=+{\bf p}+{\bf
  q}\ ,~~~~~~~ {\bf p}_2=-{\bf p}+{\bf
  q}\ ,\label{loffq}\ee respectively for up and down species and
  the sum also implies an integration over the ${\bf p}$ variables and sum
  over spin. Clearly we have
  \bea \Xi({\bf r})&=& \sum_{N,M}c_N^*c_M\langle N|\psi({\bf r},t)
  C\psi({\bf r},t)|M\rangle\cr
  &=&\sum_N c^*_N c_{N+2}\langle N|\psi({\bf r},t) C\psi({\bf r},t)
  |N+2\rangle=\cr
  &=& \sum_N c^*_N c_{N+2} e^{2i{\bf q}_N\cdot{\bf r}}
  \langle N|\psi(0)C\psi (0)|N+2\rangle\ .\eea
The homogeneous solution discussed in the previous subsection
corresponds to the choice (Cooper pairs)
  \be  {\bf q}_N=0~~~~~(\text{for ~all~} N)\ ,\ee while ${\bf q}_N\neq 0$ corresponds
  to the inhomogeneous state. Let us now assume
that the interaction favors the formation of pairs with non zero
total momentum and suppose that the values ${\bf q}_1,\,{\bf
q}_2,\,...{\bf q}_P$ are possible. Clearly this hypothesis has to
be tested by comparing the values of the free energies for the
normal, homogeneous and non homogeneous state. In any event,
under such hypothesis, since the gap is proportional to $\Xi({\bf
r})$, one would get
 \be \,\Delta({\bf
r})=\sum_{m=1}^P\Delta_m\,e^{2i{\bf q}_m\cdot{\bf r}}\
.\label{LOgeneral} \ee
\begin{center}
\begin{figure}[ht]
\epsfxsize=6truecm \centerline{\epsffile{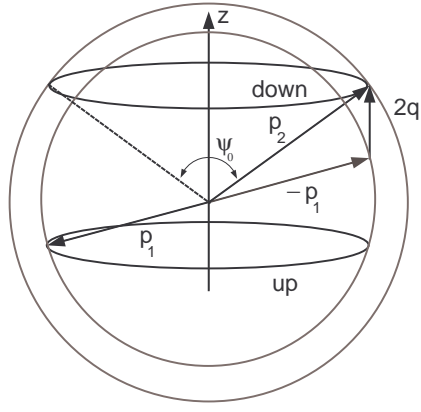}} \noindent
\caption{\it Kinematics of the LOFF state in the case of one
plane wave behavior of the condensate. The Cooper pair has a
total momentum $2{\bf q}\neq 0$.\label{fig2}}
\end{figure}
\end{center}
We will call the phase with $\Delta({\bf r})$ given by
(\ref{LOgeneral}) inhomogeneous or LOFF superconducting.
 At the moment we shall assume the existence of a single ${\bf q}$ and therefore
\be \Delta({\bf r})=\Delta \,e^{2i{\bf q}\cdot{\bf r}}\ .\ee This
is the simplest hypothesis, the one considered in  \cite{FF}, see
Fig. \ref{fig2}; it is therefore called FF state. The paper
\cite{LO} examines the more general case (\ref{LOgeneral}); we
will come to it below. The assumption (\ref{loffq}) with $q\neq
0$ produces a shift in energy: \be \xi_{{\bf p}}\pm\delta\mu=
{\bf v}_F\cdot({\bf p}-{\bf p}_F)\pm\delta\mu \to\ {\bf
v}_F\cdot({\bf p}\mp{\bf q}-{\bf p}_F)\pm\delta\mu= \xi_{{\bf
p}}\,\pm\,\bar\mu_{{\bf p}}\ ,\label{gap4}\ee with \be
\bar\mu_{{\bf p}}=\delta\mu-{\bf q}\cdot{\bf v}_F
=\delta\mu-qv_F\,\cos\theta\ ,\label{56}\ee where the upper
(resp. lower) sign refers to the $d$ (resp. $u$) quasi particle.
Using the analogous result for the hole with field $\psi^c(-\vec
p)$, one can follow the same steps leading to (\ref{gap1}) from
(\ref{trf}); therefore the gap equation is still given by
(\ref{gap1}), but now the quasiparticle occupation numbers are
\be n_u({\bf p})=\frac{1}{\dd e^{(\epsilon+\bar\mu_{{\bf
p}})/T}+1}\ ,~~~~~~~ n_d({\bf p})=\frac{1}{\dd
e^{(\epsilon-\bar\mu_{{\bf p}})/T}+1}\label{n2}\ . \ee

 By  (\ref{n2}), using
 the gap equation for the BCS superconductor with gap $\Delta_0$,
 the gap equation for the inhomogeneous superconductor is written as
 \be
 \frac{g\rho}{2}\ln\frac{\Delta_0}{\Delta}
 =\frac{g}2\int\frac{d^3p}{(2\pi)^3}\,
 \left[n_u({\bf p})+n_d({\bf p})\right]\ .\ee
Differently  from the case with equal chemical potentials
($\delta\mu=0$), when there is phase space reduction at $T\neq
0$, now also at $T=0$ the blocking factors reduce the phase space
available for pairing. As a matter of fact the gap equation at
$T=0$ reads \bea \frac{g\rho}2\ln\frac{\Delta_0}{\Delta}&=&
\frac{g}2\int\frac{d^3p}{(2\pi)^3}\frac{1}{\epsilon({\bf
p},\Delta)}\left(\theta(-\epsilon-\bar\mu_{{\bf
p}})+\theta(-\epsilon+\bar\mu_{{\bf p}})
\right)\nn\\&=&\frac{g\rho}2\int_{BR} \frac{d\Omega_{{\bf
p}}}{4\pi}\,{\rm arcsinh} \frac{C(\theta)}{\Delta}\,,
\label{GAP5}\eea where \be
C(\theta)=\sqrt{q^2v_F^2(z_q-\cos\theta)^2-\Delta^2}\label{gap5.1}\ee
and \be z_q=\frac{\delta\mu}{qv_F}=\cos\frac{\psi_0}2\
,\label{eq:115}\ee where $\psi_0$ is the angle depicted in Fig.
\ref{fig2}. The angular integration is not over the whole Fermi
surface, but only over region defined by $\epsilon({\bf
p},\Delta)<|\bar\mu_{{\bf p}}|$, or\be
q^2v_F^2(z_q-\cos\theta)^2>\Delta^2\,.\ee Notice that there are
no solutions to this inequality for $qv_F+\delta\mu\le\Delta$
(compare with Eq. (\ref{gap5})). Analyzing this inequality in
terms of $\cos\theta$ we see that there are three regions,
obtained  comparing $qv_F-\delta\mu$ to $\pm\Delta$,
characterized by different  domains of angular integration. They
are displayed
 Table \ref{table1}. As pointed out in \cite{FF}, the blocking regions correspond
to regions in momentum space where fermions do not pair. In
regions E and S fermions of one type (for instance spin up) do
not pair, whereas in region D fermions of both types do not pair.
The effect of the blocking regions is to reduce the phase space
where pairing is possible. The complementary phase space is where
the pairing is possible and therefore it will be called pairing
region. It is formed by two rings that loosely speaking are
around the two circles of Fig. \ref{fig2}. Since the pairing is
possible not only on the Fermi surface, but also for modes just
below and above it, each ring has a toroidal shape. $\psi_0=2{\rm
arcos}(z_q)$ is the aperture of the cone, with vertex at the
origin of the spheres, intersecting the Fermi surfaces along the
rings.

\vskip0.5cm
\begin{table}[h]\begin{center}\begin {minipage}{6.5in}
\begin{tabular}{|c|c|c|}
\hline Region  & Definition  & Domain of integration in
$\cos\theta$\\
 \hline  E  & $qv_F-\delta\mu\le -\Delta$ & $(-1,+1)$ \\ \hline
 S  &  $-\Delta\le qv_F-\delta\mu\le  +\Delta$ &$
 \left( -1,\cos\theta_-\right)$  \\  \hline
 D  & $qv_F-\delta\mu\ge +\Delta$ & $\left(-1,\cos\theta_-\right)
 \bigcup\left( \cos\theta_+,+1\right)$ \\ \hline
\end{tabular} \end {minipage}\end{center}\caption{\it In the table the three blocking regions are
shown. Here we have defined
$\cos\theta_\pm=z_q\left(1\pm\Delta/\delta\mu\right)$
\label{table1}}
\end{table}
 Once fixed the integration domain, the remaining integral in
 $\cos\theta$ is trivial and the result can be expressed, for the three cases, in the
 following uniform way
\be {\rm ln}\frac{\Delta_0}{\Delta}=\frac{\Delta}{2qv_F}
\left[G\left(\frac{qv_F+\delta\mu}{\Delta}\right)+
G\left(\frac{qv_F-\delta\mu}{\Delta}\right)\right]\,,\label{gapFF}\ee
where the function $G(x)$ is defined as follows: \bea G(x)&=&x\,
{\rm arccosh}(x)-\sqrt{x^2-1},~~~|x|>1\,,\nn\\
G(x)&=&0,~~~|x|<1\,,\nn\\
G(x)&=&-G(-x),~~~x<0\,.\eea

\subsubsection{Second Order phase transition point\label{c5}} 
\begin{figure}[ht]
\centerline{\epsfxsize=8cm\epsfbox{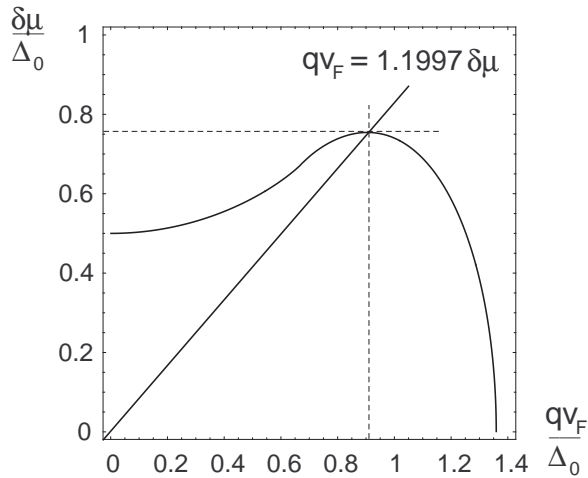}} \caption{{\it
The critical line for the LOFF phase at $T=0$ in the plane
$(qv_F/\Delta_0,\delta\mu/\Delta_0)$. Also the line determining
$qv_F$ as a function of $\delta\mu_2$ is given. \label{fig3}}}
\end{figure}

The
reduction of the available phase space
 implies a reduction of the gap, therefore one
expects in general smaller gaps in comparison with the
homogeneous
 case. In particular we see from Eq. (\ref{GAP5}) that
increasing $\delta\mu$ the effect of the blocking terms
increases; eventually a  phase transition to the normal phase
occurs when $\delta\mu$ approaches a maximum value $\delta\mu_2$.
Therefore the anisotropic superconducting phase can exist in a
window\be \delta\mu_1<\delta\mu<\delta\mu_2\ .\ee One expects
that $\delta\mu_1$ is near the Chandrasekhar-Clogston
\cite{clogston,chandrasekhar} limit $\Delta_0/{\sqrt 2}$ because
Eq. (\ref{oma1}) shows that near this point the difference in
energy between the isotropic superconducting and the normal
phases is small and one might expect that the LOFF state
corresponds to the real ground state. We shall discuss the gap
equation and prove this guess below. For the moment we determine
$\delta\mu_2$. For $\delta\mu\to\delta\mu_2$ the gap $\Delta\to
0$, and in  the blocking regions E and D the domain of
integration in $\cos\theta$ is $(-1,1)$ (the region S disappears
in the limit). Expanding the function $G(x)$ for $x\to\infty$ we
get from Eq. (\ref{gapFF}) \be {\rm
ln}\frac{\Delta_0}{\Delta}=-1+\frac 1
2\,\frac{\delta\mu}{qv_F}{\rm
ln}\frac{qv_F+\delta\mu}{qv_F-\delta\mu}-\frac 1 2\,{\rm
ln}\frac{\Delta^2}{4(q^2v_F^2-\delta\mu^2)}\,\label{LOFF_exp}\ee
which can be re-written as \be \alpha(qv_F,\delta\mu)=-1+\frac 1
2\frac{\delta\mu}{qv_F}{\rm
ln}\frac{qv_F+\delta\mu}{qv_F-\delta\mu}-\frac 1 2\,{\rm
ln}\frac{\Delta_0^2}{4(q^2v_F^2-\delta\mu^2)}=0\,.\label{92}\ee
In terms of the dimensionless variables \be
y=\frac{\delta\mu}{\Delta_0},~~~z=\frac{qv_F}{\Delta_0}\,,\ee the
condition $\alpha=0$ is equivalent to the equation \be
y+z=\frac{e}2\left(\frac{z+y}{z-y}\right)^{\frac{z-y}{2z}}\,.\ee
The critical line is plotted in Fig. \ref{fig3}.

Notice that the equation (\ref{LOFF_exp}) can be written also in
the form \be {\rm ln}\frac{\Delta_0}{2\delta\mu}=\frac 1
2f_0\left(\frac{qv_F}{\delta\mu}\right)=-1+\frac 1
2\frac{\delta\mu}{qv_F}{\rm
ln}\frac{qv_F+\delta\mu}{qv_F-\delta\mu}-\frac 1 2{\rm
ln}\frac{\delta\mu^2}{(q^2v_F^2-\delta\mu^2)}\,,\label{41}\ee
where \be f_0(x)=\int_{-1}^{+1}du\ln(1+x u)\ .\label{42}\ee We
can fix  $q$ by minimizing the function $\alpha$ with respect to
it. This is equivalent to minimize the grand potential close to
the second order phase transition. This is obtained for $x$
solution of the equation\be x\,=\,\coth\,x\,,\ee i.e. at \be
x=\frac{qv_F}{\delta\mu_2}=1.1997\equiv x_2\ .\label{69}\ee This
can be  also obtained from Fig. \ref{fig3}, intersecting the
curve $\alpha=0$ at its maximum value $\delta\mu_2/\Delta_0$ with
a straight line passing from the origin.

The value of $\delta\mu$ at which the transition occurs is
obtained by substituting this value in (\ref{41}) and solving for
$\delta\mu_2$. One gets in this way \be
\delta\mu_2=0.754\,\Delta_0 \ .\ee Since
$\delta\mu_2>\delta\mu_1\approx 0.71\Delta_0$, there exists a
window of values of $\delta\mu$ where LOFF pairing is possible.
We will prove below, using the Landau-Ginzburg approach,
 that the phase transition for the one-plane wave condensate at
$T=0$ and $\delta\mu=\delta\mu_2$ is second-order.
\section{Ginzburg-Landau approximation}
\label{ginzburg_landau}

The condensate wave function acts as an order parameter
characterized by its non vanishing value in the superconducting
phase. At the second order phase transition it vanishes and one
can apply the general Ginzburg-Landau (GL) approach there
\cite{ginzburg:1950xz}. We will begin by performing  the GL
expansion at $T=0$ for a general inhomogeneous gap function
\cite{LO,Bowers:2002xr}. From this we will derive the grand
potential measured with respect to the normal state and we will
evaluate it explicitly for several cases. Next, in Section
\ref{tricritical} we will perform an analogous expansion at
$T\not=0$ around the tricritical point that we have shown to
exist in Section \ref{homogeneous_superconductor}
\cite{combescot:2002ab,houzet:1999ab,houzet:2002ef,houzet:2000cd,
buzdin:1997ab,alexander:1978ab}. We will follow in this
discussion the Ref. \cite{combescot:2002ab}. These authors have
made a rather general analysis with the conclusion that in the
generic case the favored state  corresponds to a pair of
antipodal wave vectors.

\subsection{Gap equation in the Ginzburg-Landau approach}

\begin{center}
\begin{figure}[htb]
\epsfxsize=13truecm \centerline{\epsffile{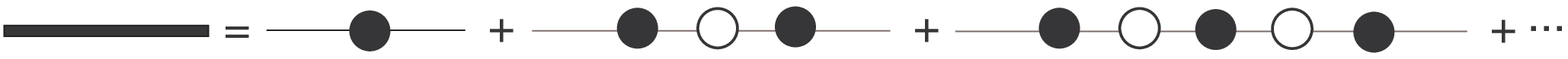}}
\caption{{\it Ginzburg-Landau expansion of the propagator; the
lines represent alternatively  ${\bf G_0^-}$ and ${\bf G_0^+}$,
see Eq. (\ref{eq:131}). Full (resp. empty) circles represent
${\bf \Delta^*}$ (resp. ${\bf \Delta}$).\label{propagator}}}
\end{figure}
\end{center}

We will start this Section by considering the Ginzburg-Landau
 expansion of the Nambu-Gor'gov equations.
Let us perform an expansion in $\Delta$ of the propagator ${\bf
F}$ in Eq. (\ref{18}). It is depicted in Fig. \ref{propagator}.

\begin{center}
\begin{figure}[htb]
\epsfxsize=10truecm \centerline{\epsffile{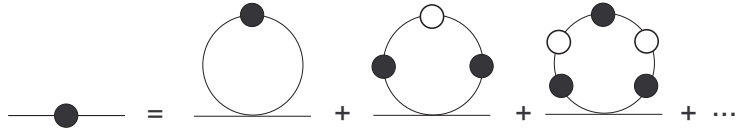}}
\caption{{\it Ginzburg-Landau expansion of the gap equation; the
lines represent alternatively  ${\bf G_0^-}$ and ${\bf G_0^+}$,
see Eq. (\ref{eq:132}). Full (resp. empty) circles represent
${\bf \Delta^*}$ (resp. ${\bf \Delta}$).\label{expansion}}}
\end{figure}
\end{center}

 Formally it is written as follows \bea {\bf F}=&+&{\bf G_0^- \Delta^{*}
G_0^+}\cr &-&{\bf G_0^- \Delta^{*} G_0^+\Delta G_0^- \Delta^{*}
G_0^+}\cr &+& {\bf G_0^- \Delta^{*} G_0^+\Delta G_0^- \Delta^{*}
G_0^+\Delta G_0^- \Delta^{*} G_0^+}\ .\label{eq:131}\eea
 The gap equation has an analogous expansion, schematically depicted
 in Fig. \ref{expansion}.

It has the form
 \bea \Delta^*&=&-\,i\,\frac{g}2\,Tr\,\int\frac{dE}{2\pi}
\Big(\int d{\bf r_1} \,G_0^-({\bf r}, {\bf r_1}) \Delta^{*} ({\bf
r_1})
 G_0^+({\bf r_1}, {\bf r})\cr
&-& \int\prod_{j=1}^3 d{\bf r_j}\, G_0^- ({\bf r}, {\bf
r_1})\Delta^{*} ({\bf r_1})G_0^+({\bf r_1}, {\bf r_2}) \Delta
({\bf r_2})G_0^- ({\bf r_2}, {\bf r}_3)\Delta^{*}({\bf r}_3)
 G_0^+({\bf r}_3, {\bf r})\cr
&+&\int \prod_{j=1}^5 d{\bf r_j} G_0^-({\bf r}, {\bf r_1})
\Delta^{*}({\bf r_1}) G_0^+({\bf r_1}, {\bf r_2}) \Delta ({\bf
r_2})G_0^- ({\bf r_2}, {\bf r}_3)\Delta^{*}({\bf r}_3)\
G_0^+({\bf r}_3, {\bf r}_4)\cr&&\times\Delta( {\bf
r}_4)G_0^-({\bf r}_4,{\bf r}_5)
 \Delta^{*} ({\bf r}_5)G_0^+({\bf r}_5, {\bf r})
 \Big)\,.\label{eq:132}\eea
Substituting (\ref{LOgeneral}) we get \bea\Delta_n^*&=& \Big(
\sum_{k}\Pi({\bf q_k},{\bf q_n})\Delta^*_k\delta ({\bf q_k}-{\bf
q_n}) \cr&+& \sum_{k,\ell,m} J({\bf q_k},{\bf q_\ell},{\bf
q_m},{\bf q_n})\Delta^*_k\Delta_\ell\Delta_m^* \delta ({\bf
q_k}-{\bf q_\ell} +{\bf q_m}-{\bf q_n})\cr
&+&\sum_{k,\ell,m,j,i}K({\bf q_k},{\bf q_\ell},{\bf q_m} {\bf
q_j},{\bf q_i},{\bf
q_n})\Delta^*_k\Delta_\ell\Delta_m^*\Delta_j\Delta_i^*
\cr&\times&\delta ({\bf q_k}-{\bf q_\ell} +{\bf q_m}-{\bf
q_j}+{\bf q_i}-{\bf q_n})\Big)\,.\eea Here $\delta({\bf q_k}-{\bf
q_n}) $ means the Kronecker delta: $\delta_{n,k} $ and
\be\Pi({\bf q_1},{\bf q_2})= \,+\,\frac{ig\rho}{2}\int
\frac{d\bf\hat
w}{4\pi}\int_{-\delta}^{+\delta}d\xi\int_{-\infty}^{+\infty}\frac{dE}{2\pi}
\prod_{i=1}^2\, f_i(E,\delta\mu,\{{\bf q}\})\ ,\label{eq:143}\ee
\be J({\bf q_1},{\bf q_2},{\bf q_3},{\bf q_4})=
\,+\,\frac{ig\rho}2 \int \frac{d\bf\hat w}{4\pi}
\int_{-\delta}^{+\delta}
d\xi\int_{-\infty}^{+\infty}\frac{dE}{2\pi} \prod_{i=1}^4\,
f_i(E,\delta\mu,\{{\bf q}\})\ ,\label{gei}\ee\be K({\bf q_1},{\bf
q_2},{\bf q_3},{\bf q_4},{\bf q_5},{\bf q_6})=
\,+\,\frac{ig\rho}2 \int \frac{d\bf\hat
w}{4\pi}\int_{-\delta}^{+\delta}
d\xi\int_{-\infty}^{+\infty}\frac{dE}{2\pi}\prod_{i=1}^6
f_i(E,\delta\mu,\{{\bf q}\}).\label{kappa}\ee We have put ${\bf
w}\,\equiv\,v_F\,\bf\hat w$ and
 \be
 f_i(E,\delta\mu,\{{\bf q}\})=\frac{1}{E+i\epsilon\,
 \text{sign}\,E-\delta\mu+(-1)^i[\xi-2\sum_{k=1}^i(-1)^{k}
\bf w\cdot{\bf q_k}]}~.\ee Moreover the condition
\be\sum_{k=1}^M(-1)^{k}{\bf q_k}=0 \ee holds, with $M=2,4,6$
respectively for $\Pi$, $J$ and $K$.

 For $\Pi(q)\equiv\Pi({\bf q},{\bf q})$
one gets \be \Pi(q) =\frac{ig\rho}2 \int \frac{d\bf\hat w}{4\pi}
\int_{-\delta}^{+\delta}d\xi\int_{-\infty}^{+\infty}\frac{dE}{2\pi}
\frac 1 {(E\,+\,i\epsilon\,\text{sign}\,E-\bar\mu)^2-
\xi^2}\,,\label{pidiq} \ee where $\bar\mu=\delta\mu-v_F{\bf
q}\cdot\hat w$ is defined in Eq. (\ref{56})
 and is
identical to the function $C(\theta)$ of Eq. (\ref{gap5.1}) with
$\Delta=0$. In performing the energy integration in (\ref{pidiq})
we use the fact that there are contributions only for
$|\xi|>|\bar\mu|$. Using the gap equation for the homogeneous
pairing to get rid of the cutoff $\delta$ we obtain the result
\be \Pi(q)= 1+\frac{g\rho}{2} \left[1+\frac 1 2
\log\frac{\Delta_0^2}{4|(qv_F)^2-\delta\mu^2|}
-\frac{\delta\mu}{2qv_F}\log\Big| \frac{q v_F+\delta\mu}{q
v_F-\delta\mu}\Big|\right]\ .\label{piq} \ee $\Pi(q)$ can be
rewritten in terms of the function $\alpha$ introduced in
(\ref{92}) as follows:
\be\alpha(q)=2\,\frac{1-\Pi(q)}{g\rho}\label{alpha3}\
.\label{eq:141}\ee

Clearly the gap equation in the GL limit, $1=\Pi(q)$, coincides
with Eq. (\ref{92}), which was obtained in the one plane wave
hypothesis. The reason is that, since $\Pi$ depends only on $|{\bf
q}|$, it assumes the same value for all the crystalline
configurations; therefore $\Pi$ does not depend on the crystalline
structure of the condensate and  the transition point we have
determined in Sec. \ref{c5} is universal.

For the evaluation of $J$ and $K$ we have to specialize to the
different LOFF condensate choices. This will be discussed below.

 \subsection{Grand potential}
 The grand potential $\Omega$ is given in the GL approximation by
 \bea\Omega&=&-\frac 1 g
\Big( \sum_{k,n=1}^P\left[\Pi({\bf q_k},{\bf
q_n})\,-\,1\right]\Delta^*_k\Delta_n\delta_{{\bf q_k}-{\bf q_n}}
\cr&+&\frac 1 2 \sum_{k,\ell,m,n=1}^P J({\bf q_k},{\bf
q_\ell},{\bf q_m},{\bf
q_n})\Delta^*_k\Delta_\ell\Delta_m^*\Delta_n \delta_{{\bf
q_k}-{\bf q_\ell} +{\bf q_m}-{\bf q_n}}\cr &+&\frac 1
3\sum_{k,\ell,m,j,i,n=1}^P K({\bf q_k},{\bf q_\ell},{\bf q_m}
,{\bf q_j},{\bf q_i},{\bf
q_n})\Delta^*_k\Delta_\ell\Delta_m^*\Delta_j\Delta_i^*\Delta_n
\delta_{{\bf q_k}-{\bf q_\ell} +{\bf q_m}-{\bf q_j}+{\bf
q_i}-{\bf q_n}}\Big)\,.\label{gapGL} \eea where $P$ is the number
of independent plane waves in the condensate. Let us assume that
\be \Delta_k=\Delta_k^*=\Delta~~~~~~(\text{for\ any\ } k)\,,\ee
so that we can rewrite (\ref{gapGL}) as follows:
 \be \frac{\Omega}{\rho}=\,P\,\frac\alpha 2\Delta^2+\frac \beta 4 \Delta^4
 +\frac\gamma 6\Delta^6\,,\label{113}\ee
where $\alpha$ is related to $\Pi(q)$ through (\ref{alpha3}) and
\bea \beta&=& -\frac 2{g\rho}\sum_{k,\ell,m,n=1}^P J({\bf
q_k},{\bf q_\ell},{\bf q_m},{\bf q_n})\delta_{{\bf q_k}-{\bf
q_\ell}
+{\bf q_m}-{\bf q_n}}\ ,\label{beta1}\\
\gamma&=&-\frac 2{g\rho}\sum_{k,\ell,m,j,i,n=1}^P K({\bf
q_k},{\bf q_\ell},{\bf q_m} ,{\bf q_j},{\bf q_i},{\bf
q_n})\delta_{{\bf q_k}-{\bf q_\ell} +{\bf q_m}-{\bf q_j}+{\bf
q_i}-{\bf q_n}}\label{gamma1}\ .\eea

It follows from the discussion in subsection \ref{c5} that, at
$\delta\mu=\delta\mu_2$, $\alpha$ vanishes; moreover $\alpha<0$
for
 $\delta\mu<\delta\mu_2$, see below, eq. (\ref{alpha1pw}).
 Exactly as in Section \ref{phasediagramhomogeneous}
we distinguish different cases:
\begin{enumerate}
  \item $\beta>0$,  $\gamma>0$. In this case $\Delta=0$
  is a maximum for $\Omega$, and the minima occur at the points
  given in Eq.  (\ref{eq:95}), which now reads:
\be
\Delta^2=\,\frac{-\beta+\sqrt{\beta^2-4P\alpha\gamma}}{2\gamma}
\label{alpha2a}\ .\ee Near the transition point one has\be
\Delta^2\approx\,-\,\frac{P\alpha}{\beta} \label{alpha2}\ .\ee A
phase transition occurs  when $\alpha=0$, i.e. at
$\delta\mu=\delta\mu_2$. The transition is second order since the
gap goes continuously to zero at the transition point.

  \item $\beta<0$,  $\gamma>0$. Both for $\alpha<0$
and for   $\alpha>0$ $\Delta^2$ in (\ref{alpha2a}) is a minimum
for $\Omega$. In the former case it is the only minimum, as
$\Delta=0$ is a maximum; in the latter case it competes with the
solution $\Delta=0$. Therefore the LOFF phase can persist beyond
$\delta\mu_2$, the limit  for the single plane wave LOFF
condensate up to a  maximal value $\delta\mu^*$. At
$\delta\mu=\delta\mu^*$ the free energy vanishes and there are
degenerate minima at \be\Delta=0\
,~~~~~\Delta^2=\frac{-3\beta}{4\gamma}\ .\ee The critical point
$\delta\mu^*$ is obtained by Eq. (\ref{eq:101}) that in the
present case can be written as
\be\alpha(qv_F=1.1997\delta\mu^*,\delta\mu^*)=\frac{3\beta^2}{16P\gamma}
\label{120} \ .\ee
 The phase transition from the crystalline to the normal phase
 at $\delta\mu^*$  is first order.
\item $\beta<0$,  $\gamma<0$: In this case the GL expansion
(\ref{113})
 is inadequate since $\Omega$ is not bounded from below
 and another term ${\cal O}(\Delta^8)$ is needed.
 \end{enumerate}
In the case $\beta<0$,  $\gamma>0$ we can select the most favored
structure by computing the free energy at a fixed value of
$\delta\mu$. We choose $\delta\mu=\delta\mu_2$ where the FF state
has a second order phase transition and $\alpha=0$. One has
there\be\Delta^2=-\frac{\beta}{\gamma}~,~~~~
\frac\Omega\rho=\frac{\beta^3}{12\gamma^2}\label{min}\ .\ee
\subsection{Crystalline structures}
\label{IIIC}
 For any crystalline structure the function $\alpha$ in the first term
 of the GL expansion is given by
\bea\alpha&=& -1-\frac 1 2
\log\frac{\Delta_0^2}{4|(qv_F)^2-\delta\mu^2|}
+\frac{\delta\mu}{2qv_F}\log\Big| \frac{q v_F+\delta\mu}{q
v_F-\delta\mu}\Big|\cr&=&-\log\frac{\Delta_0}{2\delta\mu}+ \frac
1 2 f_0\left(\frac{qv_F}{\delta\mu}\right) ,\label{alpha1}\eea
where we have used Eqs. (\ref{41}), (\ref{piq}) and
(\ref{alpha3}); $\alpha$ vanishes for $\delta\mu=\delta\mu_2$,
which characterizes the second order transition point at $T=0$,
 see (\ref{92}) or  (\ref{alpha2});
therefore  we can write \be \alpha=-\frac{\eta}{\delta\mu_2}\
,\label{alpha1pw} \ee where \be \eta=\delta\mu_2-\delta\mu\ee and
we have expanded $\alpha$ around $\delta\mu_2$ and
 used the property of minimum of $f_0(x)$ at $\delta\mu=\delta\mu_2$.
We observe that, for $\delta\mu<\delta\mu_2$, $\alpha$ is
negative;
 therefore the transition at $T=0$ is always second order if $\beta>0$.

As to the other terms, we can use the results of  Appendix
\ref{app:A} to get the first terms of the  GL expansion for any
crystal structure. The exception is the one-plane-wave case,
where the free energy can be computed at any desired order.
\subsubsection{One plane wave} Using the results of the
Appendix \ref{app:A} one gets for the Fulde-Ferrel one plane wave
condensate: \be J=
J_0\equiv\,-\,\frac{g\rho}{8}\,\frac{1}{(qv_F)^2-\delta\mu^2}\
,~~~ K=K_0\equiv\,-\,\frac{g\rho}{64}\,
\frac{(qv_F)^2+3\delta\mu^2}{[(qv_F)^2-\delta\mu^2]^3}\ .
\label{JK}\ee  From (\ref{JK}) we get ($x_2=qv_F/\delta\mu_2=
1.1997$): \bea
\beta&=&\frac{1}{4\delta\mu_2^2(x_2^2-1)}=\,+\,\frac{0.569}{
\delta\mu_2^2}
\,,\cr\gamma&=&\frac{3+x_2^2}{8\delta\mu_2^4(x_2^2-1)^3}=\,
+\,\frac{1.637}{\delta\mu_2^4} \ .\label{beta1pw}\eea Since
 $\beta>0$
the $\gamma$-term is ineffective near the transition point and
Eq. (\ref{alpha2}) gives \be \Delta^2=4\eta\,
\left(x_2^2-1\right)\delta\mu_2\approx 1.757\eta\,\delta\mu_2 \
.\label{eq:168}\ee We can get  $\Omega$ from Eq. (\ref{113}) with
$P=1$ and from Eqs. (\ref{alpha1pw}) and (\ref{beta1pw}). The
result is \be\Omega=-\frac{\alpha^2\rho}{4\beta}=-0.439\,
\rho\,(\delta\mu-\delta\mu_2)^2\ , \label{044}\ee  The same
result could also be obtained using Eqs. (\ref{d0bis}) and
(\ref{om1}).

\subsubsection{Generic crystals} 
\begin{figure}[ht]
\centerline{\epsfxsize=12cm\epsfbox{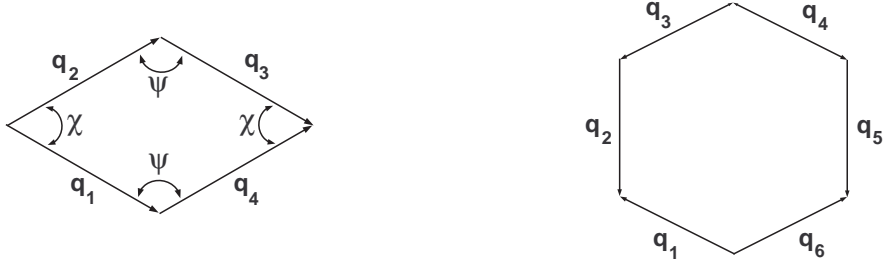}} \caption{{\it
Rhombic and hexagonal  configurations for the vectors ${\bf
q}_i$. The vectors are assumed of the same length $q$ and  such
that ${\bf q}_1-{\bf q}_2+{\bf q}_3-{\bf q}_4=0$ for the rhombus
and ${\bf q}_1-{\bf q}_2+{\bf q}_3-{\bf q}_4+{\bf q}_5-{\bf
q}_6=0$ for the hexagon. The vectors  need  not be all in the
same plane.\label{fig:rombo}}}
\end{figure}

In the general case $P\neq 1$ and
the evaluation of $J$ and $K$ is more complicated. First, one
introduces Feynman parameterizations, then the integrals over
energy, longitudinal momenta and angles are performed, along the
lines sketched in Appendix \ref{app:A}, mainly based on
\cite{Bowers:2002xr}. Next, one has to perform the integration
over the Feynman parameters. To do this it is useful to draw two
pictures: a rhombus, with lines formed by the four vectors
appearing in $J({\bf q_k},{\bf q_\ell},{\bf q_m},{\bf q_n})$,
implementing the condition ${{\bf q_k}-{\bf q_\ell} +{\bf
q_m}-{\bf q_n}}=0$, and an hexagon, with lines formed by the six
vectors appearing in $ K({\bf q_k},{\bf q_\ell},{\bf q_m} ,{\bf
q_j},{\bf q_i},{\bf q_n})$ that satisfy ${{\bf q_k}-{\bf q_\ell}
+{\bf q_m}-{\bf q_j}+{\bf q_i}-{\bf q_n}}=0$, see Fig.
\ref{fig:rombo}. Note that the rhombus and the hexagon need not
be in a plane. The simplest example is provided by two plane
waves.

\subsubsection{Two plane waves}

\begin{figure}[ht]
\centerline{\epsfxsize=8cm\epsfbox{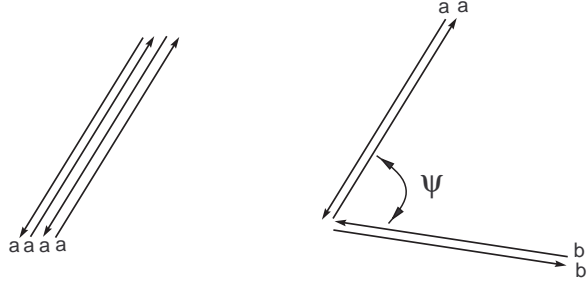}} \caption{{\it
The two rhombic structures corresponding to the integrals $J_0$
and $J_\psi$ of Eqs. (\ref{eq:171}). The indices $a$ and $b$
refer to the vectors ${\bf q}_a$ and ${\bf q}_b$
respectively.\label{fig:twowaves_quad}}}
\end{figure}

In this case  $P=2$; let the two vectors be $\bf q_a,\, q_b$,
forming an angle $\psi$; a simpler case is provided by an
antipodal pair, $\bf q_a=-q_b=q$ and $\psi=\pi$, with\be
\Delta({\bf r})=2\Delta\cos{\bf q\cdot r}\ .\ee To get $\beta$
from (\ref{138}) one may notice that the integral $J$ assumes two
different values \be J_0=J({\bf q_a},{\bf q_a},{\bf q_a},{\bf
q_a})\ ,~~~ J_\psi=J({\bf q_a},{\bf q_a},{\bf q_b},{\bf
q_b})\label{eq:171}\ee corresponding to Fig.
\ref{fig:twowaves_quad}.

\begin{center}
\begin{figure}[htb]
\epsfxsize=7truecm \centerline{\epsffile{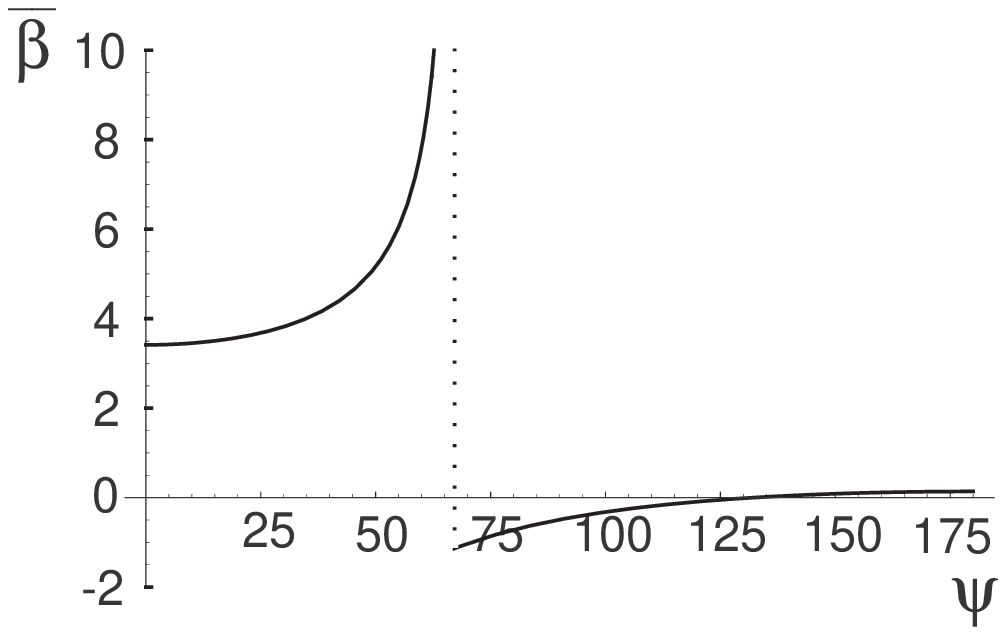}}
\caption{{\it $\bar\beta=\beta\,\delta\mu_2^2$ as a function of
the opening angle $\psi$ between the two plane wave vectors $\bf
q_a$ and $\bf q_b$; $\psi=\psi_0=67.07^0$ is the angle defining
the LOFF ring;
 $\bar\beta(\psi_0)=-1.138$. \label{betapsi}}}
\end{figure}
 \end{center}

\begin{figure}[ht]
\centerline{\epsfxsize=9cm\epsfbox{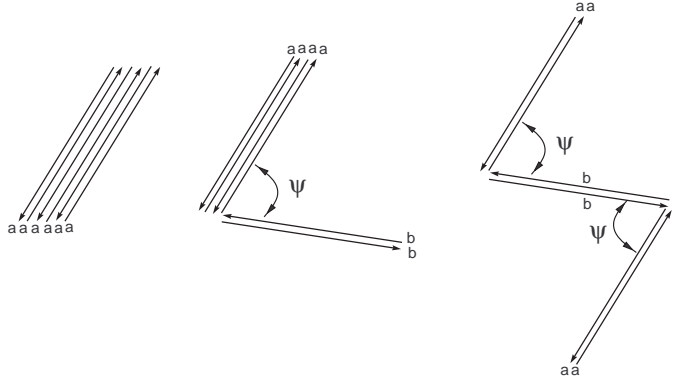}} \caption{{\it
The three hexagonal structures corresponding to the integrals
$K_0$, $K_1$ and $K_2$ of Eqs. (\ref{eq:176}). The indices $a$
and $b$ refer to the vectors ${\bf q}_a$ and ${\bf q}_b$
respectively.\label{fig:twowaves_hexa}}}
\end{figure}

 $J_0$ has been already computed, see Eq. (\ref{JK});
 on the other hand
\be J_\psi=-\frac{g\rho}{2\delta\mu^2}\,{Re}\,\frac{ \arctan
\frac{x_2\sqrt{\cos\psi-1} }{\sqrt{2-
x_2^2(1+\cos\psi})}}{2x_2\sqrt{(\cos\psi -1)
(2-x_2^2(1+\cos\psi))}} \,,\ee which for $\psi=\pi$ gives \be
J_\pi=-\frac{g\rho}{8\delta\mu^2_2} \ . \ee Using rotation and
parity symmetry of the integrals one gets\be \beta(\psi)
=-\frac{2}{g\rho}\left(2J_0+4J_\psi\right)\,.\ee The result for
$\beta(\psi)$ as a function of $\psi$ is reported in Fig.
\ref{betapsi}. In the case of the antipodal pair
 (${\bf q},-{\bf q}$), when $\psi=180^o$, one gets
\be \beta=-\frac{2}{g\rho}\left(2J_0+4J_\pi\right)=
\frac{1}{\delta\mu^2_2}\left(\frac{1}{2(x_2^2-1)}-1\right)=
\,+\,\frac{0.138}{\delta\mu^2_2}\ .\ee
For $ K$ we have three possibilities (see Fig.
\ref{fig:twowaves_hexa}): \be K_0=K({\bf q_a},{\bf q_a},{\bf
q_a},{\bf q_a},{\bf q_a},{\bf q_a})\,,~~~ K_1(\psi)=K({\bf
q_a},{\bf q_a},{\bf q_a},{\bf q_a},{\bf q_b},{\bf q_b})~,~~~
K_2(\psi)=K({\bf q_a},{\bf q_a},{\bf q_b},{\bf q_b},{\bf
q_b},{\bf q_b})\ . \label{eq:176}\ee 

Therefore we have \be \gamma(\psi)=-\frac
2{g\rho}\left(2K_0+12K_1(\psi)+6K_2(\psi)\right)\,. \ee $K_0$ has
been already computed in (\ref{JK}), whereas $K_1$ and $K_2$ can
be evaluated using the results given in Appendix A.
$\gamma(\psi)$ is plotted in Fig. \ref{gammapsi}. In the case of
the antipodal pair, when $\psi=180^o$, the result for $\gamma$ is
in Table 2.
\begin{center}
\begin{figure}[htb]
\epsfxsize=8truecm \centerline{\epsffile{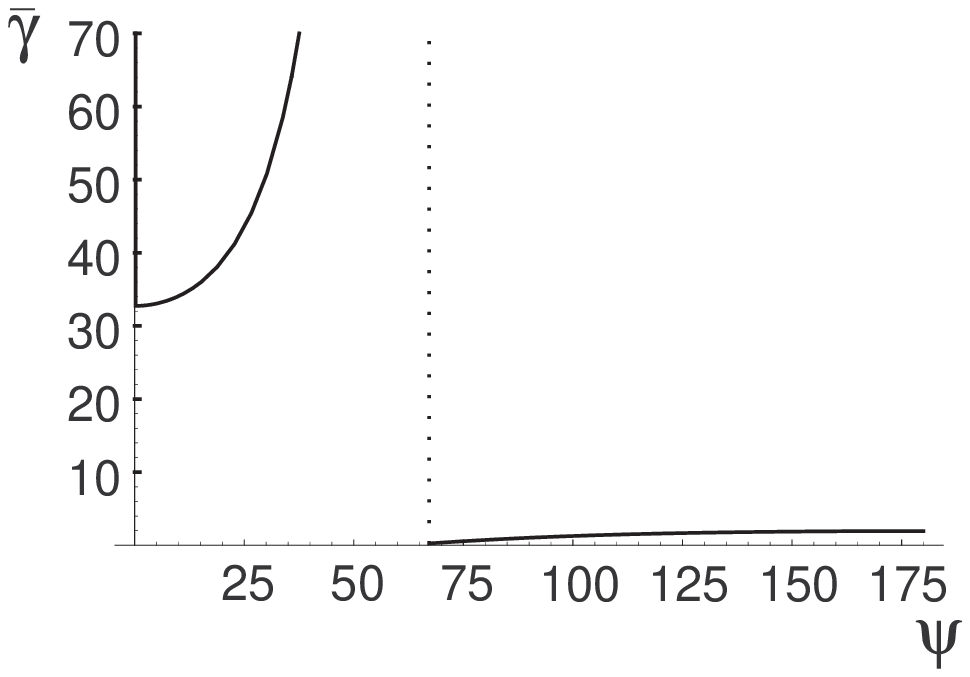}}
\caption{{\it $\bar\gamma=\gamma\,\delta\mu_2^4$ as a function of
the opening angle $\psi$ between the two plane wave vectors $\bf
q_a$ and $\bf q_b$; $\psi=\psi_0=67.07^{\,o}$ is the angle
defining the LOFF ring;
 $\bar\gamma(\psi_0)=0.249$. \label{gammapsi}}}
\end{figure}
\end{center}
Figs. \ref{betapsi} and \ref{gammapsi} show a divergence at
\be\psi=\psi_0=67.07^{\,o}=2\arccos\frac{\delta\mu_2}{qv_F}\,.\ee
$\psi_0$ is the opening angle depicted in Fig. \ref{fig2}. In
this case, differently from the one plane wave situation, we have
two different rings for each Fermi surface. For $\psi>\psi_0$ the
two rings do not intersect, at $\psi=\psi_0$ they are contiguous,
while for $\psi<\psi_0$  they overlap. The structure with
$\psi<\psi_0$ is energetically disfavored because, being $\beta$
large and positive, the free energy would be smaller according to
Eq. (\ref{alpha2}). According to the discussion in
\cite{Bowers:2002xr}, this behavior seems universal, i.e.
structures with overlapping  rings are energetically disfavored
in comparison with structures without overlaps. We will use this
result in Section \ref{Cubic structure}.

For $\psi_0<\psi<132^o$ $\beta$ is negative. Therefore according
to the discussion above we are in presence of a second order
phase transition ($\gamma$ is always positive as it can be seen
from Fig. \ref{gammapsi}). As it is clear from Eq. (\ref{min})
the most favorable case from the energetic point of view occurs
when
 $\gamma$ assumes its smallest value  and $|\beta|$ its largest, i.e.
at $\psi=\psi_0$, when the rings are tangent. The values
 for this case are reported in  Table \ref{tabII}.
 \begin{table}[htb]
\begin{center}
\begin{minipage}{6.5in}
\begin{tabular}{|c|c|c|c|c|c|}\hline
 {Structure}
& $P$ & {$\bar\beta$} & {$\bar\gamma$} & {$\bar\Omega_{\min}$} &
{$\dm^*/\Delta_0$}
\\
\hline
 FF state    &1      &  0.569 & 1.637 & 0 & 0.754 \\
 antipodal plane waves  & 2  & 0.138 & 1.952 & 0 & 0.754 \\
 Two plane waves ($\psi=\psi_0$)  & 2  & -1.138 & 0.249 & -1.126 & 1.229 \\
 Face centered cube  & 8  & -110.757 & -459.242 & -  & - \\ \hline
\end{tabular}

\end{minipage}\end{center}
\caption{\label{tabII}\it Candidate crystal structures with $P$
plane waves. $\bar\beta = \delta\mu_2^2\,\beta$, $\bar\gamma =
\delta\mu_2^4\,\gamma$, $\bar\Omega=\Omega/(\rho\Delta_0)$, with
$\rho = p_F^2/(\pi^2v_F)$, is the (dimensionless) minimum free
energy computed at $\dm = \dm_2$, obtained from (\ref{min}). The
phase transition (first order for $\bar\beta<0$ and
$\bar\gamma>0$, second order for $\bar\beta>0$ and
$\bar\gamma>0$) occurs at $\dm^*$, given, for first order
transitions,  by Eq. (\ref{120}).}
\end{table}

 For comparison, at
$\psi=90^0$ we have $\delta\mu_2^2\,\beta(90^o)=-0.491$,
$\delta\mu_2^4\,\gamma(90^o)=1.032$; the first order transition
takes place at $\delta\mu^*=0.771\Delta_0$, only marginally
larger than $\delta\mu_2$, and the dimensionless free energy
$\bar\Omega=\Omega/(\rho\Delta_0^2)$ assumes at
$\delta\mu=\delta\mu_2$ the value $\bar\Omega=-0.005$, which is
larger than the value obtained for $\psi=\psi_0$, see Table
\ref{tabII}.
\subsubsection{Other structures}
\label{IIIC.4} One could  continue in the same way by considering
other structures. An extensive analysis can be found in
\cite{Bowers:2002xr} where 23 different crystalline structures
were considered. We refer the interested reader to Table I in
this paper, as well as to its Appendix where the technical
aspects of the
 integration over the
Feynman parameters of the $K$ integrals for the more complicated
structures
 are worked out. From our previous discussion we know
 that the most energetically favored crystals are those which present
 a first order phase transition between the LOFF and the normal phase.
 Among the regular structures, with $\gamma>0$, examined in
 \cite{Bowers:2002xr} the favored one seems to be the octahedron ($P=6$),
  with $\delta\mu^*=3.625\Delta_0$.
Special attention, however,  should be given to the face centered
cube; we have reported the values of its parameters, as computed
in \cite{Bowers:2002xr}, in our Table \ref{tabII}. We note that
$\gamma<0$ for this structure. The condensate in this case is
given by \be \Delta({\bf r}) = \sum_{k=1}^8 \Delta_k({\bf r}) =
\sum_{k=1}^8\,\Delta\, \exp(2i q {\bf \hat n_k} \cdot {\bf r} )
\,, \ee where ${\bf\hat n_k}$ are the eight unit vectors defining
the vertices of the cube:\bea &&
 {\bf  \hat n_1}=\frac{1}{\sqrt 3}(+1,+1,+1),~~~ {\bf\hat n_2}=\frac{1}{\sqrt
3}(+1,-1,+1),~\cr && {\bf\hat n_3}=\frac{1}{\sqrt
3}(-1,-1,+1),~~~ {\bf\hat n_4}=\frac{1}{\sqrt 3}(-1,+1,+1),\cr &&
{\bf\hat n_5}=\frac{1}{\sqrt 3}(+1+,1,-1),~~~ {\bf\hat
n_6}=\frac{1}{\sqrt 3}(+1,-1,-1), \cr && {\bf\hat
n_7}=\frac{1}{\sqrt 3}(-1,-1,-1),~~~ {\bf\hat n_8}=\frac{1}{\sqrt
3}(-1,+1,-1)~ .\label{eq:169}\eea Strictly speaking, since both
$\beta$ and $\gamma$ are negative, nothing could be said about
the cube and one should compute the eighth order in the GL
expansion, given by $\delta\Delta^8/8$; the transition would be
first order if
 $\delta>0$. However \cite{Bowers:2002xr} argue that,
 given the large value
 of $\gamma$, this structure would necessarily dominate.
 Reasonable numerical examples discussed by the authors confirm
 this guess.
\subsection{LOFF around the tricritical point} \label{tricritical}

 The LOFF phase can be studied
analytically around the tricritical point
\cite{combescot:2002ab,buzdin:1997ab} that we have considered in
Section \ref{homogeneous_superconductor}. Here we will follow the
treatment of Ref. \cite{combescot:2002ab}. The tricritical point
is the place where one expects the LOFF transition line to start.
Close to it one expects that also the total pair momentum
vanishes, therefore one can perform a simultaneous expansion in
the gap parameter and in the total momentum. Starting from the
expressions given in Section \ref{ginzburg_landau} (see Eqs.
(\ref{eq:143}), ({\ref{gei}) and (\ref{kappa})) and proceeding as
in Section \ref{homogeneous_superconductor} we find \be
\Omega=\sum_{\bf q}\tilde\alpha({\bf q})\, |\Delta_{\bf
q}|^2+\frac 1 2 \sum_{{\bf q}_i}\tilde\beta({\bf
q}_i)\Delta_{{\bf q}_1} \Delta_{{\bf q}_2}^*\Delta_{{\bf
q}_3}\Delta_{{\bf q}_4}^*+\frac 1 3 \sum_{{\bf
q}_i}\tilde\gamma({\bf q}_i)\,\Delta_{{\bf q}_1} \Delta_{{\bf
q}_2}^*\Delta_{{\bf q}_3}\Delta_{{\bf q}_4}^*\Delta_{{\bf
q}_5}\Delta_{{\bf q}_6}^*\,. \label{eq:160}\ee Here we have used
the momentum conservation in the fourth order and in the sixth
order terms \be {\bf q}_1+{\bf q}_3={\bf q}_2+{\bf q}_4,~~~{\bf
q}_1+{\bf q}_3+{\bf q}_5={\bf q}_2+{\bf q}_4+{\bf q}_6\,,\ee with
\bea \tilde\alpha({\bf q})&=&\alpha+\frac 2 3\, \beta\, Q^2+\frac
8 {15}\, \gamma\,
Q^4\,,\nn\\
\tilde\beta({\bf q}_i)&=&\beta+\frac 4
9\,\gamma\,(Q_1^2+Q_2^2+Q_3^2+Q_4^2+{\bf Q}_1\cdot{\bf Q}_3+{\bf
Q}_2\cdot{\bf Q}_4)\,,\nn\\\tilde\gamma({\bf q}_i)&=&\gamma
\label{eq:162}\eea where $\alpha$, $\beta$ and $\gamma$ were
defined in Eqs. (\ref{eq:alpha_f}), (\ref{eq:beta2}) and
(\ref{eq:gamma2}), and \be{\bf Q}={\bf q} v_F\,.\ee In Appendix
\ref{appendixb} we show, as an example, how $\tilde\alpha$ can be
obtained from the expansion of $\Pi(q)$ around ${\bf Q}=0$.
 In order to get a coherent expansion one
has to consider the modulus of the pair total momentum of the
same order of the gap. In fact, as we shall see, the optimal
choice for $Q$ turns out to be of order $\Delta$. Correspondingly
one has to  expand the coefficient of the quadratic term in the
gap up to the fourth order in the momentum and the fourth order
term in the gap up to the second order in the momentum. In the
form given in Eq. (\ref{eq:160}) one can easily apply the general
analysis around the tricritical point used in Section
\ref{homogeneous_superconductor}. In particular for vanishing
total momenta of the pairs we are back to the case of the
homogeneous superconductor studied in Section
\ref{homogeneous_superconductor}.

It is interesting to write the expression for the grand potential
in configuration space, because it shows that around the critical
point the minimization problem boils down to solve a differential
equation, whereas in a generic point the Ginzburg-Landau
equations are integral ones. By Fourier transformation we get
from Eq. (\ref{eq:160}) \bea \Omega&=&\int d^3{\bf
r}\left[\alpha|\Delta({\bf r})|^2+\frac 2
3\beta|\vec\nabla\Delta({\bf r})|^2+\frac 8
{15}\gamma|\vec\nabla^2\Delta({\bf r})|^2\right]\nn\\&+&\int
d^3{\bf r}\left[\beta|\Delta({\bf
r})|^4+\frac{2}9\gamma(2(\vec\nabla|\Delta({\bf
r})|^2)^2+3(\vec\nabla\Delta^2({\bf
r}))(\vec\nabla\Delta^{*2}({\bf r}))\right]+\frac 1 8\gamma\int
d^3{\bf r}|\Delta({\bf r})|^6\,.\label{eq:195}\eea

Let us now recall from  Section \ref{homogeneous_superconductor},
see Eq. (\ref{eq:101}), that the first order phase transition is
given by:\be \beta_{\rm
first}=-4\sqrt{\frac{\alpha\gamma}3}\,,\label{eq:165}\ee with a
discontinuity in the gap given by  \be
\Delta^2=-4\frac\alpha\beta=-\frac{3}{4}\frac\beta\gamma\,,\ee see
Eq. (\ref{eq:102}). Let us now consider the possibility of a
second order transition in the general LOFF case. Only the
quadratic term in the gap is necessary for the discussion, and we
have to look at its zero, given by $\tilde\alpha=0$. Since we are
considering only the quadratic term we can choose an optimal value
for $Q^2$ by minimizing this term with respect to $|\bf Q|$. We
find \be Q^2=-\frac {5}{8}\frac\beta\gamma\,,\label{eq:182}\ee
requiring $\beta<0$. The corresponding value for $\alpha$ turns
out to be \be \alpha=\frac
5{24}\frac{\beta^2}{\gamma}\,,\label{eq:183}\ee or \be \beta_{\rm
second}=-\sqrt{\frac {24}5\alpha\gamma}\,.\ee The LOFF second
order transition line is higher than the first order transition
line of the homogeneous case, since $\beta_{\rm second}>\beta_{\rm
first}$ (see Fig. \ref{fig7} showing the relevant lines in the
plane $(\delta\mu/\Delta_0,T/\Delta_0)$). Therefore the second
order transition to the LOFF case overcomes the first order
transition to the homogeneous symmetric  phase as it can be
checked by evaluating the   grand potential for the LOFF state
along the first order transition line.

The situation considered before corresponds to the physics of the
problem only when the second order transition is a true minimum
of the grand potential. This is not necessarily the case and we
will explore in the following this possibility.
\begin{figure}[ht]
\centerline{\epsfxsize=11cm\epsfbox{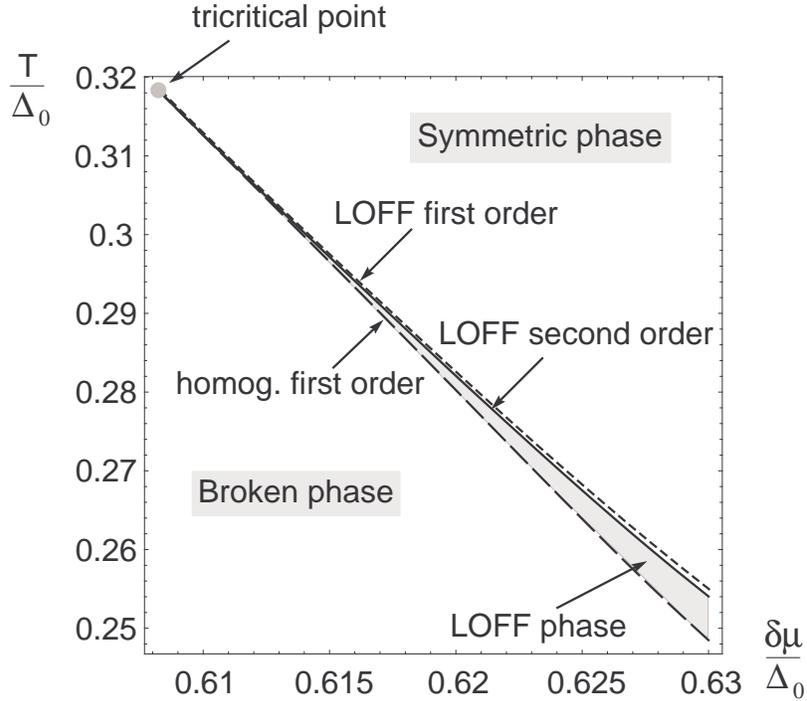}} \caption{{\it
The graph shows, in the plane $(\delta\mu/\Delta_0,T/\Delta_0)$,
the first order transition line (dashed line) from the
homogeneous broken phase to the symmetric phase. The solid line
corresponds to the second order transition from the LOFF phase to
the symmetric one. The lines start from the tricritical point and
ends when the Landau-Ginzburg expansion is not valid any
more.\label{fig7}}}
\end{figure}

\subsubsection{The LO subspace}

The second order term in the grand potential requires that the
vectors ${\bf Q}$ have  the same length along the second order
transition line. It is natural to consider the subspace (LO)
spanned by plane waves corresponding to momenta with the same
length $ Q_0$: \be\Delta({\bf r})=\sum_{|{\bf Q}|=
Q_0}\Delta_{{\bf q}}\,e^{2\,i\,{\bf q}\cdot{\bf r}}\,.\ee The
authors of Ref. \cite{LO} have restricted their considerations to
periodic solutions, but this is not strictly necessary, although
the solution found in \cite{combescot:2002ab} is indeed periodic.
We will see that within this subspace the usual LOFF transition
(the one corresponding to a single plane wave) is not a stable
one. We will show that there is a first order transition that
overcomes the second order line. In order that the LOFF line,
characterized by Eqs. (\ref{eq:182}) and (\ref{eq:183}) is a true
second order transition the coefficient of the fourth order term
should be positive. However, in the actual case, the mixed terms
in the scalar products of the vectors ${\bf Q}$ could change this
sign. In \cite{combescot:2002ab} the mixed terms are studied by
defining the following quantity \be 2\,b\, Q_0^{\,2}\sum_{{\bf
q}_i}\Delta_{{\bf q}_1}\Delta_{{\bf q}_2}^*\Delta_{{\bf
q}_3}\Delta_{{\bf q}_4}^*= \sum_{{\bf q}_i}\Delta_{{\bf
q}_1}\Delta_{{\bf q}_2}^*\Delta_{{\bf q}_3}\Delta_{{\bf
q}_4}^*({{\bf Q}_1}\cdot{{\bf Q}_3}+{{\bf Q}_2}\cdot{{\bf
Q}_4})\,.\ee Clearly \be -1\le b\le 1\,,\ee where $b=1$ is reached
in the case of a single plane wave. With this definition and for
the optimal choice of $Q_0$ (see Eq. (\ref{eq:182})), we get for
the coefficients appearing in the expression of the grand
potential (see Eqs. (\ref{eq:160}) and (\ref{eq:162})):
\be\tilde\alpha=\alpha-\frac 5{24}\frac{\beta^2}{\gamma},~~~
\tilde\beta=-\frac 1 9\beta(5b+1),~~~\tilde\gamma=\gamma\,.\ee
Therefore, for any order parameter such that\be b<-\frac 1 5\,,\ee
it follows \be \tilde\beta<0\ee and the LOFF line becomes unstable
(we recall that close to the second order line $\beta<0$). In
fact, since $\tilde\alpha=0$ along this line and $\tilde\beta<0$,
a small order parameter is sufficient to make $\Omega$ negative.
In other words one gains by increasing the order parameter as long
as the sixth order term does not grow too much. But then we can
make $\Omega=0$ (equal to its value in the symmetric phase) by
increasing $\alpha$. Therefore we have a new transition line in
the plane $(\alpha,\beta)$ (or, which is the same in the plane
$(\delta\mu/\Delta_0,T/\Delta_0)$) to the right of the LOFF line.
\cite{combescot:2002ab} also  shows that necessarily \be b\ge
-\frac 1 3\,.\ee The equality is reached for any real order
parameter $\bar\Delta({\bf r})$. In order to get a better feeling
about the parameter $b$ it is convenient to consider the following
quantity \be 2\,c\,Q_0^{\,2}\sum_{{\bf q}_i}\Delta_{{\bf
q}_1}\Delta_{{\bf q}_2}^*\Delta_{{\bf q}_3}\Delta_{{\bf
q}_4}^*=\sum_{{\bf q}_i}({\bf Q}_1-\bar{\bf Q}_2)^2\Delta_{{\bf
q}_1}\Delta_{{\bf q}_2}^*\Delta_{{\bf q}_3}\Delta_{{\bf q}_4}^*\,.
\ee Expanding the right hand side of this equation and using ${\bf
Q}_1\cdot{\bf Q}_2={\bf Q}_3\cdot{\bf Q}_4$ we find \be
c=1-b\,.\ee To minimize $b$ is equivalent to maximize $c$. To this
aim it is convenient to have opposite ${\bf Q}_1$ and ${\bf Q}_2$
because then $({\bf Q}_1-{\bf Q}_2)^2$ reaches its maximum value
equal to $4Q_0^{\,2}$. In this case we  have \be\Delta_{{\bf
q}}^*=\Delta_{-{\bf q}}\,.\ee This is equivalent to require that
the order parameter is real in configuration space. Of course, it
is not necessary that the amplitudes of  the different pairs of
plane waves are equal.

To proceed further one can introduce a measure of the size of the
order parameter \be\frac 1{\pi^3}\int d^3{\bf r}|\Delta({\bf
r})|^2=\sum_{{\bf q}_i}\Delta_{{\bf q}_1}\Delta_{{\bf
q}_2}^*\equiv N_2\bar\Delta^2\ee and \be\frac 1{\pi^3}\int
d^3{\bf r}|\Delta({\bf r})|^4=\sum_{{\bf q}_i}\Delta_{{\bf
q}_1}\Delta_{{\bf q}_2}^*\Delta_{{\bf q}_3}\Delta_{{\bf
q}_4}^*\equiv N_4\bar\Delta^4\,,\ee \be\frac 1{\pi^3}\int d^3{\bf
r}|\Delta({\bf r})|^6=\sum_{{\bf q}_i}\Delta_{{\bf
q}_1}\Delta_{{\bf q}_2}^*\Delta_{{\bf q}_3}\Delta_{{\bf
q}_4}^*\Delta_{{\bf q}_5}\Delta_{{\bf q}_6}^*\equiv
N_6\bar\Delta^6\,.\ee In the case of a single plane wave we get
\be N_2=N_4=N_6=1\,.\ee For a real gap the set of the vectors
${\bf Q}_i$ is made of $N/2$ pairs. If all the plane wave have
the same amplitude one can show \cite{combescot:2002ab} that the
quantities $N_2$, $N_4$ and $N_6$ assume the following values \be
N_2=2N,~~~N_4=3N(N-1),~~~N_6=5N(3N^2-9N+8)\,.\label{eq:197}\ee
With these notations the grand potential becomes \be
\Omega=N_2\left(\alpha+\frac 23 \beta Q_0^{\,2}+\frac 8{15}
\gamma Q_0^{\,4}\right)\bar\Delta^2+\frac 1 2N_4\left(\beta+\frac
83\,a\,\gamma\,Q_0^{\,2}\right)\bar\Delta^4+\frac 13
N_6\gamma\bar\Delta^6\,,\ee where \be a=\frac {b+2}3,~~~~\frac
13\le a\le 1\,.\ee Minimizing this expression with respect to $
Q_0$ we find\be Q_0^{\,2}=-\frac 5 8\frac\beta\gamma-\frac
54\,a\,\frac {N_4}{N_2}\,\bar\Delta^2\,.\ee Therefore a non zero
solution for $Q_0$ is obtained if \be \bar\Delta^2\le -\frac
1{2a}\frac\beta\gamma\frac{N_2}{N_4}=\bar\Delta^2_{\rm max}\,.\ee
The corresponding expression for $\Omega$ becomes \be \Omega=
\alpha N_2\left(1-\frac
5{24}\frac{\beta^2}{\alpha\gamma}\right)\bar\Delta^2+\frac 1
2\beta N_4\left(1-\frac 53 a \right)\bar\Delta^4+\frac 1 3\gamma
N_6\left( 1
-\frac{5a^2}{2}\frac{N_4^2}{N_2N_6}\right)\bar\Delta^6\,.\ee In
order to have a transition from the symmetric phase we must allow
$\Omega/\bar\Delta^2$ to become negative. The zero of $\Omega$ is
reached for \be \alpha=\frac 5{24}\frac{\beta^2}\gamma+\frac
3{16}\frac{\beta^2}\gamma\frac{\left(1-\displaystyle{\frac{5a}3}
\right)^2} {\left(\displaystyle{ \frac{N_2N_6}{N_4^2}}-\frac 52
a^2\right)}\,.\ee The value of $\bar\Delta$ corresponding to the
zero of $\Omega$ is given by \be \bar\Delta^2=-\frac 3
 2\frac\beta\gamma\frac{N_2}{N_4}\frac{\left(1-
 \displaystyle{\frac{5a}3}\right)}{\left(\displaystyle{
\frac{N_2N_6}{N_4^2}}-\frac 52 a^2\right)}\,,\ee showing that
this is a first order transition. We have shown that it is
convenient to have $b$ as small as possible and that the minimum
is reached for $b=-1/3$ or $a=5/9$. In correspondence with this
value we have $\bar\Delta^2\ge 0$ and \be \alpha> \frac
5{24}\frac{\beta^2}\gamma\,.\ee We get \be \alpha=\frac
5{24}\frac{\beta^2}\gamma+\frac 1
{972}\frac{\beta^2}\gamma\displaystyle{\frac
1{\displaystyle{\frac{N_2N_6}{N_4^2}}-\displaystyle{\frac{125}{162}}}}\,.\ee
We see that it is convenient to take $N_2N_6/N_4^2$ as small as
possible. However, notice that \be \frac {N_2 N_6}{N_4}\ge 1\ee
as it follows from the Schwartz inequality \be \frac{\left[\int
d^3{\bf r}\,\bar\Delta^2({\bf r})\right]\left[\int d^3{\bf
r}\,\bar\Delta^6({\bf r})\right]}{\left[\int d^3{\bf
r}\,\bar\Delta^4({\bf r})\right]^2}\ge 1\,.\ee

Let us look for example at the case of pairs of plane waves with
opposite ${\bf q}$  without any further constraint. In this case
we get \be \frac{N_2 N_6}{N_4^2}=\frac 5
9\frac{3N^2-9N+8}{(N-1)^2}\,.\ee This expression has a minimum at
$N=2$ (here $N$ is even), where it holds 10/9 and then it
increases monotonically up to the value 15/9. Notice that for
crystalline structures the situation could be different. For
instance, in the case of the cube the values of $N_4$ is not the
one given by Eq. (\ref{eq:197}), i. e. 168, but rather 216.
However, in Ref. \cite{combescot:2002ab} it is shown that this
expression gets indeed its minimum value for $N=2$. This result
is obtained assuming that the plane waves form a generic set of
antipodal vectors, which means that the only way to satisfy the
momentum conservation is through the cancellation of each
momentum with the opposite one in the same pair. This excludes
special configurations where other arrangements of vectors could
give a zero result. The authors of Ref. \cite{combescot:2002ab}
argue that this result should hold in general, but a complete
proof is lacking.

 The
value obtained for $\alpha$ for the two plane wave case is \be
\alpha=\frac{\beta^2}\gamma\left( \frac
5{24}+\frac{1}{330}\right)=\frac{\beta^2}\gamma\left(\frac
5{36}+3\times 10^{-3}\right)\,.\ee In this case we have three
lines, the first order line just found, the second order LOFF
transition for $\alpha=5\beta^2/({24}\gamma)$ and the first order
transition to the homogeneous broken phase for
$\alpha=3\beta^2/(16\gamma)$. The distance of these two last
lines is given by \be\frac{\beta^2}\gamma\left(\frac 3{16}-\frac
5{24}\right) =-\frac 1{48}\frac{\beta^2}\gamma=-2.1\times
10^{-2}\frac{\beta^2}\gamma\,.\ee We know that these two lines
stay close one to the other up to zero temperature. It turns out
that the same is true for the new first order line as it has been
shown in Ref. \cite{Matsuo:1998ab}. These results are illustrated
in Fig. \ref{fig7}. We notice that whereas in the expansion
around the tricritical point  the favored state seems to be the
one corresponding to a pair of plane waves,  with a first order
transition between the LOFF and the normal state,  at zero
temperature one has a second order phase transition. Therefore
the first order transition line must change into a second order
line at low temperatures. In ref. \cite{Matsuo:1998ab} it has
been shown that this happens at a temperature $T/T_{BCS}=0.075$.

It is also interesting to see how things change varying the
spatial dimensions. In fact it has been found in
\cite{Burkhardt:1994ab} that the first order transition found
previously is second order in two spatial dimensions. This result
is confirmed by
\cite{buzdin:1983cd,buzdin:1987ef,machida:1989ab,buzdin:1997ab}
which show that the transition is second order in one spatial
dimension and furthermore it can be given an exact solution in
terms of the Jacobi elliptic functions. This solution has the
property that along the second order transition line it reduces
to the two plane wave case considered here. It is very simple to
obtain the dependence on the number of dimensions. In fact the
only place where the dimensions enter is in the angular
integration as, for instance, in Eq. (\ref{eq:191}). In general
this is an average over the D-dimensional sphere and we need the
following equations for the terms of order $Q^2$ and $Q^4$
respectively: \be \int\frac{d\hat{\bf w}}{S_D} \hat w_i \hat
w_j=\frac{\delta_{ij}} D,~~~~\int\frac{d\hat{\bf w}}{S_D} \hat
w_i \hat w_j\hat w_k\hat w_l=\frac 1
{D(D+2)}(\delta_{ij}\delta_{kl}+\delta_{ik}\delta_{jl}+
\delta_{il}\delta_{jk})\,,\ee where $S_D$ is the surface of the
sphere with unitary radius in D-dimensions \be
S_D=\frac{2\pi^{D/2}}{\Gamma(D/2)}\,.\ee Therefore  the terms
proportional to $Q^2$ are multiplied by $3/D$ and the ones
proportional to $Q^4$ by $15/D(D+2)$. We get \bea
\tilde\alpha({\bf q})&=&\alpha+\frac 2 D\, \beta\, Q^2+\frac 8
{D(D+2)}\, \gamma\,
Q^4\,,\nn\\
\tilde\beta({\bf q}_i)&=&\beta+\frac 4
{3D}\,\gamma\,(Q_1^2+Q_2^2+Q_3^2+Q_4^2+{\bf Q}_1\cdot{\bf
Q}_3+{\bf Q}_2\cdot{\bf Q}_4)\,,\nn\\\tilde\gamma({\bf
q}_i)&=&\gamma \,.\eea Proceeding as before the grand potential
evaluated at the optimal value of  $Q^2$ is \be \Omega= \alpha
N_2\left(1-\frac
{D+2}{8D}\frac{\beta^2}{\alpha\gamma}\right)\bar\Delta^2+\frac 1
2\beta N_4\left(1-\frac {D+2}D a \right)\bar\Delta^4+\frac 1
3\gamma N_6\left( 1
-\frac{3(D+2)}{2D}a^2\frac{N_4^2}{N_2N_6}\right)\bar\Delta^6\,.\ee
We see from here that the critical dimension for the transition to
change from first to second order is at the zero of the second
term, that is \be D=\frac{2a}{1-a}\ee and for $a=5/9$, $D=2.5$.
This means that we have a first order transition for $D>2.5$ and a
second order one for $D<2.5$ (remember that $\beta<0$). The
location of the transition is at ($a=5/9$)\be
\alpha=\frac{D+2}{8D}\frac{\beta^2}\gamma+\frac 3
{20}\frac{\beta^2}{\gamma}\frac{(2D-5)^2}{D(7D-10)}\,.\ee The
value of the gap along the transition line is given by \be
\Delta^2=-\frac 65\frac\beta\gamma\frac{2D-5}{7D-10}\,.\ee We see
that $\Delta^2>0$ for $D>2.5$.

\cite{combescot:2002ab} have  considered also the possibility of
solutions around the tricritical point not belonging to the LOFF
subspace. In fact,  the antipodal solution does not satisfy the
Euler-Lagrange equation obtained from Eq. (\ref{eq:195}), due to
the non-linear terms. If these are small it is reasonable to look
for solutions close to $\Delta\cos({\bf q}\cdot{\bf r})$. This
assumption simplifies the problem because the antipodal solution
is essentially a one-dimensional solution characterized by the
direction of ${\bf q}$. Then \cite{combescot:2002ab}  have found
that the corrections at $\Delta\cos({\bf q}\cdot{\bf r})$
expressed in terms of higher harmonics are indeed very small. Of
course this is only a consistency argument, but it is an
indication that the  choice of the LOFF subspace is a good
approximation to the full problem.

To conclude this Section let us say that in our opinion the
status of the LOFF phase is not  yet settled. Up to now we have
considered the Ginzburg-Landau expansion both at $T=0$ and at the
tricritical point. The results in the three-dimensional case can
be summarized as follows:
\begin{itemize}
\item {\bf Zero temperature point}: In \cite{LO} it is found that
the favored phase has a gap with a phase modulation $\cos({\bf
q}\cdot{\bf r})$ corresponding to a structure with two antipodal
vectors. This phase and the normal one are separated by a second
order-transition line. However in \cite{Bowers:2002xr}, where a
rather complete study of the possible crystalline structures has
been done, it is argued that the most favorable structure would be
the face-centered cube. The transition between the corresponding
phase and the normal one should be first-order. \item {\bf
Tricritical point}: In \cite{buzdin:1997ab} the non-uniform phase
has been studied in different dimensions with the result that the
space modulation related to a single wave vector (i.e.
$\exp(2i{\bf q}\cdot{\bf r}$)) is always unfavorable. These
authors find also hat the solution with two antipodal wave vectors
is the preferred one. In 1 and 2 space dimensions the transition
to the normal state is second-order, whereas it is first-order in
3 space dimensions. Analogous results have been found in
\cite{houzet:1999ab} (see also \cite{Agterberg:2000ab}), where the
study has been extended to space modulations such as
$\cos(qx)+\cos(qy)$ or $\cos(qx)+\cos(qy)+\cos(qz)$. These authors
argue that there could be various transition lines at temperatures
lower than the tricritical point. Finally, as thoroughly discussed
in this Section, in \cite{combescot:2002ab} the results of
\cite{buzdin:1997ab} are confirmed.
\end{itemize}

There have been also numerical investigations about the full
phase space. In particular \cite{Burkhardt:1994ab}  proved that
in the two-dimensional case (layered superconductors) the phase
transition from the normal phase to the one characterized by two
antipodal vectors is second order. The second-order transition
line from the phase with a single plane wave to the normal phase
has been studied in \cite{sarma:1963ab,SjSarma}. The
two-dimensional case for type II superconductors has been studied
in \cite{Shimahara:1998ab}. The author has considered states
corresponding to single wave vectors and antipodal pairs together
with configurations corresponding to triangular, square and
hexagonal states. It is found that, according the temperature,
all these states may play a role. Finally  \cite{Matsuo:1998ab},
as already mentioned, make a numerical analysis based on the use
of quasi-classical Green's functions. They find  that in 3
dimensions the transition line between the normal phase and the
antipodal vectors phase starts being first-order at the
tricritical point and becomes second-order at
$T=0.0075\,T_{BCS}$. The two-dimensional case will be discussed
again in Section \ref{VID}.

In conclusion the question of the preferred non-uniform state
cannot be considered settled down yet. Let us discuss by way of
example the 3 dimensional case. We have seen that there are
strong indications that the favored state around the tricritical
point is the one corresponding to two antipodal vectors. This
being the case the natural question is: how the transition line
extends down to zero temperature? If at $T=0$  the preferred
state were the antipodal pair a further tricritical point in the
plane $(\delta\mu,T)$ would arise. In fact, recall that the
transition is first order at the tricritical point and
second-order at $T=0$. However, if the conjecture in
\cite{Bowers:2002xr} is correct, the cubic phase would emerge in
the path going to $T=0$. A possibility is that one goes from one
structure to another in analogy to what suggested by
\cite{Shimahara:1998ab} for the two-dimensional case. The other
logical possibility is that the Combescot and Mora result at the
tricritical point might be evaded by an exceptional arrangement
of the wave vectors as, for instance, in the case of the
face-centered cube. Therefore we think that more theoretical work
is necessary in order to fill-in these gaps in our understanding
of the non-uniform superconducting phase.
\section{Superconductivity in Quantum Chromodynamics}
\label{colorsuperconductivity}
Color superconductivity (CSC) is an old subject
\cite{collins:1975ab,Barrois:1977xd,Frautschi:1978rz,Bailin:1984bm}
that  has recently become one of the most fascinating research
fields  in Quantum Chromo Dynamics (QCD); these developments can
be found in
\cite{Alford:1998zt,Alford:1998mk,Alford:1999pa,Schafer:1998ef,
Schafer:1999jg,Schafer:1999pb,Schafer:1998na,Carter:1998ji,Rapp:1998zu,
Alford:1999pb,Pisarski:1998nh,Pisarski:1999bf,Agasian:1999id,
Shuster:1999tn,Hong:1999fh}; for reviews see
\cite{Rajagopal:2000wf}, \cite{Hong:2000ck},
\cite{Alford:2001dt}, \cite{Hsu:2000sy}. It
  offers a clue to the behavior of strong interactions at
 high baryonic densities, an issue of paramount relevance both for
 the understanding of heavy ion collisions and the physics of
 compact stars.
Color superconductivity  arises because for sufficiently high
baryon chemical potential $\mu$ and small temperature,
 the color interaction favors the formation of a
quark-quark condensate in the color antisymmetric channel
${\mathbf{\bar 3}}$. In the asymptotic regime it is also possible
to understand the structure of the condensates. In fact, consider
the matrix element
\begin{equation}
  \langle 0|\psi_{ia}^\alpha\psi_{jb}^\beta|0\rangle \label{lab2}
\end{equation}
where $\alpha,\beta=1,2,3$ are color indices, $a,b=1,2$ are spin
indices and  $i,j=1,\cdots, N$ are flavor indices. Its color,
spin and flavor structure is completely fixed by the following
considerations.
\begin{itemize}
\item Antisymmetry in color indices $(\alpha,\beta)$ in order to
have attraction. \item Antisymmetry in spin indices $(a,b)$ in
order to get a spin zero condensate. The isotropic structure of
the condensate is favored since a larger portion of the phase
space around the Fermi surface is available. \item Given the
structure in color and spin, Pauli principles requires
antisymmetry in flavor indices.
\end{itemize}
Since the quark and spin momenta in the pair are opposite, it
follows that the left(right)-handed quarks can pair only with
left(right)-handed quarks. In the case of 3 flavors the favored
condensate is
\begin{equation}
\langle 0|\psi_{iL}^\alpha\psi_{jL}^\beta|0\rangle=-\langle
0|\psi_{iR}^\alpha\psi_{jR}^\beta|0\rangle=
\Delta\sum_{C=1}^3\epsilon^{\alpha\beta C}\epsilon_{ijC}\,.
\end{equation}
This gives rise to the so-called color--flavor--locked (CFL)
phase \cite{Alford:1998mk,Schafer:1998ef}. However at moderate
densities other less attractive channels could play a role
\cite{Alford:2002rz}. The reason for the name is that
simultaneous transformations in color and in flavor leave the
condensate invariant. In fact, the symmetry breaking pattern
turns out to be
$$
  SU(3)_c\otimes SU(3)_L\otimes SU(3)_R\otimes U(1)_B\to SU(3)_{c+L+R}
  \otimes Z_2\,,
$$
where $SU(3)_{c+L+R}$ is the diagonal subgroup of the three
$SU(3)$ groups. Both the chiral group and the  color symmetry are
broken but a diagonal $SU(3)$ subgroup remains unbroken. The $Z_2$
group arises from the invariance of the condensate when the quark
fields are multiplied by -1. We have 17 broken generators; since
there is a broken gauge group,
 8
of these generators correspond to 8 longitudinal degrees of the
gluons, because the gauge bosons acquire a mass; there are 9
Nambu Goldstone bosons (NGB) organized in an octet associated to
the breaking of the flavor group and in a singlet associated
 to the breaking of the baryonic number.
 The effective theory describing the NGB for the CFL model
 has been studied in \cite{Casalbuoni:1999wu}.

This  is the typical situation when the chemical potential is much
larger than the quark masses $m_u$, $m_d$ and $m_s$ (in these
considerations one should discuss about density depending masses).
However one can ask what happens when decreasing the chemical
potential. At intermediate densities we have no more the support
of asymptotic freedom, but all the model calculations show that
one still has a sizeable color condensation. In particular if the
chemical potential $\mu$ is much less than the strange quark mass
one expects that the strange quark decouples, and the
corresponding condensate should be
\begin{equation}
\langle 0|\psi_{iL}^\alpha\psi_{jL}^\beta|0\rangle=
\Delta\epsilon^{\alpha\beta 3}\epsilon_{ij}\,,\label{condensates}
\end{equation}
since due to the antisymmetry in color the condensate must
necessarily choose a direction in color space. Notice that now
the symmetry breaking pattern is completely different from the
three-flavor case, in fact we have
$$
  SU(3)_c\otimes SU(2)_L\otimes SU(2)_R\otimes U(1)_B\to SU(2)_c
  \otimes SU(2)_L\otimes SU(2)_R\otimes U(1)\otimes Z_2\,.
$$

The chiral group remains unbroken, while the original color
symmetry group is broken to $SU(2)_c$, with generators $T^A$
corresponding to the generators $T^1,T^2,T^3$ of $SU(3)_c$. As a
consequence, three gluons remain massless whereas the remaining
five acquire a mass.   Even though the original $U(1)_B$ is
broken there is an unbroken global symmetry that plays the role
of $U(1)_B$. As for $U(1)_A$, this axial symmetry is broken by
anomalies, so that in principle there  is no Goldstone boson
associated to its breaking by the condensate; however at high
densities explicit axial symmetry breaking is weak and therefore
there is a light would be Goldstone boson associated to the
breaking of the axial $U(1)_A$. One can construct an effective
theory to describe the emergence of the unbroken subgroup
$SU(2)_c$ and the low energy excitations, much in the same way as
one builds up chiral effective lagrangian with effective fields
at zero density. For the two flavor case this development can be
found in
 \cite{Casalbuoni:2000cn,Rischke:2000cn}.

It is natural to ask what happens in the intermediate region of
$\mu$. It turns out that the interesting case is for $\mu\approx
M_s^2/\Delta$. To understand this point let us consider the case
of two fermions, one massive, $m_1=M_s$ and the other one
massless, at the same chemical potential $\mu$. The Fermi momenta
are of course different
\begin{equation}
  p_{F_1}=\sqrt{\mu^2-M_s^2},~~~~p_{F_2}=\mu\,.
\end{equation}
The grand potential for the two unpaired fermions is
\begin{equation}
 \Omega_{\rm unpair.}=2\int_{0}^{p_{F_1}}\frac{d^3p}{(2\pi)^3}\left(\sqrt{{{\bf p}\,}^2+M_s^2}-\mu\right)+
 2\int_{0}^{p_{F_2}}\frac{d^3p}{(2\pi)^3}\left(|{\bf
 p}\,|-\mu\right)\,.
\end{equation}
 For the two fermions to pair they have to reach
some common momentum $p_{\rm comm}^F$, and the corresponding
grand potential is
 \be
 \Omega_{\rm pair.}=2\int_{0}^{p_{\rm comm}^F}\frac{d^3p}{(2\pi)^3}
 \left(\sqrt{{{\bf p}\,}^2+M_s^2}-\mu\right)+
 2\int_{0}^{p_{\rm comm}^F}\frac{d^3p}{(2\pi)^3}\left(|\vec
 p\,|-\mu\right)-\frac{\mu^2\Delta^2}{4\pi^2}\,,
 \label{omega_pair}\ee
where the last term is the energy necessary for the condensation
of a fermion pair, see Eq. (\ref{eq:70b}). The common momentum
$p^F_{\rm comm}$ can be determined by minimizing $\Omega_{\rm
pair.}$ with respect to $p^F_{\rm comm}$. The result is
(expanding in $M_s$)\be p^F_{\rm
comm}=\mu-\frac{M_s^2}{4\mu}\,.\ee It is now easy to evaluate the
difference $\Omega_{\rm unpair.}-\Omega_{\rm pair.}$ at the order
$M_s^4$, with the result \be \Omega_{\rm pair.}-\Omega_{\rm
unpair.}\approx\frac
1{16\pi^2}\left(M_s^4-4\Delta^2\mu^2\right)\,.\ee We see that in
order to have condensation the condition \be
\mu>\frac{M_s^2}{2\Delta}\ee must be realized. The problem of one
massless and one massive flavor has been studied by
\cite{Kundu:2001tt}. However, one can simulate this situation by
letting the two quarks being both massless but with two different
chemical potentials, which is equivalent to have two different
Fermi spheres. The big advantage here is that one can use the
LOFF analysis discussed in Section \ref{section4}

Color superconductivity due to the non vanishing of the
condensates (\ref{lab2}) or (\ref{condensates}) results from a
mechanism analogous to the formation of an electron Cooper pair
in a BCS superconductor and, similarly to the BCS
superconductivity the only relevant fermion degrees of freedom
are those near the Fermi surface. Therefore a two-dimensional
effective field theory has been developed. We shall briefly
review it below, but our main interest is to delineate another
development of color superconductivity, i.e. the presence of a
LOFF superconducting phase.  Also in this case the condensation
is generated by the attractive color interaction in the
antitriplet channel. This phase of QCD has been mainly studied at
small temperatures, see e.g.
\cite{Alford:2000ze,Alford:2000sx,Bowers:2001ip,Leibovich:2001xr,
Rajagopal:2001yd,Bowers:2002xr}.  Similarly to the CFL and 2SC
phases, the QCD LOFF phase can be studied by the effective
theory, as shown in
\cite{Casalbuoni:2001gt,Nardulli:2001iv,Casalbuoni:2002pa,
Casalbuoni:2002hr,Casalbuoni:2002my}. This description is useful
to derive the effective lagrangian for the Goldstone bosons
associated to the breaking of the space symmetries, i.e. the
phonons. It is based on an analogy with the Heavy Quark Effective
Theory and is called High Density Effective Theory (HDET). To
describe these developments we organize this Section as follows.
In Subsection \ref{IVA} we give an outline of the HDET. We
specialize the formalism to the CFL phase in Subsection \ref{IVB}
and to the 2SC phase in Subsection \ref{sect:2sc}. The final
Subsections are devoted to the  LOFF phase in  QCD. In Subsection
\ref{LOFF_QCD}, after a general introduction to the subject, we
consider a Nambu-Jona Lasinio coupling  for a QCD liquid formed
by quarks with two flavors. Given the similarities with the BCS
four-fermion interaction arising from the electron phonon
interactions in metals, we can apply the same formalism discussed
in previous Sections. In the present case, however the two
species we consider are quarks of different flavors, $up$ and
$down$, with different chemical potentials $\mu_u$, $\mu_d$.  We
limit or analysis to the FF one plane wave state. However the
results of Subsection \ref{IIIC} are valid also for the QCD LOFF
state; in particular the guess on the favored structure at $T=0$
discussed in \cite{Bowers:2002xr} and reviewed in Subsection
\ref{IIIC.4} should point to the cubic structure as the most
favorable LOFF crystal. In Subsection \ref{IVE} we discuss the
differences induced by considering the one-gluon interaction
instead of the effective four fermion interaction. LOFF
superconductivity in QCD can be induced not only by a difference
in the quark chemical potential but also by mass differences
among the quarks. This situation is discussed in Subsection
\ref{IVF} that shows the role the strange quark mass can play in
favoring the LOFF phase.

\subsection{High Density Effective Theory}
\label{IVA}
 At very high baryonic chemical potential $\mu$ and
very small temperature ($T\to 0$)  it is useful to adopt an
effective description of QCD known as High Density Effective
Theory (HDET), see
\cite{Hong:1998tn,Hong:1999ru,Beane:2000ms,Casalbuoni:2000na},
and, for reviews, \cite{Casalbuoni:2001dw,Nardulli:2002ma}. Let
us consider the fermion field
 \be \psi(x)=\int\frac{d^4 p}{(2\pi)^4}e^{-ip\cdot x}\psi(p)\ .\ee
Since the relevant degrees of freedom are those near the Fermi
surface, we decompose the fermion momentum as
 \be p^\mu=\mu v^\mu+\ell^\mu\,,\label{eq:2}\ee where
$v^\mu=(0,{\bf v})$, $\bf v$ the Fermi velocity (for massless
fermions $|{\bf v}|=1$) and  $\ell^\mu$ is a residual momentum.
We also use $ V^\mu=(1,\,{\bf v})$ ,$\tilde V^\mu=(1,\,-{\bf
v})$.

We now introduce the velocity-dependent positive-energy
$\psi_{\bf v}$ and negative-energy $\Psi_{\bf v}$  left-handed
fields {\it via} the decomposition \be\psi(x)=\int\frac{d\bf
v}{4\pi} e^{-i\mu v\cdot x}
 \left[\psi_{\bf v}(x)+\Psi_{\bf v}(x)\right]\ .\ee
Here \be \psi_{\bf v}(x)=e^{i\mu v\cdot
x}P_+\psi(x)=\int_{|\ell|<\delta}\frac{d^4\ell}{(2\pi)^4}e^{-i\ell\cdot
x}P_+\psi(\ell)\label{eq:413}\ee and \be \Psi_{\bf v}(x)=e^{i\mu
v\cdot
x}P_-\psi(x)=\int_{|\ell|<\delta}\frac{d^4\ell}{(2\pi)^4}e^{-i\ell\cdot
x}P_-\psi(\ell)\,.\ee $P_\pm$ are projectors that for massless
quarks are defined by \be
 P_\pm\equiv \,P_\pm({\bf v})\,=\, \frac{1}2 \left(1\pm{\bm{\alpha}\cdot {\bf v}}\right)
 \ .\ee
The extension to massive quarks is discussed in
\cite{Casalbuoni:zx,Casalbuoni:2003cs}.
  The cut-off $\delta$ satisfies $\delta\ll\mu$ while being
  much larger than the energy gap.

Using the identities\bea \bar\psi_{\bf v}\gamma^\mu\psi_{\bf
v}&=&V^\mu\bar\psi_{\bf v}\gamma^0\psi_{\bf v}\ ,~~~~~~~~
\bar\Psi_{\bf v}\gamma^\mu\Psi_{\bf v}=\tilde V^\mu\bar\Psi_{\bf
v}\gamma^0\Psi_{\bf v}\ ,\cr \bar\psi_{\bf v}\gamma^\mu\Psi_{\bf
v}&=& \bar\psi_{\bf v}\gamma^\mu_\perp\Psi_{\bf v}\ ,~~~~~~~~~~~
\bar\Psi_{\bf v}\gamma^\mu\psi_{\bf v}=\bar\Psi_{\bf
v}\gamma^\mu_\perp\psi_{\bf v} \ ,\eea and substituting into the
Dirac part of the QCD lagrangian we obtain \be {\cal
L}_D=\int\frac{d\bf v}{4\pi} \Big[\psi_{\bf v}^\dagger iV\cdot
D\psi_{\bf v}+\Psi_{\bf v}^\dagger(2\mu+ i\tilde V\cdot
D)\Psi_{\bf v}\,+\, (\bar\psi_{\bf v}i\slash D_\perp\Psi_{\bf v}
+ {\rm h.c.} )\Big]\ ;\label{eq:257}\ee $\slash
D_\perp=D_\mu\gamma^\mu_\perp$ and $D_\mu$ is the covariant
derivative: $D^\mu=\partial^\mu+ig A^\mu$. We note that here
quark fields
 are   evaluated at the same Fermi velocity; off-diagonal terms
are subleading due to the Riemann-Lebesgue lemma, as they
 are cancelled by the rapid oscillations of the
exponential factor in the $\mu\to\infty$ limit (Fermi velocity
superselection rule). A similar behavior occurs in QCD in the
$m_Q\to\infty$ limit, when one uses the Heavy Quark Effective
Theory
\cite{Isgur:1989vq,Isgur:1990ed,Eichten:1990zv,Georgi:1990um},
and for reviews,
\cite{Neubert:1994mb,Manohar:2000dt,Casalbuoni:1997pg}.

We can get rid of the negative energy solutions by integrating
out the $\Psi_{\bf v}$ fields in the generating functional; at
tree level this corresponds to solve the equations of motion,
which gives \be iV\cdot D\, \psi_{\bf v}=0\ee and\be \Psi_{\bf
v}=-\frac{i}{2\mu+i\tilde V\cdot D}\,\gamma_0 \, \slash D_\perp\,
\psi_{\bf v}\ ,\ee  which shows the decoupling of $\Psi_{\bf v}$
in the $\mu\to\infty$ limit. In the resulting effective theory
for $\psi_{\bf v}$ only the energy and the momentum parallel to
the Fermi velocity are relevant and the effective theory is
two-dimensional.

It is useful to introduce  two separate fields\be\psi_\pm\equiv
\psi_{\pm\bf v}\ ;\ee therefore the average over the Fermi
velocities is defined as follows: \be \sum_{\bf v}=\int\frac{d\bf
v}{8\pi}\ . \label{sumvel} \ee The extra factor $1/ 2$ occurs
here because, after the introduction of the field with opposite
velocity $\psi_-$, one doubles the degrees of freedom, which
implies that the integration is only over half solid angle.

In conclusion, if ${\cal L}_{0}$ is the free quark lagrangian and
${\cal L}_{1}$ represents the coupling of quarks to one gluon,
the high density effective lagrangian can be written as \be {\cal
L}_D={\cal L}_{0}+{\cal L}_1+{\cal L}_2\ +\ (L\to
R)\,,\label{dirac} \ee where
 \bea {\cal L}_{0}& =&
\sum_{\bf v} \Big[\psi_+^\dagger iV\cdot
\partial\psi_+\ +\
\psi_-^\dagger i\tilde V \cdot \partial\psi_-\Big]\ ,\\
{\cal L}_1& =&i\,g\, \sum_{\bf v} \Big[\psi_+^\dagger iV\cdot
A\psi_+\ +\ \psi_-^\dagger i\tilde V \cdot A\psi_-\Big]\
\label{l1}\ ,\eea and \be {\cal L}_2=-\sum_{\bf v}\,P^{\mu\nu}\,
\Big[ \psi_+^\dagger\frac{1}{2\mu+i\tilde V\cdot D}D_\mu D_\nu
\psi_+ + \psi_-^\dagger\frac{1}{2\mu+iV\cdot D}D_\mu D_\nu
  \psi_-\Big]\ .\label{l2}
  \ee ${\cal L}_{2}$ is a non local lagrangian
 arising when one integrates over the  $\Psi_{\bf v}$
 degrees of freedom in the functional integration.
  It contains couplings of two quarks to any number of gluons
  and gives contribution to the gluon
 Meissner mass. We have put \be P^{\mu\nu}=g^{\mu\nu}-\frac 1
2\left[V^\mu\tilde V^\nu+V^\nu\tilde V^\mu\right]\
.\label{219}\ee
 This construction  is  valid for any theory
describing massless fermions at high density provided one
excludes
 degrees of freedom  far from the Fermi surface.

\subsection{CFL phase}
\label{IVB} Even though we shall consider the LOFF phase only for
two flavors, for completeness we present HDET for the 3-flavor
Color Flavor Locking phase as well. In the CFL phase the symmetry
breaking is induced by the condensates
 \be
\langle\psi_{i\alpha }^{L\,T} C\psi_{j\beta  }^L\rangle=
-\langle\psi_{i\alpha }^{R\,T}C\psi_{j\beta
}^R\rangle=\frac{\Delta}2\, \epsilon_{\alpha\beta
I}\epsilon_{ijI}~,\label{3.2.1}\ee where $\psi^{L,\,R}$ are Weyl
fermions and $C=i\sigma_2$. Eq. (\ref{3.2.1})  corresponds to the
invariant coupling ($\psi\equiv \psi_L$): \be -\frac{\Delta} 2
\sum_{I=1,3}\psi^T C\epsilon_I\psi\epsilon_I~-(L\to R)\ +
h.c.,\ee and $\left(\epsilon_I\right)_{ab}=\epsilon_{Iab}$.
Neglecting the negative energy components, for the Dirac fermions
$\psi_\pm$ we introduce the compact notation \be
\chi=\frac{1}{\sqrt 2}\left( \matrix{\psi_+\cr
C\psi^*_-}\right)\label{chi}\ee  in a way analogous to Equation
(\ref{ngspinor}). We also use a different basis for quark
fields:\be\psi_{\pm{\bf v}, i\alpha}=
\sum_{A=1}^9\frac{(\lambda_A)_{ i\alpha}}{\sqrt 2}\psi_{\pm}^A
~.\ee The CFL fermionic lagrangian has therefore the form:  \bea
{\cal L}_D&=& {\cal L}_{0}\,+\, {\cal L}_1\,+\, {\cal
L}_\Delta\,=\sum_{\vec v}\sum_{A,B=1}^9 \chi^{A\dagger}\left(
\matrix{iTr[T_A\,V\cdot D\,T_B] & -\Delta_{AB}\cr -\Delta_{AB} &i
Tr[T_A\,\tilde V\cdot D^*\,T_B]}\right)\chi^B\,+\, (L\to R)\,,
\label{cflcomplete0} \eea where \be \Delta_{AB}
=\,\Delta\,Tr[\epsilon_IT_A^T\epsilon_IT_B]\label{2.31} \ee and
\be T_A=\frac{\lambda_A}{\sqrt 2}\ . \ee Here
$\lambda_9=\lambda_0=\sqrt{\frac 2 3 }\times\bf 1$. We use the
identity ($g$ any 3$\times$3 matrix): \be
\epsilon_Ig^T\epsilon_I\,=\,g\,-\,Tr\,g \ ; \label{identity}\ee
we obtain \be \Delta_{AB}\,=\,\Delta_A\delta_{AB}\,,
\label{2.38}\ee where \be \Delta_1=\cdots=\Delta_8=\Delta
\label{2.39}\ee and \be \Delta_9=-2\Delta\ . \label{2.40}\ee The
CFL free fermionic lagrangian assumes therefore the form: \be
{\cal L}_0+\, {\cal L}_\Delta\,=\sum_{\bf v}~\sum_{A=1}^9
\chi^{A\dagger}\left(\matrix{iV\cdot \partial & -\Delta_A\cr
-\Delta_A &i\tilde V\cdot \de}\right)\chi^A\ +\ (L\to R)\
.\label{cflcomplete}\ee Clearly the equations of motion following
from this lagrangian are of the same type as the NG equations,
see Eq. (\ref{smenouno}). For  applications of HDET to the CFL
phase we refer the reader to \cite{Casalbuoni:2000na}.

\subsection{2SC phase}
\label{sect:2sc}
 For the two flavor case, which encompasses
both the 2SC model and the existing calculation in the LOFF
phase, we follow a similar approach. The symmetry breaking is
induced by the condensates
 \be
\langle\psi_{ i\alpha}^{L\,T}C\psi_{ j\beta}^L\rangle=
-\langle\psi_{ i\alpha}^{R\,T}C\psi_{
j\beta}^R\rangle=\frac{\Delta}2\, \epsilon_{\alpha\beta
3}\epsilon_{ij3}~,\ee and the invariant coupling is ($\psi\equiv
\psi^L$): \be {\cal L }_\Delta=-\frac{\Delta}2\, \psi^T
C\epsilon\psi\epsilon\ -(L\to R)+{\rm h.c.}~,\ee where
\be\epsilon=i\sigma_2\ .\ee We use a different basis for the
fermion fields  by writing the positive energy effective fields
$\psi_{\pm{\bf v}, i\alpha}$ as follows:\be \psi_{\pm{\bf v},
i\alpha}= \sum_{A=0}^5\frac{(\tilde\lambda_A)_{ i\alpha}}{\sqrt
2}\psi_{\pm}^A ~.\label{basis}\ee The $\tilde\lambda_A$ matrices
are defined in terms of the usual $\lambda$ matrices as follows:
\be\dd \tilde\lambda_0=\frac 1{\sqrt 3}\lambda_8\,+\, {\sqrt\frac
2 3}\lambda_0,~~~ \dd \tilde\lambda_A=\lambda_A\,(A=1,2,3),~~~\dd
\tilde\lambda_4=\frac{\lambda_{4-i5}}{\sqrt 2},~~~\dd
\tilde\lambda_5=\frac{\lambda_{6-i7}}{\sqrt 2}\,.\ee We also
define $\tilde \epsilon=i\lambda_2$. After the introduction,
analogously to (\ref{chi}), of the fields $\chi^A$, the 2SC
fermionic lagrangian assumes the form:  \bea {\cal L}_D&=& {\cal
L}_{0}\,+\, {\cal L}_1\,+\,\, {\cal L}_\Delta\cr&=& \sum_{\bf
v}\sum_{A,B=0}^5 \chi^{A\dagger}\left(\matrix{iTr[\tilde
T_A\,V\cdot D\,\tilde T_B ]& -\Delta_{AB}\cr -\Delta_{AB}
&iTr[\tilde T_A\,\tilde V\cdot D^*\,\tilde T_B
]}\right)\chi^B\,+\, (L\to R)\ . \label{2sccomplete0} \eea Here
\bea \Delta_{AB} &=&\,\frac{\Delta}{2}\,Tr[\tilde\epsilon
\tilde\lambda_A^T\tilde\epsilon \tilde\lambda_B]
\hskip1cm(A,B=0,...3)\,,\cr \Delta_{AB}
&=&\,0\hskip3.cm(A,B=4,5)\ . \eea and \be \tilde
T_A=\frac{\tilde\lambda_A}{\sqrt 2}\hskip1cm(A=0,...,5)\ .\ee
 Analogously to
(\ref{identity}) we use the identity: \be \tilde\epsilon
g^T\tilde\epsilon\,=\,g\,-\,Tr\,g \ ; \ee we obtain \be
\Delta_{AB}\,=\,\Delta_A\delta_{AB}\,, \ee where
  $\Delta_{A}$ is defined as follows:
\be \Delta_A=\left(
+\,\Delta,\,-\Delta,\,-\Delta,\,-\Delta,\,0,\,0\right)\
.\label{delta2sc}\ee Therefore the effective lagrangian for free
quarks in the 2SC model
 can be written as follows
 \be {\cal L}_0 \,+\,{\cal L}_\Delta=\sum_{\bf v}~\sum_{A=0}^5
\chi^{A\dagger}\left(\matrix{iV\cdot \partial &- \Delta_A\cr-
\Delta_A &i\tilde V\cdot \de}\right)\chi^A\ +\ (L\to R)\
.\label{2sccomplete}\ee
\subsection{LOFF phase in QCD} \label{LOFF_QCD}

We shall assume here that in the most interesting phenomenological
applications, i.e. in compact stars (see Section \ref{stars}),
there is a significant difference between the Fermi momenta of
different flavors. Since this produces a difference in the
densities, the BCS phase may be disrupted
\cite{Alford:1999pa,Schafer:1999pb} and a phase analogous to the
LOFF phase might arise. The case of a LOFF phase in QCD was also
discussed in Ref. \cite{Son:2000xc,Splittorff:2000mm} in the
context of quark matter at large isospin density. Differences in
the Fermi momenta in these examples arise both from the difference
in the chemical potential, due to the weak equilibrium, and from
the mass difference between the strange and the up and down
quarks. A complete study requires to take into account both
effects. This has been made in Ref. \cite{Kundu:2001tt}. We will
discuss this paper below. Here we will consider a simpler case
where all quark are massless but have  different chemical
potentials \cite{Alford:2000ze}. To simplify further the problem
we will restrict ourselves to the case of two massless quarks with
chemical potentials $\mu_u$ and $\mu_d$ given by \be
\mu_u=\mu+\delta\mu,~~~\mu_d=\mu-\delta\mu\,.\ee These equations
are the same as (\ref{eq:2.2}) but now up and down refer to
flavor.

Everything  goes according to the discussion made in Sections
\ref{homogeneous_superconductor} and \ref{section4} except that
now the density of gapped  states at the Fermi surface is
multiplied by a factor 4, coming from the two colors and the two
flavors. In fact, the condensate has the form \be
\langle\psi^\alpha_i\psi^\beta_j\rangle\propto\epsilon^{\alpha\beta
3}\epsilon_{ij}\,,\ee where $\alpha,\beta=1,2,3$ and $i,j=1,2$
are respectively color and flavor indices. Other differences are
in the value of the Fermi velocity, which is $v_F=1$, since we
deal with massless fermions, and in the Fermi momentum which is
given by $p_F=\mu$. As a consequence the density of states is now
\be\rho=4\,\frac{\mu^2}{\pi^2}\,.\ee It follows that the first
order transition from the homogeneous phase to the normal one, in
the weak coupling limit, is given, using (\ref{oma1}), by
\be\Omega_{\Delta}(\delta\mu)-\Omega_0(\delta\mu)=\frac\rho
4(2\delta\mu^2-\Delta_0^2)=\frac{\mu^2}{\pi^2}
(2\delta\mu^2-\Delta_0^2)=0\,.\ee Applying the results obtained
in the  Section \ref{section4} to color superconductivity
requires some care. For instance, although only two colors are
gapped, in order to describe the mixed phase it is necessary the
use of a proper treatment of the two ungapped quarks
\cite{Bedaque:1999nu}. Another situation that can be present in
QCD but not in condensed matter is the case of equal chemical
potentials  with different Fermi momenta due to unequal masses.
This is discussed in Ref. \cite{Alford:1999pa,Schafer:1999pb}.
However in the realistic case different chemical potentials must
be considered.

We will describe also the LOFF phase using the formalism of
fields close to the Fermi surface, although in the present case
the corrections to the leading order are expected to be larger
since we are not considering asymptotic values of the chemical
potential. This formalism is very close to the NG formalism
developed  in Section \ref{section2}. We  consider a four-fermion
interaction modelled on one-gluon exchange, that is \be {\cal
L}_I=-\frac 3 8 G\bar\psi\gamma^\mu
\lambda^a\psi\,\bar\psi\gamma^\mu
\lambda^a\psi\label{eq:291}\,,\ee where $\lambda^a$ are Gell-Mann
matrices. We then introduce the fields $\psi_{i+}^\alpha$ through
the procedure outlined in Section \ref{sect:2sc}. We perform the
same transformation $\exp(-i\mu v\cdot x)$ for both flavors. For
simplicity, in the rest of this Section we will denote the fields
$\psi_{i+}^\alpha$ by $\psi_{i}^\alpha$. Separating the
left-handed  and the right-handed modes the previous interaction
can be written as \be {\cal L}_I=-\frac G
2\,(3\delta^\alpha_\delta\delta^\gamma_\beta-
\delta^\alpha_\beta\delta^\gamma_\delta)\,\epsilon_{\dot a \dot
c}\,\epsilon_{bd}\,\psi^{\dagger i}_{\alpha \dot
a}\,\psi^{\dagger j}_{\gamma \dot c}\,\psi^{\beta}_{ib}\,
\psi^{\delta}_{jd}\equiv
V_{\beta\delta}^{\alpha\gamma}\,\psi^{\dagger
i}_\alpha\,\psi^{\dagger
j}_\gamma\,\psi_i^\beta\psi_j^\delta\,,\ee where in the last
expression the sum over the spin indices, $\dot a,\dot c, b,d$ is
understood and\be V_{\beta\delta}^{\alpha\gamma}=-\frac G
2\,(3\delta^\alpha_\delta\delta^\gamma_\beta-
\delta^\alpha_\beta\delta^\gamma_\delta)\,.\ee In obtaining this
result we have used the  identities \be\sum_{a=1}^8
(\lambda^a)_{\alpha\beta}(\lambda^a)_{\delta\gamma}=\frac 2
3(3\delta_{\alpha\gamma}\delta_{\beta\delta}-
\delta_{\alpha\beta}\delta_{\gamma\delta})\ee and \be
(\sigma_\mu)_{\dot a b}(\tilde\sigma^\mu)_{d \dot
c}=2\epsilon_{\dot a\dot c}\epsilon_{bd}\,.\ee Here
\be\sigma^\mu=(1,\bm{\sigma}),~~~\tilde\sigma^\mu=(1,-\bm{\sigma})\,,\ee
with $\bm{\sigma}$ the Pauli matrices. As in Section
\ref{section2} we divide ${\cal L}_I$ in two pieces
 \be {\cal L}_{cond}=
V_{\beta\delta}^{\alpha\gamma}(\psi_\alpha^{\dagger
i}\psi_\gamma^{\dagger j}\langle\psi_i^\beta\psi_j^\delta\rangle+
\psi_i^\beta\psi_j^\delta\langle\psi_\alpha^{\dagger
i}\psi_\gamma^{\dagger j}\rangle)\,+\,(L\to R)\ee and \be{\cal
L}_{int}= V_{\beta\delta}^{\alpha\gamma}(\psi_\alpha^{\dagger
i}\psi_\gamma^{\dagger j}-\langle\psi_\alpha^{\dagger
i}\psi_\gamma^{\dagger j}\rangle)(\psi_i^\beta\psi_j^\delta-
\langle\psi_i^\beta\psi_j^\delta\rangle)\,+\,(L\to R)\,.\ee  The
first piece can be written as \be{\cal L}_{cond}=-\frac 1
2\epsilon_{\alpha\beta
3}\epsilon^{ij}(\psi_i^\alpha\psi_j^\beta\Delta e^{2i{\bf
q}\cdot{\bf r}}\,+\,{\rm c.c.})\,+\,(L\to R)\,,\ee where we have
defined \be \Gamma_S\, e^{2i{\bf q}\cdot{\bf r}}=-\frac 1
2\,\epsilon^{\alpha\beta 3}\epsilon_{ij}
\langle\psi_\alpha^i\psi_\beta^j\rangle\ee and \be
\Delta=G\Gamma_S\,.\ee The quadratic part of the lagrangian
${\cal L}_{cond}$ in terms of the NG fields can be written as
\be{\cal L}_{cond}^{(2)}=\frac 1 2\sum_{\alpha,
i}\chi_i^{\dagger\alpha} (S^{-1})_{\alpha j}^{\beta
i}\chi_\beta^j\,,\ee where, in momentum space \be
(S^{-1})_{\alpha j}^{\beta
i}=\left(\matrix{\delta_{\alpha\beta}(\delta_{ij} V\cdot
\ell+\delta\mu(\sigma_3)_{ij}\delta(\ell-\ell^\prime)) &
-\epsilon_{\alpha\beta
3}\epsilon_{ij}\Delta\delta(\ell-\ell^\prime+2 q)\cr
-\epsilon_{\alpha\beta
3}\epsilon_{ij}\Delta\delta(\ell-\ell^\prime-2 q)&
\delta_{\alpha\beta}(\delta_{ij} \tilde V\cdot
\ell+\delta\mu(\sigma_3)_{ij}\delta(\ell-\ell^\prime))}\right)\,,\ee
and $q^\mu=(0,{\bf q})$. Using this expression and performing the
same derivation as  in Section \ref{section2} we find the gap
equation \be \Delta=2ig\Delta\int\frac{d{\bf
v}}{4\pi}\frac{d\ell_0}{2\pi} d\ell_\parallel
\frac{\mu^2}{2\pi^2} {\rm
Tr}\frac{1}{(V\cdot(\ell+q))+\delta\mu\sigma_3)(\tilde
V\cdot(\ell-q))+\delta\mu\sigma_3)-\Delta^2}\,.\ee Comparing this
result with Eq. (\ref{gapt0}), one notices a factor 2 coming from
the trace on the color; the remaining trace is on the flavor
indices where the matrix $\sigma_3$ acts. Performing the trace
and making explicit the $i\epsilon$ prescriptions for the energy
integration we find \be
\Delta=i\frac{g\rho}2\Delta\int\frac{d{\bf
v}}{4\pi}\int_{-\delta}^{+\delta}\,d\xi\, \frac{dE}{2\pi}
\frac{1}{(E-\bar\mu+i\epsilon\,{\rm
sign}E)^2-\xi^2-\Delta^2}\,,\ee where $\rho=4\mu^2/\pi^2$ is the
relevant density at the Fermi surface and we have defined
$E=\ell_0$ and $\xi=\ell_\parallel$ to emphasize the similarity
with the original LOFF equation (see Section \ref{section4}).
Moreover \be\bar\mu=\delta\mu-{\bf v}\cdot{\bf q}\,.\ee
Performing the integration over the energy we get \be
1=\frac{g\rho}2\int\frac{d{\bf
v}}{4\pi}\int_0^{\delta}\frac{d\xi}{\sqrt{\xi^2+\Delta^2}}\,
\theta(\epsilon-|\bar\mu|)\,.\ee Since
\be\theta(\epsilon-|\bar\mu|)=1-\theta(-\epsilon-\bar\mu)-\theta(-\epsilon+
\bar\mu)\,,\ee we get exactly the LOFF gap equation (compare with
Eq. (\ref{GAP5})), except for the different definition of the
density of states.

We have already shown that in the present case there is a first
order transition in $\delta\mu$, between the homogenous state
(that from now on will be referred to as the BCS state) and the
normal state. Furthermore, from Section \ref{section4} we know
that there is a second order transition between the LOFF state
and the normal one. These results are valid also in the present
case with the only change in the density of gapped states at the
Fermi surface, which, as already stressed, is now a factor of
four larger than the one for electrons. We recall from that
analysis that around the second order critical point
$\delta\mu_2=0.754\Delta_0$ (with $\Delta_0$ the BCS gap) we have
(see Eq. (\ref{eq:168})) \be
\Delta_{LOFF}=\sqrt{1.757\,\delta\mu_2(\delta\mu_2-\delta\mu)}=
1.15\,\Delta_0\sqrt{\frac{\delta\mu_2-\delta\mu}{\Delta_0}}\,.\ee
As for the grand potential, we have from Eq. (\ref{oma1})
\be\Omega_{BCS}-\Omega_{normal}=\frac 1
4\rho(2\delta\mu^2-\Delta_0^2)\ee and from Eq. (\ref{044}) \be
\Omega_{LOFF}-\Omega_{normal}=-0.439\,\rho(\delta\mu-\delta\mu_2)^2\,.\ee
These results are summarized in Fig. \ref{fig:17}, where we plot
the grand potentials for the different phases.
\begin{center}
\begin{figure}[htb]
\epsfxsize=7truecm \centerline{\epsffile{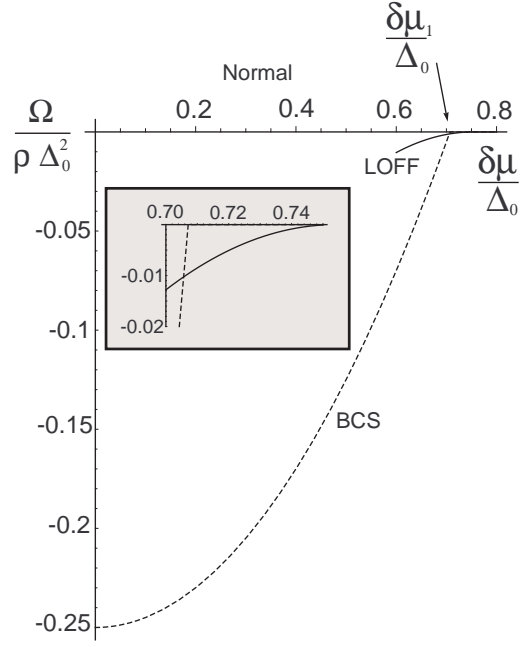}}
\caption{{\it The figure shows the differences between the grand
potential for the BCS and the normal state and that between the
LOFF and the normal state plotted vs. $\delta\mu$. The grand
potentials are normalized to $\rho\Delta_0^2$. The inset shows
the intersection of the two curves close to $\delta\mu_1$. The
solid lines correspond to the LOFF case, whereas the dotted ones
to the BCS case.\label{fig:17}}}
\end{figure}
\end{center}
Since the interval $(\delta\mu_1,\delta\mu_2)$ is rather narrow
there is practically no difference between the values of
$\delta\mu$ corresponding to the BCS-normal transition
($\delta\mu_1=\Delta_0/\sqrt{2}$) and the value corresponding to
the BCS-LOFF transition. This can be visualized easily from Figs.
\ref{fig:17} and \ref{fig:18}. The figures have been obtained by
using the previous equations in the Ginzburg-Landau expansion
around $\delta\mu_2$, but they are a very good approximation of
the curves obtained numerically
\cite{Takada:1969ab,Alford:1999pb}
\begin{center}
\begin{figure}[htb]
\epsfxsize=8truecm \centerline{\epsffile{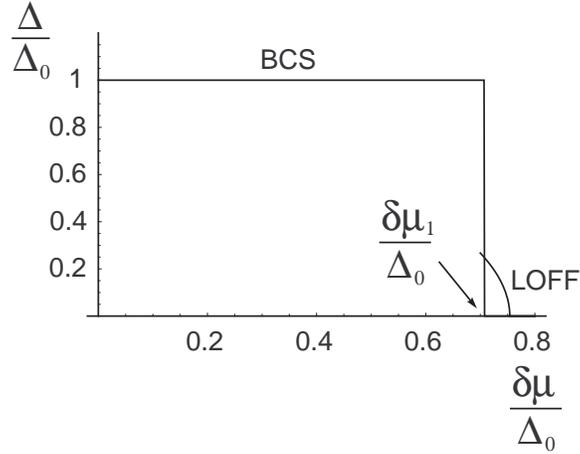}}
\caption{{\it The figure shows the condensates of the  BCS and
LOFF phases vs. $\delta\mu$.\label{fig:18}}}
\end{figure}
\end{center}
 All the discussion
here has been done in the weak coupling limit. For a more correct
treatment see \cite{Alford:1999pb} where the results from the
numerical integration of the gap equation are given. In
particular we want to stress the results on the size of the
window obtained by these authors. If $\delta\mu_1$ and
$\delta\mu_2$ are evaluated for a general coupling and at the
same time one takes into account corrections from the chemical
potential in the measure of integration, the windows get smaller
and smaller for increasing BCS gap $\Delta_0$. The corrections in
the chemical potential arise from the momentum integration, which
is made on a shell of height 2$\delta$ but with an integration
measure given by $p^2 dpd\Omega$, rather than $p_F^2 dp d\Omega$
as usually done in the treatment of the BCS gap in the weak
coupling limit. The results are illustrated in Fig. \ref{fig:21},
where the behavior of the critical points vs. the BCS gap are
shown. The curves are plotted for a range of values of the cutoff
$\Lambda$ ranging from 0.8 to 1.6 GeV. The cutoff dependence is
not very strong, in particular for $\delta\mu_2$, moreover the
window closes for $\Delta_0$ between 80 to 100 MeV, according to
the chosen value of $\Lambda$.
\begin{center}
\begin{figure}[htb]
\epsfxsize=7.5truecm \centerline{\epsffile{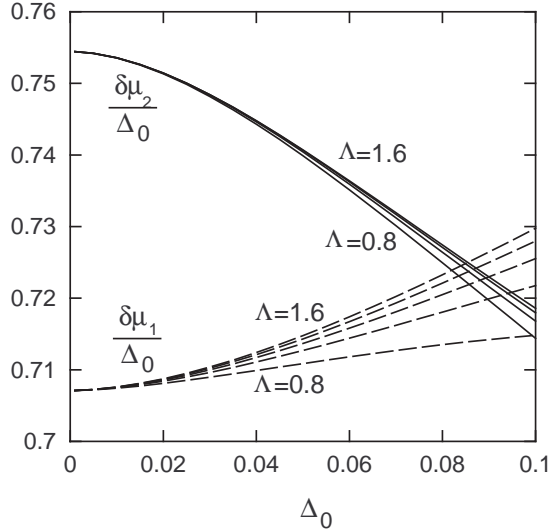}}
\caption{{\it The figure shows the critical points $\delta\mu_1$
(solid lines) and $\delta\mu_2$ (dashed lines)  in $\Delta_0$
units vs. $\Delta_0$, for cutoff values $\Lambda=\mu+\delta$
ranging from 0.8 GeV up to 1.6 GeV and for $\mu=0.4$ GeV. This
figure is taken from \cite{Alford:2000ze}. \label{fig:21}}}
\end{figure}
\end{center}
In \cite{Alford:1999pb} the presence of a vector condensate in
the QCD LOFF phase is also discussed.  The reason why this
condensate can be formed in QCD but not in condensed matter is
the following. Both in the BCS and in the LOFF phases the
coupling is between fermions of the same helicity. In the BCS
phase the fermions have also opposite momentum giving rise to a
$J=0$ pair. On the other hand in the LOFF phase momenta are not
exactly aligned, therefore a small component of $J=1$ condensate
may arise. A spin 1 state is symmetric in the spin indices and
therefore  Fermi statistics forbids it for electron pairing. On
the other hand in QCD with two flavors one can form a state
antisymmetric in color and symmetric in flavor, and the Pauli
principle is satisfied. Therefore the structure of the vector
condensate is \cite{Alford:1999pb} \be
\langle(\sigma_1)_{ij}\epsilon_{\alpha\beta 3}\psi^\alpha_{i
L}\sigma^{0i}\psi^\beta_{j L}\rangle=-2i\frac {q^i}{|{\bf
q}|}\Gamma_V\, e^{2i{\bf q}\cdot{\bf r}}\,.\ee The ratio
$\Gamma_V/\Gamma_S$ is practically constant within the LOFF
window varying between 0.121 at $\delta\mu_1$ and 0.133 at
$\delta\mu_2$. However this condensate does not contribute to the
grand potential in the present case \cite{Alford:1999pb};
therefore it does not change the original LOFF results. The
situation is different if, instead of using the NJL interaction
(\ref{eq:291}), use is made of the following interaction \be
{\cal L}_I=-\frac 3 8 \left[G_E(\bar\psi\gamma^0
\lambda^a\psi)(\bar\psi\gamma^0\lambda^a\psi)-G_M
(\bar\psi\gamma^i
\lambda^a\psi)(\bar\psi\gamma^i\lambda^a\psi)\right]\,.\ee This
expression is not Lorentz invariant, but since we are trying to
model QCD at finite density, there is no reason to use a Lorentz
invariant effective action. For instance, at high density the
electric gluons are expected to be screened, whereas the magnetic
ones are Landau damped. In particular, it has been shown by
\cite{Son:1998uk} that  at high density the magnetic gluon
exchange is dominating in the pairing mechanism, which can be
simulated assuming $G_E\ll G_M$.

For the following discussion it is convenient to introduce the
quantities  \be G_A=\frac 1 4(G_E+3G_M),~~~G_B=\frac
14(G_E-G_M)\,.\ee For $G_E=G_M=G$, as discussed above one has \be
G_A=G, ~~~~G_B=0\,.\ee At zero density we expect $G_B=0$, whereas
at high density we expect $G_E=0$ or $G_B/G_A=-1/3$. Therefore
the relevant physical region for $G_B/G_A$ should be given by \be
-\frac 1 3\le\frac{G_B}{G_A}\le 0\,.\label{eq:318}\ee The gap
parameters are now defined by \be \Delta=G_A\Gamma_S,
~~~~\Delta_V= G_B\Gamma_V\,.\ee Since the grand potential  and
the quasi particle energy are determined by the gap, for the
Lorentz invariant case, $G_B=0$, there is no contribution from
the vector condensate. For $G_B\not =0$, one has to solve two
coupled gap equations \cite{Alford:1999pb}. The most interesting
result found by these authors is about the LOFF window which is
modified by the presence of the $J=1$ gap. The result is shown in
Fig. \ref{fig:19}.
\begin{center}
\begin{figure}[htb]
\epsfxsize=8truecm \centerline{\epsffile{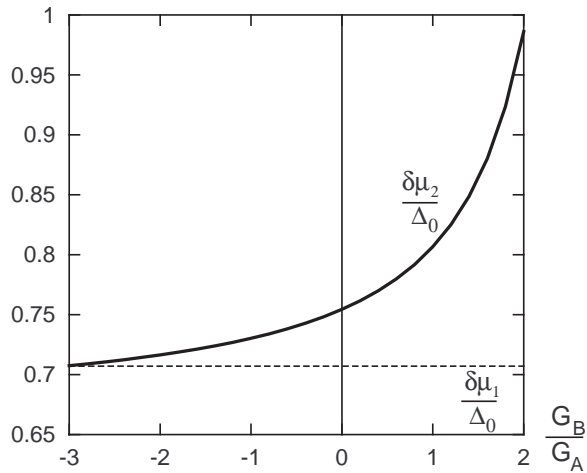}}
\caption{{\it The figure shows the variation of the critical
points $\delta\mu_1/\Delta_0$ (dotted line) and
$\delta\mu_2/\Delta_0$ (solid line) with $G_B/G_A$. This figure
is taken from \cite{Alford:2000ze}. \label{fig:19}}}
\end{figure}
\end{center}
The LOFF window closes at $G_B/G_A=-3$ and increases with
increasing $G_B/G_A$. For $G_B/G_A$ inside the physical region
(see Eq. (\ref{eq:318})) the maximal opening is for $G_B=0$.
However, inside the physical region the variation of the window
is rather small. Since $\delta\mu_1$ is essentially defined by
the BCS-normal state transition, it is given by
$\Delta_0/\sqrt{2}$ independently of the vector condensate.
However, the second order critical point $\delta\mu_2$ is rather
sensitive to $G_B/G_A$, the reason being that, for $G_A=0$, the
$J=1$ channel is attractive for $G_B>0$ and repulsive for
$G_B<0$. Therefore the stability of the LOFF state is reinforced
by the vector condensate in the region $G_B>0$.

Let us close this Section considering a different pairing
discussed in
\cite{Deryagin:1992rw,Shuster:1999tn,Park:1999bz,Rapp:2000zd}. It
is a  quark-hole pairing with non-zero momentum at large baryon
density. This produces a $\langle \bar q q\rangle$ condensate
varying in space with a wave number $2\mu$, to be contrasted with
$2|{\bf q}|\approx 2|\delta\mu|$. This state is energetically
favorable only for very large values of the number of colors
\cite{Deryagin:1992rw}. \cite{Shuster:1999tn,Park:1999bz} found
that $N_c$ should be larger than about 1000.

\subsection{One-gluon exchange approximation}
\label{IVE}
 The previous results have been obtained in the case
of a NJL interaction. In \cite{Leibovich:2001xr} the case of the
one-gluon exchange interaction has been studied. Of course this
would be a realistic case only at very high densities
\cite{Rajagopal:2000rs} where, presumably, the CFL phase
dominates over the LOFF phase. However the study of different
interactions allows to understand the model dependence of the
LOFF window. In \cite{Leibovich:2001xr} the standard QCD vertex
is used in conjunction with the following propagator for the
gluon \be
D_{\mu\nu}=\frac{P_{\mu\nu}^T}{p^2-G(p)}+\frac{P_{\mu\nu}^L}{p^2-F(p)}\,,\ee
where \bea &P_{ij}^T=\delta_{ij}-\dd{\frac{p_i p_j}{|{\bf
p}|^2}},~~~P^T_{00}=P^T_{0i}=0\,,\nn\\
&P_{\mu\nu}^L=-g_{\mu\nu}+\dd{\frac{p_\mu
p_\nu}{p^2}}-P_{\mu\nu}^T\,.\eea Here \be G(p)=\frac\pi 4
m^2\frac{p^0}{|{\bf p}|}\ee describes the Landau damping and \be
F(p)=m^2\,,\ee where $m^2$ is the Meissner mass evaluated for 2
flavors \be m^2=g^2\frac{\mu^2}{\pi^2}\,.\ee The expressions for
$F(p)$ and $G(p)$ are obtained in the hard-loop approximation
\cite{LeBellac:1996ab} and evaluated here for $p_0\ll |{\bf
p}|\approx \mu$ (we recall that $\mu$ is the average chemical
potential). Solving the gap equation it is found that the LOFF
window is enlarged of about a factor 10 at average chemical
potential $\mu=400$ MeV. In fact, whereas $\delta\mu_1$, as
already noticed, is essentially fixed by the BCS-normal state
transition at the value $\delta\mu_1=\Delta_0/\sqrt{2}$,
$\delta\mu_2$ increases in a dramatic way. At $\mu=400$ MeV the
authors of Ref. \cite{Leibovich:2001xr} find \be
\delta\mu_2=1.24\Delta_0 \Rightarrow
\delta\mu_2-\delta\mu_1=0.55\Delta_0\,,\ee to be compared with
the NJL case, where \be\delta\mu_2=0.754\Delta_0 \Rightarrow
\delta\mu_2-\delta\mu_1=0.05\Delta_0\,.\ee At $\mu=10^3$ MeV the
window is about 60 times larger than the window for the
point-like case. In general, by increasing $\mu$, $\delta\mu_2$
increases as well. The interpretation of these results, according
to \cite{Leibovich:2001xr}, goes as follows. For weak coupling
the $q-q$ scattering via one-gluon exchange is mostly in the
forward direction. This implies that, after the scattering has
taken place, quarks remain close at the angular position
possessed before the scattering, meaning that the theory is
essentially 1+1 dimensional. In fact, in this case the only
possible value for $2|{\bf q}|$ is $\mu_d-\mu_u=2\delta\mu$. This
is not exactly the case in 3+1 dimensions. In fact, as it can be
seen from Fig. \ref{fig2}, $2|{\bf q}|$ is generally bigger than
$2\delta\mu$. Furthermore, it is known from the 1+1 dimensional
case \cite{buzdin:1983cd,buzdin:1987ef} that in the weak coupling
limit $\delta\mu_2/\Delta_0\to\infty$. Both these features have
been found in \cite{Leibovich:2001xr}.

A  similar analysis has been done in Ref. \cite{Giannakis:ab}.
The results found here are somewhat different from the ones
discussed before. In particular it is found that at weak
coupling: \be
\delta\mu_2=0.968\Delta_0\Rightarrow\delta\mu_2-\delta\mu_1=0.26\Delta_0\,,\ee
with an enhancement of the window of a modest factor 5 with
respect to the point-like interaction. However, the evaluation
made in this paper consists in an expansion around the
tricritical point (called by these authors $\delta_{on}$)
implying, in particular, an expansion in $|{\bf q}|$. In order to
compare the results of these two groups one should extrapolate
the results of Ref. \cite{Giannakis:ab} to zero temperature. It
is not evident, at least to us, that this can be safely made. Of
course, also the physical interpretation is different. According
to \cite{Giannakis:ab} in 3+1 dimensions, increasing $\delta\mu$
implies a  reduction of the phase space and therefore a smaller
gap and  a smaller $\delta\mu_2$. This reduction effect,
according to these authors, overcomes the enhancement due to the
1+1 dimensional effect discussed before.

In our opinion the case of the one-gluon exchange in the LOFF
phase deserves further studies. In fact a sizeable increase of
the LOFF window  would make the LOFF state very interesting as
far as the applications to compact stellar objects are concerned.
\subsection{Mass effects}
\label{IVF} In Ref. \cite{Kundu:2001tt} the combined effect of
having two quarks with different chemical potentials and one of
the two quarks being massive has been studied. In the free case
the Fermi momenta are given by (assuming that the pair contains
an up and a strange quark) \be p_F^u=\mu-\delta\mu,
~~~~p_F^s=\sqrt{(\mu+\delta\mu)^2-M_s^2}\,.\ee Assuming both
$\delta\mu/\mu$ and $m_s/\mu$ to be much smaller than one, one
finds \be |p_F^u-p_F^s|\approx
2\Big|\delta\mu-\frac{M_s^2}{4\mu}\Big|\,.\ee The effect of
$M_s\not=0$ amounts to something more than the  simple shift
$\delta\mu\to\delta\mu-M_s^2/4\mu$. In fact, let us recall that
at $M_s=0$ the BCS condensate is not changed by $\delta\mu$ as
long as $\delta\mu<\delta\mu_1$. However $\Delta_0$ decreases
with $M_s$, see the results in \cite{Kundu:2001tt} and
\cite{Casalbuoni:zx}. Furthermore, for $M_s^2/\mu^2\ll 1$ the
decreasing is practically linear. This produces corrections in
the grand potential of order $\Delta_0^2(0) M_s^2$. For small
values of $\delta\mu-M_s^2/4\mu$ we have BCS pairing, whereas for
large values there is no pairing and the system is in the normal
phase. Therefore the BCS-normal transition is $M_s$ dependent and
occurs for $\delta\mu$ approximately given by \be
\Big|\delta\mu-\frac{M_s^2}{4\mu}\Big|=\frac{\Delta_0(M_s)}{\sqrt{2}}\,.
\label{eq:330}\ee It can be noted that differently from the case
$M_s=0$ this condition is not symmetric for
$\delta\mu\to-\delta\mu$ and the LOFF phase can exist in two
different windows in $\delta\mu$, above
($\delta\mu_2>\delta\mu_1$) and below ($\delta\mu_2<\delta\mu_1$)
the BCS region; in any case, for $M_s=0$ one gets back the
Clogston-Chandrasekar limit. In order to discuss the size of the
window the correct variable is \be
\frac{\delta\mu_2(M_s)-\delta\mu_1(M_s)}{\Delta_0(M_s)}\,.\ee At
weak coupling (small $\Delta_0(0)$) it is found that the window
is essentially the same as for the case $M_s=0$. Otherwise the
window generally increases with $M_s$ as shown in  Fig.
\ref{fig:20} for various values of $M_s$. We have plotted both
the cases $\delta\mu>0$ (left panel) and $\delta\mu<0$. This
shows that the LOFF phase is rather robust for $M_s\not=0$.
\begin{center}
\begin{figure}[htb]
\epsfxsize=13truecm \centerline{\epsffile{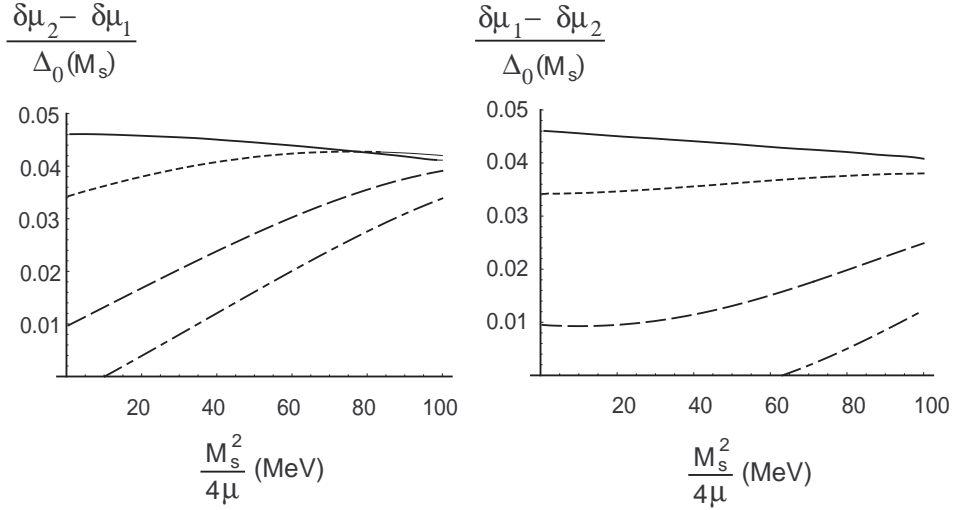}}
\caption{{\it In the left (right) panel  the width of the LOFF
window above (below) the BCS region is reported. The four curves
correspond to the following values of $\Delta_0(0)(MeV)$: 10
(solid line), 40 (dotted line), 80 (dashed line), 100
(dash-dotted line). This figure is taken from
\cite{Kundu:2001tt}. \label{fig:20}}}
\end{figure}
\end{center}
\section{Phonon and gluon effective lagrangians}
\label{phonons} Translational and rotational invariance are
spontaneously broken within a LOFF phase. The energy gap is not
uniform and actually is expected to vary according to some
crystalline structure, as result of the analysis developed in the
previous Sections. The crystal defined by the space modulation of
the gap can fluctuate and its local deformations define phonon
fields $\phi^{(i)}$ that are the Nambu-Goldstone bosons
associated to the breaking of the translational symmetry. The
number of the phonon fields is equal to the number of the broken
generators of the translation group. The existence of long
wavelength oscillations with phonon dispersion law was already
noted in \cite{FF}. More recently an effective lagrangian for
phonons in a QCD medium has been developed in
\cite{Casalbuoni:2001gt,Casalbuoni:2002pa,Casalbuoni:2002hr,Casalbuoni:2002my}
and we wish to review it in this Section, dominantly dedicated to
the QCD LOFF phase. For color superconductivity only the $T\to 0$
case is physically interesting and we shall consider only this
limit. However the theory developed in this Section could be
extended to $T\neq 0$ as well as  to other physical cases (solid
state, nuclear physics).

Being long wavelength oscillations of the crystalline LOFF
structure, the phonons exist only if the quarks of the Cooper
pair are in the pairing region. This is a portion of the phase
space around the Fermi surface and  is formed by a few annular
rings, that are likely to be contiguous, according to the
discussion in \cite{Bowers:2002xr}, see also the discussion in
Section \ref{IIIC}. The effective theory for the phonon fields
$\phi^{(i)}$ has to display this behavior and therefore the
phonon-quark coupling must vanish outside the pairing region. The
mathematical formalization of this behavior is rather involved
and some approximation is needed.
\cite{Casalbuoni:2002pa,Casalbuoni:2002my} write the phonon-quark
interaction using the HDET discussed in Section \ref{IVA}. They
introduce effective velocity dependent fermion fields and the
lagrangian as a sum of terms, each characterized by its own Fermi
velocity ${\bf v}$. Also the quark-phonon coupling constant
becomes velocity-dependent and is proportional to
\be\Delta_{eff}\propto\Delta\sum_{k}\sum_{\bf v}\, \frac\pi
R\,\delta_R[h({\bf v\cdot\hat n}_k)]\label{eq:5.1}\ .\ee Here
${\bf\hat n}_k $ are the vectors defining the LOFF crystal, $R$
is a parameter and $\delta_R[h(x)]$ is a function that vanishes
outside the pairing region. More precisely, it reaches its
maximum when the pairing quarks are on the Fermi surface and
decreases when they leave it. By this approximation an evaluation
of the phonon effective lagrangian is possible. In Subsection
\ref{hdet} we consider the HDET for the inhomogeneous LOFF state
and write the quark-phonon lagrangian. Below, we discuss two
crystalline structures. First we consider the Fulde-Ferrell one
plane wave, which is the benchmark case for the whole LOFF
theory; next we shall examine the cubic structure, already
studied in Subsection \ref{IIIC.4}, because this seems the most
favored crystalline structure according to \cite{Bowers:2002xr}.
On the basis of symmetry arguments one can write down the
effective phonon lagrangians for the two cases. This is done in
Subsections \ref{VB} and \ref{Cubic structure}, whereas in
Subsections \ref{opw} and \ref{VE} we show how the parameters of
the effective lagrangian can be computed by the HDET. Let us
mention here that the parameter $R$ appearing in (\ref{eq:5.1})
should be fixed by a comparison of the gap equation computed in
the HDET and the approach discussed in Subsection \ref{section4}
for the FF state and in Section \ref{ginzburg_landau} for generic
structures. This comparison has not yet been done and therefore,
in the discussion below, we leave $R$ as a parameter, even
though, in the case of a cubic structure, the requirement that
the annular rings are contiguous can be used to fix its value. We
conclude this Section with a discussion in \ref{VF} on the
modifications induced by the LOFF pairing of quarks on the gluon
lagrangian.

\subsection{Effective lagrangian for the LOFF phase\label{hdet}}
Let us begin by writing the gap term in the lagrangian in
presence an inhomogeneous condensate. As in Section
\ref{section4}  we write the following formula for the LOFF
condensate
 \be \Delta({\bf
r})=\sum_{m=1}^P\Delta_m\,e^{2i{\bf q}_m\cdot{\bf
r}}\,.\label{1}\ee We will consider only two cases below:
\begin{itemize}
    \item[a)] One plane wave: $P=1$
    \item [b)] Cubic structure: $P=8$.
\end{itemize}
In the former case we shall take into account the possibility of
having both a $J=0$ and a $J= 1$ condensate as discussed above.
In the case of the cubic structure we will consider only spin
zero condensate; we will take  $\Delta_m\equiv\Delta$, real,
${\bf q}_m=\hat {\bf n}_m\,q$
    with $\hat {\bf n}_m$ the eight unit vectors defined in Eq. (\ref{eq:169}).
 To describe the quark condensate in the case of the single plane wave,
  we consider the lagrangian term: \be
{\cal L}_\Delta={\cal L}_\Delta^{(s)}+{\cal
L}_\Delta^{(v)}=-\frac{ e^{2i{\bf
 q}\cdot{\bf
r}} }2\,\epsilon_{\alpha\beta 3}
\psi_{i\alpha}^T(x)C\left(\Delta^{(s)}\epsilon_{ij}
+{\bm{\alpha}}\cdot\hat {\bf n} \Delta^{(v)} \sigma^1_{ij}\right)
\psi_{i\beta}(x)\,-(L\to R)+{\rm h.c.}\ , \label{ldeltaeff}\ee
which includes both the scalar and the vector condensate.

We introduce velocity-dependent fields as in Eq. (\ref{eq:413}),
with factors $\exp(i\mu_i v_i\cdot x)$, and we take into account
only the positive energy part that we write as $\psi_{{\bf
v_i};\,i\alpha}$  for a quark with flavor $i$ and color $\alpha$;
we keep track of the velocities of the two quarks that are not
opposite in the LOFF phase. We have: \be {\cal L
}_\Delta=-\frac{1} 2\, \sum_{\bf v_i, v_j} \exp\{i{\bf r\cdot{\bf
f}}({\bf v_i},\,{\bf v_j},\,{\bf q_k} )\}\epsilon_{\alpha\beta
3}\psi^T_{-\,{\bf v_i};\,i\alpha}(x)C
\left(\Delta^{(s)}\epsilon_{ij} +{\bm{\alpha}}\cdot\hat {\bf n}
\Delta^{(v)} \sigma^1_{ij}\right) \psi_{-\,{\bf v_j};\,j\beta}(x)
-(L\to R)+{\rm h.c.}\,,\label{loff6}\ee where \be{\bf f}({\bf
v_i},\,{\bf v_j},\,{\bf q})=2{\bf q}-\mu_i{\bf v_i}-\mu_j{\bf
v_j}\,.\label{fv}\ee We also define
\be\mu=\frac{\mu_1+\mu_2}{2}~,~~~~~~~~~\delta\mu=\,-\,\frac{\mu_1-\mu_2}{2}
.\label{dec2}\ee Since $\bf q$= ${\cal O}(\Delta_{2sc})\ll\mu$,
the condition\be {\bf p_1} +{\bf p_2}= 2{\bf q}\ee gives in the
$\mu\to\infty$ limit\be {\bf v_1}+ {\bf v_2}={\mathcal
O}\left(\frac\delta\mu\right)\ .\label{appr}\ee Taking into
account that $P_+({-\bf v})C\alpha^k P_+({\bf v})=v^kP_+({-\bf
v})C P_+({\bf v})$ we can rewrite (\ref{loff6}) as follows \be
{\cal L }_\Delta=-\frac{1} 2\, \sum_{\bf v_i, v_j} \exp\{i{\bf
r\cdot{\bf f}}({\bf v_i},\,{\bf v_j},\,{\bf q_k}
)\}\epsilon_{\alpha\beta 3}\psi^T_{{\bf v_j};\,i\alpha}(x)C
\left(\Delta^{(s)}\epsilon_{ij} -{\bf v_j}\cdot{\hat{\bf n}}
\Delta^{(v)} \sigma^1_{ij}\right) \psi_{-\,{\bf v_j};\,j\beta}(x)
-(L\to R)+{\rm h.c.}\ .\label{loff6deuxieme}\ee

 These equations  can be easily generalized
to the case of the face centered cube. We shall discuss this
generalization below.

\subsection{One plane wave structure}
\label{VB}

Let us rewrite (\ref{loff6deuxieme}) as follows: \be {\cal L
}_\Delta=-\frac{1} 2\,e^{2i{\bf r}\cdot{\bf q}} \sum_{\bf v_i,
v_j} e^{-i(\mu_i{\bf v_i}+\mu_j{\bf v_j})\cdot \bf
r}\,\epsilon_{\alpha\beta 3}\psi^T_{{\bf v_j};\,i\alpha}(x)C
\left(\Delta^{(s)}\epsilon_{ij} -{\bf v_j}\cdot\hat{\bf n}
\Delta^{(v)} \sigma^1_{ij}\right) \psi_{-\,{\bf v_j};\,j\beta}(x)
-(L\to R)+{\rm h.c.}\ ,\label{ldeltaeffb}\ee There are
 two
sources of space-time symmetry  breaking in (\ref{ldeltaeffb}),
one arising from the exponential term $\exp(2i{\bf r}\cdot{\bf
q})$, which breaks both translation and rotation invariance, and
another one in the vector condensate breaking rotation
invariance. On the other hand
  the factor $\exp(-i\mu_i{\bf v_i}-i\mu_j{\bf
v_j})$  breaks no space symmetry, since it arises from a field
  redefinition in a lagrangian which was originally invariant. For definiteness' sake
  let us
take the $z$-axis pointing along the direction of $\bf q$. As a
consequence of the breaking of translational invariance along the
$z$-axis, Goldstone's theorem predicts the existence of one
scalar massless particle, the Nambu-Goldstone Boson (NGB)
associated to the spontaneous symmetry breaking (SSB). The
symmetry breaking associated to the vector condensate is not
independent of the SSB arising from the exponential term
$\exp(2i{\bf r}\cdot{\bf q})$, because the direction of $\bf q$
coincides with the direction $\bf\hat n$ of the vector
condensate. For this reason, while there are in general three
phonons associated to the breaking of space symmetries here one
NGB is sufficient. The argument is sketched in Fig.
\ref{fig:phonon3} and follows from the fact that rotations and
translations are not independent transformations, because the
result of a translation plus a rotation is locally equivalent to
a pure translation.
\begin{center}
\begin{figure}[htb] \epsfxsize=8truecm
\centerline{\epsfxsize=5truecm\epsffile{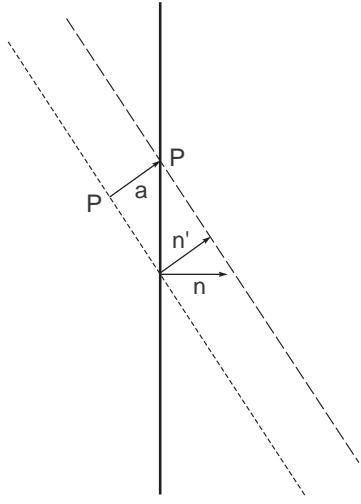} }
 \caption{{\it In the
point P the effect of the rotation ${\bf n}\to{\bf n}^\prime$ and
the effect of the translation ${\bf  r}\to {\bf  r}+{\bf a}$ tend
to compensate each other.\label{fig:phonon3}}}
\end{figure}
\end{center}
The lagrangian (\ref{ldeltaeffb}) induces a lattice structure
given by parallel planes perpendicular to $\hat {\bf n}$:\be \hat
{\bf n} \cdot{\bf r}\,=\,\frac{\pi k}{q}\hskip 1cm (k\,=0,\, \pm
1,\,\pm 2,...) \ .\label{planes}\ee We can give the following
physical picture of the lattice structure of the LOFF phase: Due
to the interaction with the medium,  the Majorana masses of the
red and green quarks oscillate in the direction $\bf\hat n$,
reaching on subsequent planes maxima and minima.  The NGB is a
long wavelength small amplitude variation of the condensate
$\Delta({\bf r})$; formally it is described by the
substitution\be\Delta({\bf r})=e^{2iq\hat {\bf n} \cdot {\bf
r}}\Delta\to e^{i\Phi/f}\Delta\,,\ee with
 \be  \frac{\Phi}{f}=2q(\hat {\bf n}+\delta{\bf n})\cdot({\bf r}+
 \delta{\bf r})\equiv\frac{\phi}{f}+2q\hat {\bf n}\cdot{\bf
 r}~\ee and $\langle\phi\rangle=0$.
 We assume \bea |\hat {\bf n}+\delta{\bf n}|&=&1~,\label{n00}
 \\\langle\delta{\bf
 n}\rangle_0&=&0\,.\label{n0}\eea
 Let
 us  introduce the auxiliary functions $\bf R$ and $T$,
  \be {\bf R}=\hat {\bf n}+\delta{\bf n}~,~~~~~ T\,=\,2\,q\,{\bf R}\cdot\delta{\bf r}\,.\ee
 The lattice fluctuation $\phi$ describes, in second quantization, the phonon field.
 Since it must be
small, $T$ and $\bf R$ are not independent fields and $T$ must
depend functionally on $\bf R$, i.e. $T=F[{\bf R}]$,
 which means\be \frac{\Phi}{f}
 =2q{\bf R}\cdot{\bf r}+F[{\bf R}]\equiv G[{\bf R},\,{\bf r}]\ .\ee
 The solution of this functional relation has the form
 \be {\bf R}={\bf
 h}[\Phi]\,,\ee where $\bf h$ is a vector built out of the scalar function
 $\Phi$. By this function one can only\footnote{In principle
there is a second vector,  $\bf r$, on which $\bf R$ could depend
linearly, but this possibility is excluded because $\bf R$ is a
vector field transforming under translations as $\bf R({\bf
r})\to {\bf R^{\,\prime}}({\bf r^{\,\prime}})={\bf R({\bf r})}$}
form the
 vector ${\bm{\nabla}} \Phi$ ; therefore  we
 get
 \be
{\bf R }=\frac{ {\bm{ \nabla}}\Phi}{|{\bm{ \nabla}}\Phi|}\,
,\label{r3}\ee which satisfies (\ref{n00}). In terms of the
phonon field $\phi$ the vector field ${\bf R}$ is given (up to
the second order terms in $\phi$) by the expression \be {\bf
R}=\hat{\bf n}+\frac{1}{2fq}\left[{\bm{\nabla}}\phi-\hat{\bf
n}(\hat{\bf n}\cdot{\bm{ \nabla}}\phi)\right]+\frac{\hat{\bf
n}}{8f^2 q^2}\left[3(\hat{\bf
n}\cdot{\bm{\nabla}}\phi)^2-|{\bm{\nabla}}\phi|^2\right]-\frac{{\bm{\nabla}}\phi}
{4f^2q^2} (\hat{\bf n}\cdot{\bm{\nabla}}\phi)\,.\label{63}\ee
 We stress  that the only dynamical field is $\phi$, $\Phi$ is an
auxiliary field with a non vanishing vacuum expectation value
$\langle\Phi\rangle_0=2\bf q\cdot\bf r$; as to $\delta\hat {\bf
n}$, $\bf R$ and $\delta\bf r$, they can all be expressed in
terms of $\phi$. In conclusion, the interaction term with the NGB
field   is contained in \be {\cal L }_{int}=-\frac{1}
2\,e^{i\Phi/f} \sum_{\bf v_i, v_j} e^{-i(\mu_i{\bf v_i}+\mu_j{\bf
v_j})\cdot{\bf r}}\,\epsilon_{\alpha\beta 3}\psi^T_{{\bf
v_j};\,i\alpha}(x)C \left(\Delta^{(s)}\epsilon_{ij} -{\bf
v_j}\cdot{\bf R} \Delta^{(v)} \sigma^1_{ij}\right) \psi_{-\,{\bf
v_j};\,j\beta}(x) -(L\to R)+{\rm h.c.}\ ,\label{ldeltaeffbis}\ee
where the fields $\Phi$ and ${\bf R}$ have been introduced in
such a way to reproduce Eq. (\ref{ldeltaeffb}) in the ground
state. At the first order in the fields one gets the following
three-linear coupling:
 \bea
 {\cal L}_{\phi\psi\psi}&=&-\frac{i\phi}{2f}
\sum_{{\bf v_i},{\bf v_j}} e^{i{\bf r\cdot f}({\bf v_i},\,{\bf
v_j},\,{\bf q})}\left[\Delta^{(s)}\epsilon_{ij}- {\bf
v_j}\cdot\hat {\bf n}
\Delta^{(v)}\sigma^1_{ij}\right]\epsilon_{\alpha\beta
3}\psi^T_{{\bf v_j};\,i\alpha}\,C\,\psi_{-\,{\bf
v_j};\,j\beta}\cr&&\cr&&-\frac{1}{4fq} \sum_{{\bf v_i},{\bf v_j}}
e^{i{\bf r\cdot f}({\bf v_i},\,{\bf v_j},\,{\bf q})}( -{\bf
v_j})\cdot \left[ {\bm{ \nabla}}\phi-\hat {\bf n}(\hat {\bf
n}\cdot{\bm{\nabla}}\phi)\right]
\Delta^{(v)}\sigma^1_{ij}\epsilon_{\alpha\beta 3}\psi^T_{{\bf
v_j};\,i\alpha}\,C\,\psi_{-\,{\bf v_j};\,j\beta}-(L\to R)\ +\
h.c. \,. \label{Trilineare}\eea We  also write down the
quadrilinear coupling:
 \bea
 {\cal L}_{\phi\phi\psi\psi}&=&\frac{\phi^2}{4f^2}\,
\sum_{{\bf v_i},{\bf v_j}} e^{i{\bf r\cdot f}({\bf v_i},\,{\bf
v_j},\,{\bf q})}\left[\Delta^{(s)}\epsilon_{ij}- {\bf
v_j}\cdot\hat {\bf n}
\Delta^{(v)}\sigma^1_{ij}\right]\epsilon_{\alpha\beta
3}\psi^T_{{\bf v_j};\,i\alpha}\,C\,\psi_{-\,{\bf
v_j};\,j\beta}\cr&&\cr&&-\frac{i\phi}{2f}\,\sum_{{\bf v_i},{\bf
v_j}} e^{i{\bf r\cdot f}({\bf v_i},\,{\bf v_j},\,{\bf q})}(-{\bf
v_j})\cdot \left[ {\bm{ \nabla}}\phi-\hat{\bf n}(\hat {\bf
n}\cdot{\bm{\nabla}}\phi)\right]
\Delta^{(v)}\sigma^1_{ij}\epsilon_{\alpha\beta 3}\psi^T_{{\bf
v_j};\,i\alpha}\,C\,\psi_{-\,{\bf v_j};\,j\beta}\cr&&\cr&&
-\frac{1}{8f^2q^2}\,\sum_{{\bf v_i},{\bf v_j}} e^{i{\bf r\cdot
f}({\bf v_i},\,{\bf v_j},\,{\bf q})}\left[-\frac {{\bf v_j}\cdot
\hat {\bf n}} 2 \left(3(\hat {\bf
n}\cdot{\bm{\nabla}}\phi)^2-|{\bm{\nabla}}\phi|^2\right)+ ({\bf
v_j}\cdot{\bm{ \nabla}}\phi)(\hat {\bf
n}\cdot{\bm{\nabla}}\phi)\right]\times \cr&&\cr&&\times
\Delta^{(v)}\sigma^1_{ij}\epsilon^{\alpha\beta 3}\psi^T_{{\bf
v_j};\,i\alpha}\,C\,\psi_{-\,{\bf v_j};\,j\beta} -(L\to R)\ +\
h.c.
 \label{quadrilineare}\eea
Through a bosonization procedure  one can derive an effective
lagrangian for  the NGB field. This will be done below. For the
moment we derive the general properties of the phonon effective
lagrangian. It must contain only derivative terms. Polynomial
terms are indeed forbidden by translation invariance, since $\phi$
is not an invariant field. In order to write the kinetic terms it
is better to use the auxiliary field $\Phi$ which behaves as a
scalar under both rotations and translation. To avoid the presence
of polynomial terms  in the phonon lagrangian one has to exclude
polynomial terms in the auxiliary field $\Phi$ as well; therefore
the lagrangian should be constructed only by derivative terms. The
most general invariant lagrangian will contain a tower of
space-derivative terms \cite{Casalbuoni:2001gt}. In fact, since
$\langle\bm{\nabla}\Phi\rangle=2{\bf q}$ is not a small quantity,
we cannot limit the expansion in the spatial derivatives of $\Phi$
to any finite order. Therefore we write
 \be{\cal
L}(\phi,\partial_\mu\phi)=\frac{f^2}
2\left[\dot\Phi^2-\sum_{n=1}^\infty c_n
(|{\bm{\nabla}}\Phi|^{2})^n\right]\ .\label{Phi}\ee
 In this lagrangian $\Phi$
must be thought as a function of the phonon field $\phi$.

Since \be |{\bm{\nabla}}\Phi|^2=4q^2+\frac{4q}{f}\hat {\bf
n}\cdot{\bm{\nabla}}\phi+\frac{1}{f^2}|{\bm{\nabla}} \phi|^2\
,\ee with similar expression for higher powers.  At the lowest
order in the derivatives of the phonon field $\phi$ we get,
neglecting a constant term: \be{\cal
L}(\phi,\partial_\mu\phi)=\frac{1}{2}\left[\dot\phi^2-
v_\parallel^2|{\bm{\nabla}}_\parallel\phi|^2-
v^2\left(4qf{\bm{\nabla}}_\parallel\phi+|{\bm{\nabla}}\phi|^2\right)
\right]\ , \label{phonon}\ee where
${\bm{\nabla}}_\parallel\phi=\hat {\bf n}\cdot{\bm{\nabla}}\phi$,
and $v_\parallel^2$,  $v^2$ are constants.
\subsection{Parameters of the phonon effective lagrangian:
one plane wave\label{opw}} In order to derive the parameters of
the phonon lagrangian (\ref{phonon}) it is useful to make an
approximation. We assume that $\delta\mu\sim
\Delta_{2sc}\ll\delta\ll\mu$. Clearly we cannot take simply the
$\mu\to\infty$ limit in the exponential term $\exp\{i{\bf
r\cdot{\bf f}}({\bf v_i},\,{\bf v_j},\,{\bf q_k} )\}$ in Eq.
(\ref{loff6deuxieme}); therefore we consider a smeared amplitude
as follows:
 \be\lim_{\mu\to\infty}\exp\{i{\bf r\cdot{\bf
f}}({\bf v_i},\,{\bf v_j},\,{\bf q_k} )\} \equiv
\lim_{\mu\to\infty}\int d{\bf r^{\,\prime}}\, \exp\{i{\bf
r^{\,\prime}\cdot{\bf f}}({\bf v_i},\,{\bf v_j},\,{\bf q_k}
)\}g({\bf r},{\bf r^{\,\prime}}) \ .\label{1bis}\ee We assume the
following
 smearing function:  \be g({\bf r},{\bf r^{\,\prime}})=g({\bf r}-{\bf r^{\,\prime}})=
 \prod_{k=1}^3
\frac{\sin\left[\dd\frac{\pi
q(r_k-r^\prime_k)}{R}\right]}{\pi(r_k-r^\prime_k)} \label{gII}\ee
and we evaluate (\ref{1bis}) in the $\mu\to\infty$ limit by
taking ${\bf q}$ along the $z-$axis, and using the following
identity: \be \int d^3{\bf r^{\,\prime}} \exp\{i{\bf
r^{\,\prime}\cdot{\bf f}}\}g({\bf r}-{\bf r^{\,\prime}})=
\exp\{i{\bf r\cdot{\bf
f}}\}\left(\frac{\pi}R\right)^3\delta_R^3\left(\frac{\bf
f}{2q}\right)\,,\ee where \be \delta_R(x)= {\dd\Bigg \{ }
\begin{array}{cc} {\dd\frac{R}{\pi}} & \textrm{~~for~~}
|x| < \dd{\frac{\pi}{2R}} \cr&\cr
0 & \textrm{elsewhere.} \\
\end{array}
\label{deltaII}\ee For the components $x$ and $y$ of $\bf f$ we
get \be |(\mu_1 v_1+\mu_2 v_2)_{x,y}|<\frac{\pi q}R\,,\ee i.e.
approximately (for $\delta\mu\ll\mu$) \be |( v_1+
v_2)_{x,y}|<\frac{\pi q}{R\mu}\,.\label{25b}\ee From this, in the
high density limit, it follows \be{\bf v_1}= -{\bf v_2}+{\cal
O}(\delta\mu/\mu)\,.\ee We used already this result in Eq.
(\ref{eq:257}), in connection with the Riemann-Lebesgue lemma and
in Eq. (\ref{appr}). A more accurate result is as follows. If
$\theta_1$ and $\theta_2$ are the angles of $\bf v_1$ and $\bf
v_2$ with respect to the $z$-axis one gets \be
\theta_1\,=\,\theta_2\,+\,\pi\,+\,\frac{2\delta\mu}{\mu}\tan\theta_2\,.\ee
For the $z$ component we get \be
f_z\,=\,2\,q\,h(\cos\theta_2)\,,\ee where\be h(x)=1+\frac
{x\mu_2} {2q}\left(-1+\sqrt{1-\frac{4\mu\,\delta\mu}{\mu_2^2\,
x^2}}\right)\,, \label{29}\ee and, neglecting corrections of
order $\delta\mu/\mu$,\be
h(x)=1-\frac{\cos\theta_q}x\,,~~~~~~~~~\cos\theta_q=
\frac{{\delta\mu}}q\,.\label{hdix}\ee Notice that
$\theta_q=\psi_0/2$ where $\psi_0$ is the angle depicted in Fig.
\ref{fig2}, see also Eq. (\ref{eq:115}).  The two factors $\pi/R$
arising from the $x$ and $y$ components are absorbed into a wave
function renormalization of the quark fields, both in the kinetic
and in the gap terms. As for the $z$ component one remains with
the factor \be \frac{\pi}{R}e^{i2qhz}\,\delta_R[h({\bf
v}\cdot\hat {\bf n})]\approx\frac{\pi}{R} \delta_R[h({\bf
v}\cdot\hat {\bf n})]\label{dr}\ee in the gap term, whereas for
the kinetic term we get a factor of 1. We have assumed
$\exp[i2qhz]=1$ in Eq. (\ref{dr}) owing to the presence of the
$\delta_R$ function, that, in the $R/\pi\to\infty$ limit,
enhances the domain of integration where $h=0$. We  will discuss
this approximation below.

Eq. (\ref{deltaII}) defines a region where $\delta_R\neq 0$, i.e.
a domain where pairing between the two quarks can occur; it
correspond to the {\it pairing region} in the analysis of
\cite{FF} and \cite{Bowers:2001ip}, in contrast with the {\it
blocking region}, where $\delta_R=0$. The pairing region
intersects the Fermi surface with a 'ring' whose size depends on
the value of $R$. As we noticed above, $R=\infty$ implies the
vanishing of the pairing region and therefore one expects
$R\to\infty$ at the second order phase transition
\cite{Casalbuoni:2002my}.  The precise value of $R$ should be
fixed by the gap equation; since this calculation is still
missing, for the purpose of this paper we leave $R$ as a
parameter.

In conclusion we can approximate Eq. (\ref{loff6deuxieme}) as
follows \be {\cal L }_\Delta=-\frac{1} 2\, \sum_{\bf
v}\,\frac{\pi}{R} \delta_R[h({\bf v}\cdot\hat {\bf n})]\,
\epsilon_{\alpha\beta 3}\psi^T_{{\bf v};\,i\alpha}(x)C
\left(\Delta^{(s)}\epsilon_{ij} -{\bf v}\cdot{\hat {\bf n}}
\Delta^{(v)}\,\sigma^1_{ij}\right) \psi_{-\,{\bf v};\,j\beta}(x)
-(L\to R)+{\rm h.c.}\ . \label{loff6deuxiemeapproximate} \ee
Using the same notations as in Section \ref{sect:2sc} we can
write the effective lagrangian as follows: \be  {\cal
L}_{0}+{\cal L}_1+ {\cal L}_\Delta=\sum_{\bf v}\sum_{A,B=0}^5
\chi^{A\dagger}\left(\matrix{iTr[\tilde T_A^\dagger\,V\cdot
D\,\tilde T_B ]& -\Delta^{\dagger}_{AB}\cr -\Delta_{AB}
&iTr[\tilde T_A^\dagger\,\tilde V\cdot D^*\,\tilde T_B
]}\right)\chi^B\,+\, (L\to R)\ . \label{2sccomplete0b} \ee Here
\be \chi^A=\frac 1{\sqrt 2}\left(\matrix{\psi^A_+\cr
C\psi^{A*}_-}\right)\label{chiA}\ee and \be \tilde
T_A=\frac{\tilde\lambda_A}{\sqrt 2}\hskip1cm(A=0,...,5)\
.\label{380}\ee The matrix $\Delta_{AB}$ vanishes for $A\,{\rm
or} \,B=4\,{\rm or} \,5 $, while, for $A,B=0,...,3$, is given by
\be \Delta_{AB}=  \left( \Delta^{(s)}_{eff}\,\tau_{AB}-\,{\bf
v}\cdot\hat {\bf n}\, \Delta^{(v)}_{eff}\, \sigma_{AB}\right)\
,\label{eq:38} \ee with\be \tau_{AB}= \left(\begin{array}{cccc}
  1 & 0 & 0 & 0 \\
  0 & -1 & 0 & 0 \\
  0 & 0 & -1& 0 \\
 0 & 0 & 0 & -1
\end{array}\right)~,~~~~~~~~~~~~~~~~ \sigma_{AB}=
\left(\begin{array}{cccc}
  0 & 0 & 0 & -1 \\
  0 & 0 & -i & 0 \\
  0 & +i & 0& 0 \\
 +1 & 0 & 0 & 0
\end{array}\right)\,,
\ee and \be\Delta^{(s)}_{eff}=\frac{\Delta^{(s)}\pi}
R\delta_R[h({\bf v}\cdot\hat {\bf
n})]\,,~~~~~~~~~~\Delta^{(v)}_{eff}=\frac{\Delta^{(v)}\pi}
R\delta_R[h({\bf v}\cdot\hat{\bf n})]\ .\label{deltaeffs}\ee In
the present approximation the quark propagator is given by
 \be
\displaystyle D_{AB}(\ell,\ell^{\prime\prime})=
(2\pi)^4\delta^4(\ell-\ell^{\prime\prime})\times\sum_C\left(
\begin{array}{cc} \displaystyle
 \frac{ \tilde V\cdot\ell\,\delta_{AC}}{\tilde D_{CB}(\ell)}\,
& ~~~\displaystyle \frac{\Delta_{AC}^\dag}{D_{CB}(\ell)}
\\ \\
\displaystyle \frac{\Delta_{AC}}{\tilde D_{CB}(\ell)}
 &  ~~~  \displaystyle
 \frac{ V\cdot\ell\,\delta_{AC}}{D_{CB}(\ell)}\,
\end{array}
\right)\,,\label{propagatore}\ee where \be D_{CB}(\ell)=
\left(V\cdot \ell\,\tilde V\cdot
\ell\,-\,\Delta\Delta^\dag\right)_{CB}\,,~~~~~~\tilde
D_{CB}(\ell)= \left(V\cdot \ell\,\tilde V\cdot
\ell\,-\,\Delta^\dag\Delta\right)_{CB}\ . \ee On the other hand
 the propagator for  the fields $\chi^{4,5}$
does not contain gap mass terms and is given by \be
D(\ell,\ell^{\prime})=(2\pi)^4\,\delta^4(\ell-\ell^{\prime})\,\left(
\begin{array}{cc}
 (V\cdot\ell)^{-1}&0\\
 0&  (\tilde V\cdot\ell)^{-1}
\end{array}
\right)~.\ee For the other  fields $\chi^A$, $A=0,\cdots,3$, it
is useful to go to a representation where $\Delta\Delta^\dag$ and
$\Delta^\dag\Delta$ are diagonal. It is accomplished by
performing a unitary transformation which transforms the basis
$\chi^A$ into the new basis $\tilde \chi^A$ defined by \be
\tilde\chi^{A}=R_{AB}\chi^{B}\ ,\label{newbasis1}\ee with \be
R_{AB}=\frac 1 {\sqrt 2 } \left(\begin{array}{cccc}
  1 & 0 & 0 & 1 \\
  0 & 1 &- \,i & 0 \\
  0 & +i &-\,1& 0 \\
 1 & 0 & 0 & -\,1
\end{array}\right)\ .\label{newbasis2}
\ee In the new basis we have
 \bea
\left(\Delta\Delta^\dag\right)_{AB}&=&\alpha_A\delta_{AB}\,,\cr
\left(\Delta^\dag\Delta\right)_{AB}&=&\tilde\alpha_A\delta_{AB}\,,\eea
where \be\alpha_0=\alpha_2=
\tilde\alpha_2=\tilde\alpha_3=(\Delta^{(s)}_{eff}-{\bf
v\cdot}\hat {\bf n}\Delta^{(v)}_{eff})^2~,~~~~~~~~~~~~
\alpha_1=\alpha_3=\tilde\alpha_0=\tilde\alpha_1=
(\Delta^{(s)}_{eff}+{\bf v\cdot}\hat {\bf
n}\Delta^{(v)}_{eff})^2~. \ee For further reference we also
define
  \be \mu_C=(\Delta^{(s)}_{eff}-{\bf v\cdot}\hat
{\bf n}\Delta^{(v)}_{eff}, \,\Delta^{(s)}_{eff}+{\bf v\cdot}\hat
{\bf n}\Delta^{(v)}_{eff}, \,\Delta^{(s)}_{eff}-{\bf v\cdot}\hat
{\bf n}\Delta^{(v)}_{eff}, \,\Delta^{(s)}_{eff}+{\bf v\cdot}\hat
{\bf n}\Delta^{(v)}_{eff}) \label{muc} \ .\ee In the basis
$\tilde\chi$ the 3-point and 4-point couplings (\ref{Trilineare})
and (\ref{quadrilineare}) are written as follows: \be {\mathcal
L}_3\,+\,{\mathcal L}_4=
 \sum_{\bf v}\sum_{A=0}^3
\tilde\chi^{A\,\dag}\,   \left(
\begin{array}{cc}
 0 & -g_3^\dag\,-\,g_4^\dag
\\
  -g_3\,-\,g_4& 0\end{array}\right)
    \,\tilde\chi^B\ ,\label{vertex0}\ee
Here\bea
g_3&=&\,\left[\frac{i\phi\,\Delta^{(s)}_{eff}}{f}\,\tau_{AB}\,+\,\hat
O[\phi]\,\sigma_{AB}\right] \ ,\label{g3}\\
g_4&=&\,\left[-\frac{\phi^2\,\Delta^{(s)}_{eff}}{2f^2}\,\tau_{AB}\,+\,
 \left(\frac{i\phi}{f}\hat O[\phi]
 \,+\,\hat Q[\phi]\right)\,\sigma_{AB}\right] \ ,\label{g4}\eea with \bea
 \hat O[\phi]&=&\frac{1}{2fq}{\bf v\cdot} \left[ {\bm{
 \nabla}}\phi-\hat
 {\bf n}(\hat
{\bf n}\cdot{\bm{\nabla}}\phi)\right] \Delta^{(v)}_{eff}\ ,\cr
 \hat Q[\phi]&=&
\frac{\Delta^{(v)}_{eff}}{4f^2q^2}\left[\frac {{\bf v\cdot}\hat
{\bf n}} 2\left(3 ({\hat {\bf
n}\cdot\bm{\nabla}}\phi)^2-|{\bm{\nabla}}\phi|^2\right)- ({\bf
v\cdot \bm{\nabla}}\phi)({\hat {\bf
n}\cdot\bm{\nabla}}\phi)\right]\ ,
 \eea
Terms in $g_3$ and $g_4$ that are  proportional to $\tau_{AB}$
arise from the expansion of $\exp{i\phi/f}$ alone, whereas terms
proportional to $\sigma_{AB}$ get also contribution from the
expansion of $\bf R$ in the vector condensate. The effective
action for the NGB is obtained at the lowest order  by the
diagrams in Fig. \ref{figgraph}.
\begin{center}
\begin{figure}[htb]
\epsfxsize=8truecm
\centerline{\epsfxsize=6truecm\epsffile{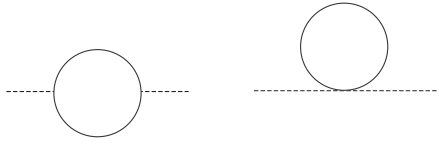}}
{\caption{\it Self-energy (a) and tadpole (b) diagrams.
\label{figgraph}}}
\end{figure}
\end{center}
 The result of the calculation of the  two diagrams
at the second order in the momentum expansion is as follows: \bea
\Pi(p)_{s.e.}&=&
 \frac{i\,\mu^2}{16\pi^3f^2} \sum_{{\bf
v}}\sum_{C=0}^3\int  {d^2\ell} \Big[\frac{4\,\alpha_C^2}{
D^2_C(\ell)}\,-\,\frac{4\alpha_C\,V\cdot\ell\,\tilde V\cdot\ell
}{D_C^2(\ell)}
 \cr
 &&
 \cr \,
 &-&\,
 4\alpha^2_C\frac{V\cdot p\,\tilde
V\cdot p}{D_C^3(\ell)}\, -\,\left(\frac{\Delta_{eff}^{(v
)}}{q}\right)^2\omega^2(\vec p)\left(
\frac{2\alpha_C}{D^2_C(\ell)}\,+\,\frac{1}{D_C(\ell)}
\right)\Big]\ ,
 \cr
 &&
 \cr
 \Pi(p)_{tad}
 &=&
 \frac{i\,\mu^2}{16\pi^3f^2}
  \sum_{\bf
v}\sum_{C=0}^3
  \int  \frac
{d^2\ell}{D_C(\ell)}\,\Big[\,4\,\alpha_C \,-\,\frac{\Delta^{(v
)}_{eff}}{q^2}\,\mu_C\,\times \cr&\times&
\left(-p_x^2-p_y^2\,+2p_z^2-\,2{\bf p}\cdot{\bf v}\,p_z
\right)\Big]\,,
 \eea
  where
  \be
D_C(\ell)=\ell_0^2-\ell_\parallel^2-\alpha_C+i\epsilon\hskip0.3cm
, \ee $\mu_c$ defined in (\ref{muc}) and \be
 \omega({\bf
p})={\bf p}\cdot{\bf v}-({\bf p}\cdot\hat {\bf n})({\bf
v}\cdot\hat {\bf n})
 \ .
\ee

To perform the calculation we will take the limit $R\to\infty$,
when the $\delta_R$ function becomes the Dirac delta. We handle
the $\delta_R$ functions according to the Fermi trick in the
Golden Rule; in the numerator, in presence of a product of two
$\delta_R$, we substitute one $\delta_R$ function with the Dirac
delta and for the  other one we take
 \be
\frac{\pi\delta_R[h(x)]} R\to\frac{\pi\delta_R(0)}R\to \,1 .\ee A
similar substitution is performed in the denominator. Moreover we
use\be \int
\frac{d^2\ell}{[D_C(\ell)]^3}\,=\,-\,\frac{i\pi}{2\alpha_{C}^2}\
.\ee
 Therefore one has\be
\frac\pi R \delta_R[h({\bf v\cdot}\hat {\bf n})]\,\frac\pi
R\delta_R[h({\bf v\cdot}\hat {\bf
n})]\to\,\frac{\pi}{R}\delta[h({\bf v\cdot}\hat {\bf
n})]=\frac\pi R\,\delta\left[1-\frac{\delta\mu}{q\,{\bf
v\cdot}\hat {\bf n}}\right]=k_R\delta\left[{\bf v\cdot}\hat {\bf
n}-\frac{\delta\mu}{q}\right]\,, \ee
 with \be
k_R=\frac{\pi|\delta\mu|}{qR} \ .\ee
 At the second order in the momentum
expansion one gets
 \be\Pi(p)\,=\,-\,\frac{\mu^2k_R}{2\pi^2 f^2}\sum_{{\bf v}}\delta\left[{\bf v\cdot}\hat
{\bf n}-\frac{\delta\mu}{q}\right]
 \Big[
 V_\mu\tilde V_\nu p^\mu p^\nu+\Omega^{(v)}(\vec p)\Big]\ . \ee
 Here
 \bea\Omega^{(v)}({\bf
p})&=&
 -\,\left(\frac{\Delta^{(v
)}}{q}\right)^2\omega^2({\bf p})\left(2-\frac 1 2
\sum_{C=0}^3{\rm
arcsinh}\frac\delta{|\mu_c|}\right)\,+\cr&&\cr&+&
\frac{\Delta^{(v)}}{2q^2} \Phi({\bf p})
\sum_{C=0}^3\mu_C\times{\rm arcsinh}\frac\delta{|\mu_c|}
 \,\approx\cr&&\cr&\approx&
-2\,\left(\frac{\Delta^{(v
)}}{q}\right)^2\left(1-\log\frac{2\delta}{\Delta^{(s)}}\right)\left(\omega^2({\bf
p})+{\bf v}\cdot\hat {\bf n}\Phi({\bf p})\right)\,,
\label{final2}\eea where we have used the result
\cite{Alford:2000ze} $\Delta^{(v)}\ll\Delta^{(s)}$ and \be
\Phi({\bf p})=\left(3p_z^2-{\bf p}^{\,2}\right){\bf v}\cdot\hat
{\bf n} \,-\,2{\bf p}\cdot{\bf v}\,p_z \,.\ee From
 \be
 {\cal L}_{eff}(p)\,=\,-\,\frac{\mu^2k_R}{2\pi^2f^2}\sum_{\bf v }
\delta\left\{{\bf v\cdot}\hat {\bf n}-\frac{\delta
\mu}{q}\right\} V_\mu\tilde V_\nu p_\mu\phi p_\nu\phi\ , \ee
after averaging over the Fermi velocities we obtain  \be {\cal
L}_{eff}=\frac 1 2\left[(\dot\phi_k)^2-v_\perp^2
  (\de_x\phi_k)^2-v_\perp^2 (\de_y\phi_k)^2-v_\parallel^2(\de_z\phi_k)^2\right]\
.\label{continuum}\ee One obtains canonical normalization for the
kinetic term provided \be
 f^2= \frac {\mu^2k_R}{4\pi^2}
  \ .\ee On the other hand
 \be v_\perp^2=\frac 1
 2\sin^2\theta_q+\left(1-3\cos^2\theta_q\right)\left(1-\log\frac{2\delta}{\Delta_0}\right)
 \left(\frac{\Delta^{(v)}}q\right)^2
 \,,~~~~~
 v_\parallel^2=\cos^2\theta_q
  \ . \ee
In conclusion we get the anisotropic phonon dispersion law\be
E(\vec p)=\sqrt{v_\perp^2(p_x^2+p^2_y)+v_\parallel^2 p^2_z}\ .\ee
Besides the anisotropy related to
 $v_\perp\neq v_\parallel$, there is another source of
 anisotropy, due to the fact that
 $p_z$, the component of the momentum perpendicular to the planes (\ref{planes}),
 differently from $p_x$ and $p_y$
 is a quasi momentum and not a real momentum. The difference can
 be better  appreciated in coordinate space, where the effective
 lagrangian reads
  \be {\cal L}=\frac 1 2\left[(\dot\phi_k)^2-v_\perp^2
  (\de_x\phi_k)^2-v_\perp^2 (\de_y\phi_k)^2-v_\parallel^2\left(\frac
q\pi\right)^2(\phi_k-\phi_{k-1})^2\right]\ .\label{discrete}\ee
The effective action for the field $\phi$, $S[ \phi ]$, is
obtained by the lagrangian  as follows \be S=\int dt\,dx\,dy\
\frac{\pi}{q}\sum_{k=-\infty}^{+\infty}{\cal
L}(\phi(t,x,y,k\pi/q)\ .\label{127}\ee In the action bilinear
terms of the type $\phi_k\phi_{k^\prime}$ with $k\neq k^\prime$
may arise. In the continuum limit these terms correspond to
derivatives with respect to the $z$ direction. However, in the
long distance limit $\ell\gg\pi/q$, the set of fields
$\phi_k(x,y)$ becomes a function $\phi(x,y,z)$ and the last term
in (\ref{discrete}) can be approximated by
$v_\parallel^2(\de_z\phi)^2$.
\subsection{Cubic structure\label{Cubic structure}}
 The space dependence of the condensate
corresponding to this lattice is as follows\be \Delta({\bf
r})=\Delta\sum_{k=1}^8\,\exp\{2iq\hat {\bf n}_k\cdot{\bf r}\}\,,
\label{8a}\ee where the eight unit vectors $\hat n_k$ are given
in (\ref{eq:169}) and  \be q=\pi/a\,.\ee In Fig. \ref{fig:cube}
some of the symmetry axes of this cube are shown: they are
denoted as $C_4$ (the three 4-fold axes), $C_3$ (the four 3-fold
axes) and $C_2$ (the six 2-fold axes).
\begin{center}\begin{figure}[htb]
\epsfxsize=8truecm
\centerline{\epsfxsize=6truecm\epsffile{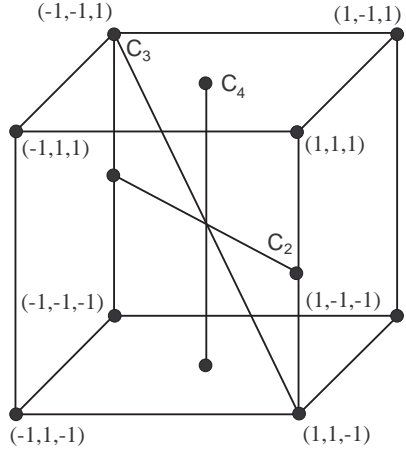}}
{\caption{\it Symmetry axes $C_2$, $C_3$ and $C_4$ of the cube.
\label{fig:cube}}}
\end{figure}
\end{center}
 To describe the quark condensate we add a term ${\cal
L}_\Delta$ completely analogous to (\ref{ldeltaeffb}). By the
same procedure used for the plane wave condensate one has \be
{\cal L }_\Delta=-\frac{\Delta} 2\sum_{k=1}^8\sum_{\bf v}\,
\frac\pi R\,\delta_R[h({\bf v\cdot}\hat {\bf
n}_k)]\epsilon_{ij}\epsilon_{\alpha\beta 3}\psi^T_{{\bf
v};\,i\alpha}(x)C \psi_{-\,{\bf v};\,j\beta}(x)\, -(L\to
R)\,+\,{\rm h.c.}\label{loff6b}\ee

 ${\mathcal L}_0 +
 {\mathcal L}_1 + {\mathcal L}_\Delta
 $ is still given by Eq. (\ref{2sccomplete0b})
but now \be\Delta_{eff}=\frac{\Delta\pi}
R\sum_{k=1}^{8}\delta_R[h({\bf v}\cdot\hat {\bf n}_k)]\ ;
\label{deltaeff3}\ee the quark propagator is given by
(\ref{propagatore}) with $\Delta_{eff}$ given by
(\ref{deltaeff3}).

An interesting point should be noted. This equation shows that
the pairing region for the cubic LOFF condensate is formed by
eight distinct rings; each ring is associated to one vertex of
the cube and has as its symmetry axis one of the threefold axes
$C_3$. According to the analysis of \cite{Bowers:2002xr}, the
LOFF vacuum state corresponds to a situation where these domains
have at most one common point. Given the symmetry of the cubic
structure we can limit the analysis to one pair of rings, for
example those associated to the vertices ${\bf n}_1,\,{\bf n}_5$.
The common point between these two rings lies on the axis $C_2$
and has ${\bf v}=1/{\sqrt 2} (1,\,1,\,0).$ Since it must also
belong to the boundary of the two pairing regions we have the
condition:\be |h({\bf v\cdot}\hat {\bf n}_1)|=\frac \pi {2R }\
,\ee which implies \be
R\,=\frac\pi{2h\left({\sqrt{2/3}}\,\right)} \ .\ee Using Eq.
(\ref{29}) one gets \be R\approx 18\,.\label{RT}\ee

 The condensate (\ref{8a}) breaks both
translations and rotations. It is however invariant under the
discrete group $O_h$, the symmetry group of the cube. This can be
seen  by noticing that the condensate is invariant under the
following coordinate transformations \bea
&&R_1:~~~ x_1\to x_1,~~~x_2\to x_3,~~ x_3\to -x_2,\nn\\
&&R_2:~~~ x_1\to -x_3,~~~x_2\to x_2,~~~ x_3\to x_1,\nn\\
&&R_3:~~~x_1\to x_2,~~~x_2\to -x_1,~~~x_3\to x_3,\nn\\ &&I:~~~~~
x_1\to -x_1,~~~x_2\to -x_2,~~~x_3\to -x_3, \label{basic}\eea that
is rotations of $\pi/2$ around the coordinate axes, and the
inversion with respect to the origin. Since the group $O_h$ is
generated by the previous 4 elements  the invariance follows at
once.

 The crystal defined by the condensate (\ref{8a}) can
fluctuate and its local deformations define three phonon fields
$\phi^{(i)}$ that are the Nambu-Goldstone bosons associated to
the breaking of the translational symmetry. They can be formally
introduced following the same procedure discussed for the single
plane wave case. One effects the substitution in (\ref{8a}) \be
2qx^i\to\frac{\Phi^{(i)}(x)}f=\frac{2\pi}a
x^i+\frac{\phi^{(i)}(x)}f\,,\label{small}\ee where the three
auxiliary scalar fields $\Phi^{(i)}$  satisfy \be
\Big\langle\frac{\Phi^{(i)}}f\Big\rangle_0=\frac{2\pi}ax_i\,,\label{vacuum}\ee
whereas for the phonon fields one has \be
\langle\phi^{(i)}(x)\rangle_0=0\,.\ee We have therefore three
fluctuating fields $\phi^{(i)}_{k_1\,k_2\,k_3}$ for any
elementary cube defined by discrete coordinates \be
x_{k_1}=\frac{k_1\pi}q~,~~y_{k_2}=
\frac{k_2\pi}q~,~~z_{k_3}=\frac{k_3\pi}q~,~~\ee
 i.e. \be
\phi^{(i)}_{k_1\,k_2\,k_3}\equiv\phi^{(i)}
(t,x_{k_1},y_{k_2},z_{k_3})\ .\label{83}\ee The interaction term
with the NGB fields is therefore given by an equation similar to
(\ref{127}):\be S_{int}=-\int dt\left( \frac{\pi}{q}\right)^3
\sum_{k_1,k_2,k_3=-\infty}^{+\infty}\, \sum_{\bf
v}\sum_{m=1}^8\Delta
\exp\{i\,\varphi^{(m)}_{k_1\,k_2\,k_3}/f\}\,\epsilon_{ij}
\epsilon_{\alpha\beta 3}\psi^T_{{\bf
v};\,i\alpha}\,C\,\psi_{-\,{\bf v};\,j\beta}-\,(L\to R)\ +\
h.c.\,, \label{external0}\ee where \be
\varphi^{(m)}_{k_1\,k_2\,k_3}=\sum_{i=1}^3\epsilon_i^{(m)}
\phi^{(i)}_{k_1\,k_2\,k_3}\ee and the eight vectors $
\bm{\epsilon}^{(m)}$ are given by \be
(\bm{\epsilon}_i^{(m)})\equiv \sqrt 3 \,\hat{\bf  n}_m . \ee
 The complete effective action for the NGB fields $\phi^{(i)}$
will be of the form \be S=\int dt\left( \frac{\pi}{q}\right)^3
\sum_{k_1,k_2,k_3=-\infty}^{+\infty}\,{\cal
L}(\phi^{(i)}(t,k_1\pi/q,k_2\pi/q,k_3\pi/q))\ .\ee In the low
energy limit, i.e. for wavelengths much longer than the lattice
spacing $\sim 1/q$, the fields $\phi^{(i)}_{k_1,k_2,k_3}$ vary
almost continuously and can be imagined as continuous functions
of three space variables $x$, $y$ and $z$.

 The
coupling of the quark fields to the NGB fields generated by the
condensate will be written as
 \be
 \Delta\psi^TC\psi\sum_{\epsilon_i=\pm}\exp\left\{
i(\epsilon_1\Phi^{(1)}+\epsilon_2\Phi^{(2)}+\epsilon_3\Phi^{(3)})\right\}
\,, \label{coupling}\ee making the theory invariant under
translations and rotations. These symmetries are broken
spontaneously in the vacuum defined by Eq. (\ref{vacuum}). In
order to write down the effective lagrangian for the phonon
fields $\phi^{(i)}$ it is useful to start with the effective
lagrangian for the auxiliary fields $\Phi^{(i)}$ that has to
enjoy the following symmetries: rotational and translational
invariance;
 $O_h$ symmetry on the fields $\Phi^{(i)}$.
The latter requirement follows from the invariance of the
coupling (\ref{coupling}) under the group $O_h$ acting upon
$\Phi^{(i)}$. The phonon fields $\phi^{(i)}(x)$ and the
coordinates $x^i$ must transform under the diagonal discrete
group obtained from the direct product of the rotation group
acting over the coordinates and the $O_h$ group acting over
$\Phi^{(i)}(x)$. This is indeed the symmetry left after the
breaking of translational and rotational invariance. The most
general low-energy effective lagrangian displaying these
symmetries is\be {\cal L}=\frac {f^2}
2\sum_{i=1,2,3}({\dot\Phi}^{(i)})^2+ {\cal L}_{\rm
s}(I_2(\bm{\nabla}\Phi^{(i)}),
I_4(\bm{\nabla}\Phi^{(i)}),I_6(\bm{\nabla}\Phi^{(i)}))\ ,\ee
where\be I_2(X_i)=X_1^2+X_2^2+X_3^2,~~I_4(X_i)=X_1^2 X_2^2 +X_2^2
X_3^2+ X_3^2 X_1^2,~~ I_6(X_i)=X_1^2 X_2^2 X_3^2, \label{8}\ee
are the three basic symmetric functions of three variables. At
the lowest order in the fields $\phi^{(i)}$ and at the second
order in the derivatives one gets \cite{Casalbuoni:2002hr} \be
{\cal L}=\frac 1 2\sum_{i=1,2,3}({\dot\phi}^{(i)})^2-\frac a 2
\sum_{i=1,2,3}|\bm{\nabla}\phi^{(i)}|^2- \frac b 2
\sum_{i=1,2,3}(\de_i\phi^{(i)})^2-
c\sum_{i<j=1,2,3}\de_i\phi^{(i)}\de_j\phi^{(j)}\,.\label{effective}\ee
which depends on three arbitrary parameters.

\subsection{Parameters of the phonon effective lagrangian: cubic
crystal} \label{VE} The parameters $a,b,c$ appearing in
(\ref{effective}) are computed by a method similar to the one
used in Section \ref{opw}. One puts \be\varphi^{(m)}(t,\vec
r)=\sum_{i=1}^3\epsilon_i^{(m)} \phi^{(i)}(t,\vec r) \ ,\ee which
allows to write the 3-point and the 4-point couplings as follows:
 \be {\mathcal L}_3\,+\,{\mathcal L}_4=
 \sum_{\vec v}\sum_{A=0}^3
\tilde\chi^{A\,\dag}\,   \left(
\begin{array}{cc}
 0 & -g_3^\dag\,-\,g_4^\dag
\\
  -g_3\,-\,g_4& 0\end{array}\right)
    \,\tilde\chi^B\ ,\label{vertex}\ee
Here \bea g_3&=&\sum_{m=1}^8\frac{\pi\Delta}{R}\,\delta_R[h({\bf
v}\cdot\hat {\bf n}_m)]\frac{i\varphi^{(m)}}f\tau_{AB}\ ,\cr&&\cr
g_4&=&\,-\sum_{m=1}^8\frac{\pi\Delta}{R}\,\delta_R[h({\bf
v}\cdot\hat {\bf
n}_m)]\frac{(\varphi^{(m)})^2}{2f^2}\tau_{AB}\,,\eea to be
compared with Eqs. (\ref{g3}) and (\ref{g4}) valid for the
one-plane wave form of the condensate (we have here neglected the
vector condensate). To perform the calculation one employs the
propagator given in Eq. (\ref{propagatore}) with $\Delta_{eff}$
given in (\ref{deltaeff3}) and the interaction vertices in
(\ref{vertex}). The result of the calculation of the two diagrams
in Fig. \ref{figgraph} at the second order in the momentum
expansion is \bea
 {\cal L}_{eff}(p)_{s.e.}&=&i\,
 \frac{4\times 4\,\mu^2}{16\pi^3f^2} \sum_{\bf
v}\sum_{m,k=1}^8\frac 1
2\left(\frac{\pi\Delta}{R}\right)^2\,\delta_R[h({\bf v\cdot}\hat
{\bf n}_m)](i\,\varphi^{(m)})\delta_R[h({\bf v\cdot}\hat {\bf
n}_k)](i\,\varphi^{(k)})\int
\frac{d^2\ell}{D(\ell)D(\ell+p)}\cr&&\cr&\times&
\Big[-2\Delta_{eff}^2+V\cdot\ell\,\tilde V\cdot(\ell+p)+\tilde
V\cdot\ell\, V\cdot(\ell+p) \Big]\ ,\label{se1}
 \\
 &&
 \cr
 &&
 \cr
{\cal L}_{eff}(p)_{tad}
 &=&
 i\,\frac{4\times 4\mu^2}{16\pi^3f^2}
  \sum_{\bf v} \,\sum_{m=1}^8 \int  \frac
{d^2\ell}{D(\ell)}\frac{\pi\Delta\Delta_{eff}}{R}\,\delta_R[h({\bf
v\cdot}\hat {\bf n}_m)](\varphi^{(m)})^2\,,
 \eea
  where
  \be
D(\ell)=\ell_0^2-\ell_\parallel^2-\Delta^2_{eff}+
i\epsilon\,,\label{tad1} \ee and, analogously to
(\ref{deltaeffs}), \be\Delta_{eff}=\frac{\Delta\pi}R\sum_{k=1}^8
\delta_R[h({\bf v}\cdot\hat {\bf n}_k)]\,.\ee
 From
(\ref{se1} and (\ref{tad1}) one can easily control that the
Goldstone theorem is satisfied and the phonons are massless. As a
matter of fact one has \bea{\cal L}_{mass}&=&{\cal
L}_{eff}(0)_{s.e.}\,+\,{\cal L}_{eff}(0)_{tad}\,
=i\,\frac{4\times
4\mu^2}{16\pi^3f^2}\,\frac{\pi\Delta}{R}\times\cr&&\cr&\times&
  \sum_{\bf v}\int  \frac
{d^2\ell}{D(\ell)}\Big[-\sum_{m,k=1}^8\frac{\pi\Delta}{R}\,\delta_R[h({\bf
v\cdot}\hat {\bf n}_m)]\varphi^{(m)}\delta_R[h({\bf v\cdot}\hat
{\bf n}_k)]\varphi^{(k)}+\Delta_{eff}\sum_{m=1}^8 \delta_R[h({\bf
v\cdot}\hat {\bf n}_m)](\varphi^{(m)})^2\Big]\ .\label{45}\eea

In the double sum in the r.h.s. of Eq. (\ref{45}) only the terms
with $m=k$ survive and one immediately verifies the validity of
the Goldstone's theorem. i.e. the vanishing of (\ref{45}). Notice
that in this approximation the masses of the Goldstone bosons
vanish because the pairing regions are not overlapping, signaling
that when they do overlap one is not at the minimum of the
free-energy, see \cite{Bowers:2002xr}.

 At the second order in the momentum
expansion one has \be {\cal L}_{eff}(p)=i\,
 \frac{4\times 4\,\mu^2}{16\pi^3f^2} \sum_{\bf
v}\sum_{m,k=1}^8\frac 1
2\left(\frac{\pi\Delta}{R}\right)^2\,\delta_R[h({\bf v\cdot}\hat
{\bf n}_m)](i\,\varphi^{(m)})\,\delta_R[h({\bf v\cdot}\hat {\bf
n}_k)](i\,\varphi^{(k)})\int d^2\ell\,\frac{2\Delta_{eff}^2\,
V\cdot p\,\tilde V\cdot p}{[D(\ell)]^3}\ . \ee Using the result
\be \int
\frac{d^2\ell}{[D(\ell)]^3}\,=\,-\,\frac{i\pi}{2\Delta_{eff}^4}\
,\ee and the absence of off-diagonal terms in the double sum, we
get the effective lagrangian in the form
 \be
 {\cal L}_{eff}(p)=-\,\frac{\mu^2}{2\pi^2 f^2}\sum_{{\bf v}}
\left(\frac{\pi\Delta}{R}\right)^2\,\sum_{k=1}^8\frac{(\delta_R[h({\bf
v}\cdot {\bf n}_k)])^2}{\Delta_{eff}^2}\,\left( V\cdot
p\right)\varphi^{(k)}\left(\tilde V\cdot p\right)\varphi^{(k)} \
.\label{seinove} \ee To perform the calculation one can exploit
the large value found per $R$, and use the same approximations of
Section \ref{opw}. The  sum over $k$ in (\ref{seinove}) gives
\bea &&\sum_{k=1}^8\delta_R[h({\bf v}\cdot{\bf
n}_k)]\varphi^{(k)}\delta_R[h(\vec v\cdot{\bf
n}_k)]\varphi^{(k)}\to\,\frac{R}{\pi}\, \sum_{k=1}^8\delta[h({\bf
v}\cdot{\bf n}_k)]\left(\varphi^{(k)}\right)^2=\cr &&=\frac
R\pi\sum_{k=1}^8\,\delta\left[1-\frac{\delta\mu}{q{\bf
v}\cdot{\bf n}_k}\right]\left(\varphi^{(k)}\right)^2=\frac{
R^2}{\pi^2}k_R\sum_{k=1}^8\delta\left[{\bf v}\cdot{\bf
n}_k-\frac{\delta\mu}{q}\right]\left(\varphi^{(k)}\right)^2\,,
\eea with  \be k_R=\frac{\pi|\delta\mu|}{qR} \ .\ee Therefore one
gets
 \be{\cal
L}_{eff}(p)\,=\,-\,\frac{\mu^2k_R}{2\pi^2f^2}\sum_{i,j=1}^3\sum_{k=1}^8\sum_{\bf
v } \delta\left\{{\bf v}\cdot{\bf n}_k-\frac{\delta
\mu}{q}\right\} V_\mu\tilde V_\nu
\epsilon^{(k)}_i\epsilon^{(k)}_j p_\mu\phi^{(i)}p_\nu\phi^{(j)}\
. \ee The integration over the Fermi velocities requires special
attention. We use the result
 \be
 \sum_{k=1}^8
 \epsilon^{(k)}_i\epsilon^{(k)}_j \,=\,8\delta_{ij}\,;\label{74}\ee
 this fixes the constant multiplying the time derivative term in
 the effective lagrangian at the value (taking into account
 (\ref{sumvel}))
 \be \frac{8\mu^2k_R}{2\times4\pi^2 f^2}\ .
 \ee Therefore
  one obtains canonical normalization for the kinetic term
provided \be
 f^2= \frac{8\mu^2k_R}{4\pi^2}
  \ .
\ee The parameters $a,\,b,\,c$ of the effective lagrangian
(\ref{effective}) can be now evaluated and one finds
\cite{Casalbuoni:2002my}:  \bea &&{\cal L}_{eff}(p)=\frac 1
2\left( {p^0}^2{\phi^{(i)}}^2-\frac 1 8\beta^{ij}_{lm}p^l
p^m\phi^{(i)}\phi^{(j)}\right)=\nn\\&=&\frac 1 2
\Big({p^0}^2{\phi^{(i)}}^2-\frac {|{\bf p}|^2}{12}
{\phi^{(i)}}^2-\frac{3\cos^2\theta_q-1}6\sum_{i<j=1,2,3}p^i\phi^{(i)}p^j\phi^{(j
)}\Big)\ , \eea i.e., comparing with Eq. (\ref{effective}),\be
a=\frac{1}{12} \ ,~~~~~ b=0\ ,~~~~~
 c=\frac{3\cos^2\theta_q-1}{12}\ .\ee
\subsection{Gluon dynamics in the LOFF phase}
\label{VF}
\subsubsection{One plane wave structure} In this Section and in
the subsequent one we wish to derive the effective lagrangian for
the gluons of the unbroken $SU(2)_c$ subgroup of the two-flavor
LOFF phase.
 To begin with we
assume the crystal structure given by a plane wave and we neglect
the vector condensate, so that we write \be
\Delta_{eff}=\frac{\Delta\pi} R\delta_R[h({\bf v\cdot}\hat {\bf
n})] \label{deltaeff1} ~.\ee

The effective action allows the evaluation of the one loop
diagrams with two external gluon lines and internal quark lines
similar to those in Fig. \ref{figgraph}. If one writes\be
\Pi^{\mu \nu}_{a b}(p)\ =\ \Pi^{\mu \nu}_{a b}(0) + \delta
\Pi^{\mu \nu}_{a b}(p) \label{pi}\,,\ee then the Meissner mass
vanishes\be \Pi_{ab}^{ij}(0)=0\ee and the Debye screening mass is
non-vanishing \be m_D = \frac{g \mu}{\pi} \sqrt{ 1 + \frac{\cos
\theta_a-\cos\theta_b}{2} } \,,\ee where $\cos\theta_a$ and
$\cos\theta_b$ ($-1\le\cos\theta_a\le\cos\theta_b\le 1$) are the
solutions of the equation\be |h(\cos\theta)|=\frac
\pi{2R}\,.\label{106}\ee

Next consider $\delta \Pi^{\mu \nu}_{a b}(p)$. The only non
vanishing contribution to $\delta \Pi^{\mu \nu}_{a b}(p)$ comes
from the pairing region, i.e. where $\Delta_{eff}\neq 0$. In the
approximation of small
 momenta ($|p\,|\ll\Delta$) one finds \cite{Casalbuoni:2002pa}
\be -\,\delta \Pi^{\mu \nu}_{a b}(p)\ =\ \delta_{ab}\frac{\mu^2
g^2}{12 \pi^2} \sum_{{\bf v};\,\, pairing}\frac{V^{\mu} V^{\nu}
(\tilde V \cdot p)^2 -
 \tilde V^{\mu} V^{\nu}( V \cdot p\ \tilde V \cdot p)
 +V\leftrightarrow\tilde V}{\Delta^2_{eff}}. \label{pi2}
\ee That is
 \be -\delta\Pi^{0 0}_{ab}(p)=\delta_{ab} \frac{g^2 \mu^2}{3
\pi^2} \sum_{{\bf v};\,pairing} \frac{v_i v_j}{\Delta_{eff}^2}
p_i p_j\, =\,\delta_{ab} \frac{g^2 \mu^2R^2}{3 \Delta^2 \pi^4}
\int_{pairing} \frac{d\cos\theta\,d\phi}{8 \pi} \frac{v_i
v_j}{\left(\delta_R[h(\cos\theta)]\right)^2}\, p_i p_j\,,
\label{pi0pbis1} \ee where ${\bf
v}=(\sin\theta\cos\phi,\sin\theta\sin\phi,\cos\theta)$. The
integration domain is defined by
$\cos\theta_a<\cos\theta<\cos\theta_b$.  Therefore we get
 \be -\delta\Pi^{0 0}_{ab}(p)\ =\
\delta_{ab} k \left( f(R) p_\bot^2+
 g(R)p^2_\| \right)~,
\label{pi00} \ee where $k$ is given by ($\Delta_0$ the
homogeneous condensate):\be k= \frac{g^2 \mu^2}{18 \pi^2
\Delta_0^2} \label{glue} \ee and \be f(R) \, = \, \frac{3 }{4}
\int_{pairing} d \cos \theta \left(1 - \cos^2\theta\right)\ , \ee
\be g(R) \, = \, \frac{3}{2} \int_{pairing} d \cos
\theta\,\cos^2\theta \ee are functions of the  parameter $R$ and
are reported in Fig. \ref{plot}.

 It is interesting to note the anisotropy of the
polarization tensor exhibited by these results. One has always
$g>f$; for large $R$, and neglecting $\delta\mu/\mu$ corrections,
one finds approximately \be
\frac{g(R)}{f(R)}\,\to\,\frac{2}{\dd\left(\frac q
{\delta\mu}\right)^2-1}\ .
 \ee

\begin{figure}[htb]
\centerline{\epsfxsize=7.5truecm\epsffile{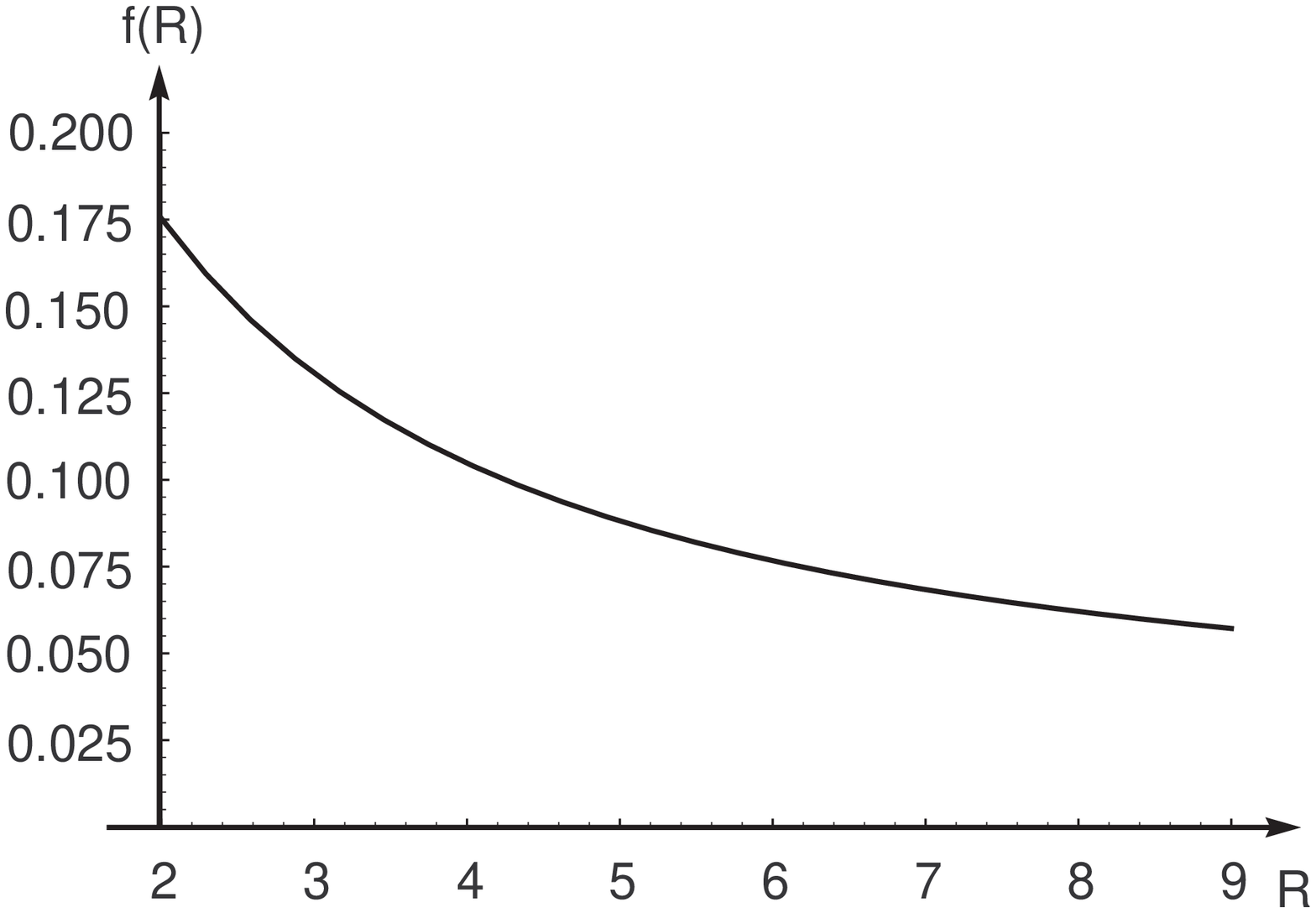}
\hskip1cm\epsfxsize=7.5truecm\epsffile{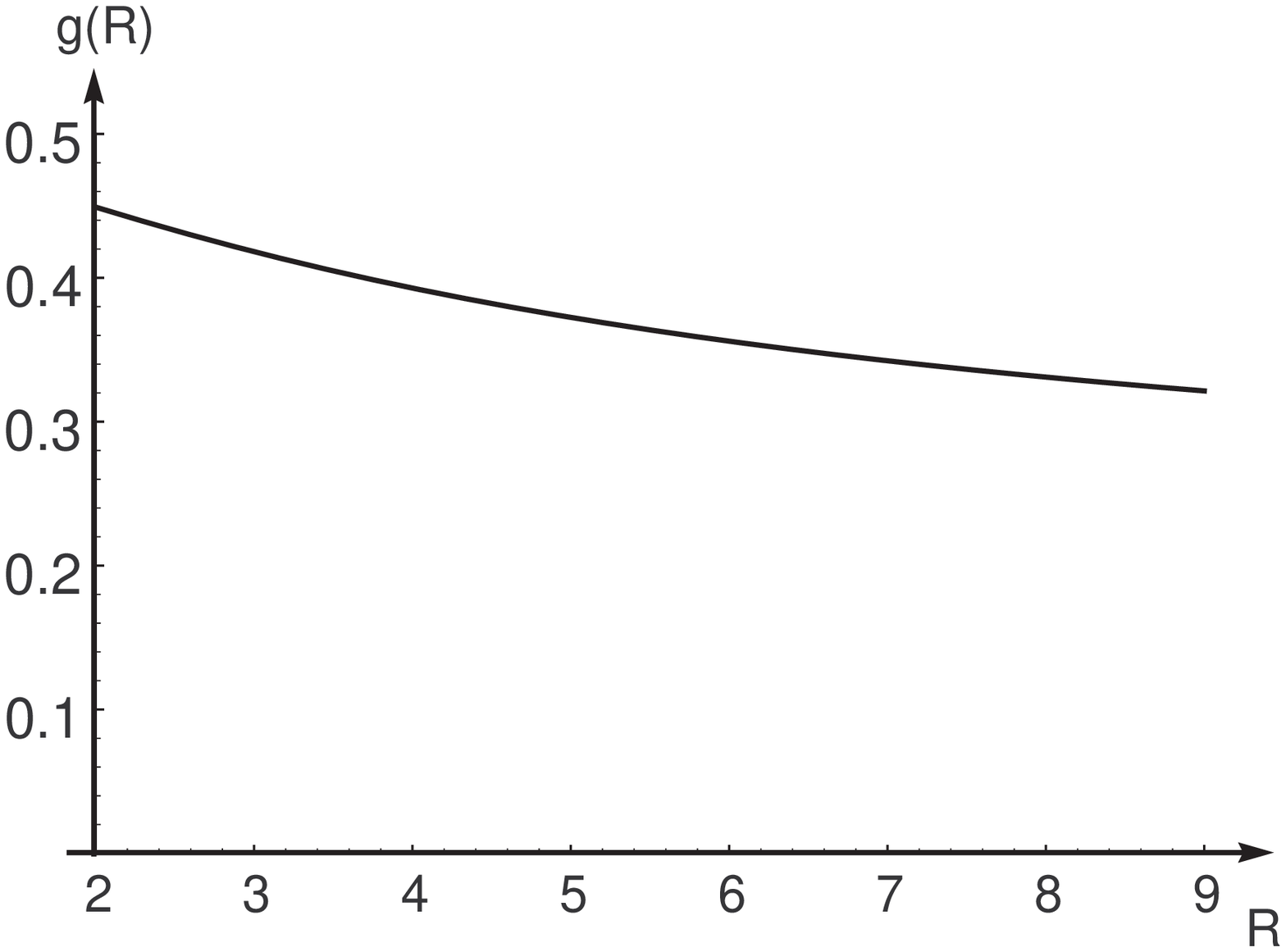}}
\vskip1cm\noindent \caption{\it \label{plot} Plots of the
functions $f(R)$ and $g(R)$.}
\end{figure}

Let us finally write down the remaining components of the
polarization tensor. From (\ref{pi2}) we get \be
-\,\Pi^{ij}_{ab}(p)\ =\ \delta_{ab} \frac{g^2 \mu^2}{3 \pi^2}
\sum_{{\bf v};\,\, pairing} \frac{v_i v_j}{ \Delta^2}  p_0^2 ~=\
k \, p_0^2\ \Big( f(R) (\delta_{i1} \delta_{j1}
+\delta_{i2}\delta_{j2})+
 g(R) \delta_{i3} \delta_{j3} \Big)
\label{piijp} \ee and \be -\,\Pi^{0 i}_{ab}(p)\ =\ k \, p_0\,p^j\
\Big( f(R) (\delta_{i1} \delta_{j1} +\delta_{i2}\delta_{j2})+
 g(R) \delta_{i3} \delta_{j3} \Big) ~.
\label{pi0ip} \ee These results complete the analysis of the LOFF
model in the one plane wave approximation. From $\Pi^{\mu \nu}_{a
b}$ we get the dispersion law for the gluons at small momenta.
The lagrangian at one loop is\footnote{We do not include here the
3 and 4-gluon vertices that however can be handled as in
\cite{Casalbuoni:2001ha}, with the result that the local gauge
invariance of the one-loop lagrangian is satisfied.}\be {\mathcal
L}\ =\ -\frac{1}{4} F^{\mu \nu}_{a} F_{\mu \nu}^{a} - \frac{1}{2}
\Pi^{\mu \nu}_{a b} A_{ \mu}^a A_{ \nu}^b\,. \label{su2lag2}  \ee
(sum over the repeated color indices $a,b=1,2,3$). Introducing
the fields $\dd E_i^a \equiv F_{0i}^a$ and $\dd B_i^a \equiv i
\varepsilon_{ijk} F_{jk}^a$, and using (\ref{pi00}),
(\ref{piijp}) and (\ref{pi0ip}) we can rewrite the lagrangian
(\ref{su2lag2}) as follows \be {\mathcal L}\ =\ \frac{1}{2}
\left(\epsilon_{ij}\,E_i^a E_j^a  - B_i^a B_i^a\right)+\frac 1 2
m_D^2 A_a^0 A_a^0 \label{lag2} ~, \ee where\be
\epsilon_{ij}=\left( \matrix{1+kf(R) & 0& 0 \cr 0 & 1+kf(R) & 0
\cr 0 & 0 & 1+k g(R)}\right) ~. \ee This means that the medium
has a non-isotropic {\it dielectric tensor} $\dd \epsilon$ and a
{\it magnetic permeability} $\dd \lambda = 1$. These results have
been obtained taking the total momentum of the Cooper pairs along
the $z$ direction. Therefore we distinguish the dielectric
constant along the $z$ axis, which is \be \epsilon_{\parallel}
=1+ kg(R) ~, \ee and the dielectric constant in the plane
perpendicular to the $z$ axis \be \epsilon_{\perp} =1+ k f(R) ~.
\ee This means that the gluon speed in the medium depends on the
direction of propagation of the gluon; along the $z$ axis the
gluon velocity is \be v_{\parallel}\simeq\frac{1}{\sqrt{k g(R)}}
~, \ee while for gluons which propagate in the $x-y$ plane we
have \be v_{\perp}\simeq\frac{1}{\sqrt{k f(R)}} \ee and in the
limit of large $R$, and neglecting $\delta\mu/\mu$ corrections,
\be v_{\parallel}\to \frac 1{\sqrt{2}}\tan\theta_q\, v_\perp
\,.\ee with $\cos\theta_q$ defined in Eq.
(\ref{hdix}).\subsubsection{Cubic structure\label{gluecube}} The
condensate in this case is given by Eq. (\ref{8a}), so that we
will use the results of Section \ref{hdet} with $\Delta_{eff}$
given by (\ref{deltaeff3}).
 The  calculations  are similar to the previous
case and, similarly, the $SU(2)_c$ gluons have vanishing Meissner
mass and exhibit partial Debye screening.  However the dispersion
law of the gluons is different.

As a matter of fact we write  the  one loop lagrangian for the
$SU(2)_c$ gluons as\be {\mathcal L}\ =\ \frac{1}{2} (E_i^a E_i^a
 - B_i^a B_i^a) +\delta{\mathcal L}~, \label{multilag1} \ee
with \be\delta{\mathcal L}=-  \frac{1}{2} \Pi^{\mu \nu}_{a b} A_{
\mu}^a A_{ \nu}^b \label{su2lag1} ~.\ee In the approximation
$|p|\ll\Delta$, $\delta\Pi^{\mu \nu}_{a b}$ is again given by Eq.
(\ref{pi2}), but now $\Delta_{eff}$ is given by
(\ref{deltaeff3}). One gets \be \delta{\mathcal L}\equiv E^a_i
E^b_j \ \delta_{ab} \frac{g^2 \mu^2}{6 \pi^2} \int_{pairing}
\frac{d\cos\theta\,d\phi}{8\pi} \frac{v_i
v_j}{\Delta_{eff}^2}+A^a_0 A^b_0 \,\delta_{ab}\frac{g^2 \mu^2}{4
\pi^2} \int_{blocking} d\cos\theta ~. \label{pi0pbis} \ee
Evaluating the integrals one finds \be {\mathcal L}\ =\
\frac{1}{2} (\tilde\epsilon_{ij}\,E_i^a E_j^a  - B_i^a B_i^a)
+\frac 1 2 {M_D}^2A_0^a A_0^a~, \label{multilag2} \ee with the
tensor $\tilde\epsilon^{ij}$ given by \be \tilde\epsilon_{ij}
=\delta_{ij}\left[1\,+ \, k \,t(R)\right]
 ~.\ee and \be M_D=\frac{g\mu}\pi\sqrt{1+8\frac{\cos\theta_a-\cos\theta_b}2}\,.\ee where
 $\cos\theta_{a,b}$ are solutions of Eq. (\ref{106}).
 The tensor $\tilde\epsilon^{ij}$ is
 isotropic. This result can be easily explained noticing that the
 lagrangian should be a quadratic function of the field strengths
  and should also satisfy the cubic symmetry. Therefore it must
  be constructed by the invariants $I_2(E_i)$ and $I_2(B_i^a)$
  that are isotropic. As shown in \cite{Casalbuoni:2002pa} $t(R)$ is given by\be t
(R)=\frac 8 3[2f(R)+g(R)] \ .\ee It should also be noted that the
 values of the parameter $R$ for the cube and the plane wave can
 be different.
 A plot of
the function $t(R)$ is in Fig. \ref{tdir}.
\begin{figure}[htb]
\centerline{ \epsfxsize=7.5truecm\epsffile{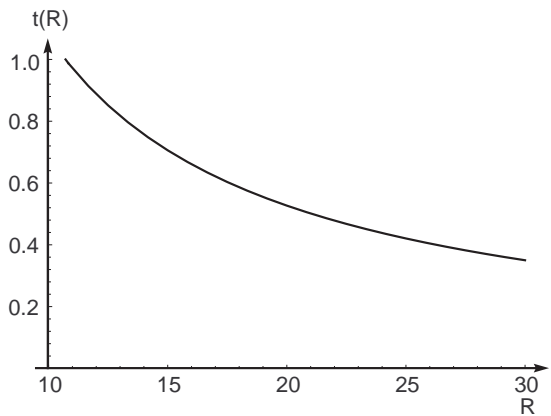}}
\caption{\it \label{tdir} Plot of the function $t(R)$.}
\end{figure}
Even if the crystalline structure is not isotropic, the
dielectric properties of the medium will be isotropic and the
velocity of propagation of the gluons will be the same in all the
directions.
\section{Inhomogeneous superconductivity in condensed matter,
nuclear physics and astrophysics} \label{phenomenology} As
observed in the introduction, the main focus of this review is on
the theoretical methods rather than  phenomenological
consequences. However, for completeness, in this Section we give
a review of the different approaches developed so far to detect
the inhomogeneous phase in superconductors. The LOFF is expected
to be ubiquitous, therefore one might expect to find it in
completely different physical systems. For obvious reasons
research in solid state physics is much more advanced
 and indeed signals of formation of the LOFF phase have been
 reported in the literature. In the first part of this Section we
 review them. In order to produce the effective exchange interaction of Eq. (\ref{exch})
  a sufficiently high magnetic field is needed to produce an appreciable difference
  between the chemical potentials. This could be done in type I superconductors, but the
  required fields are likely to destroy superconductivity altogether. This issue is discussed
   in Subsection \ref{VIA}.  To overcome this difficulty, type II
   superconductors must be used; moreover they should be free of
   impurities and have large electron mean free path; the needed
   requirements and the associated phenomenology are discussed in Subsection \ref{VIB}
   and \ref{VIC}. Another way to overcome the effects of high
   magnetic fields that are detrimental to electron
   superconductivity is to use layered superconductors and magnetic
   fields parallel to the layers. Superconductors of this type are
   rather different from the ones  considered
in the original LOFF papers; in particular organic
superconductors are compounds  with these features and are
therefore good candidates. They will be discussed in Subsection
\ref{VID}, while in Subsection \ref{VIE} we briefly discuss
future possible developments in the area of atomic physics. The
final part of this Section is devoted to phenomenological
implications of the LOFF phase in nuclear physics (Subsection
\ref{VIF}) and QCD (Subsections \ref{stars} and \ref{astro}). In
particular, in this last Subsection we discuss a possible role of
the QCD LOFF phase in the explanation of a peculiar phenomenon of
pulsars, i.e. the periodic glitches in their angular velocity.
\subsection{Type I
superconductors} \label{VIA}In the original LOFF papers
\cite{LO}, \cite{FF} the difference in chemical potentials
between spin up and spin down electrons arises from an
interaction of a magnetic field with the electron magnetic dipole
moments. The magnetic field can hardly be the external field
$H_{ext.}$, which
 exerts a stronger influence  on the
orbital motion than on the electron spin. Therefore the
inhomogeneous phase was thought to arise in nonmagnetic metals in
presence of paramagnetic impurities. Under an external field the
host impurities display ferromagnetic alignment; decreasing the
temperature the material becomes a superconductor while the
ferromagnetic alignment persists, leading to a constant
self-consistent exchange field, proportional to the average spin
of the impurities. This field  is at the origin of the modulated
order parameter. The value $\delta\mu_1\approx\Delta_0/{\sqrt 2}$
above which LOFF phase can exist corresponds to a critical value
of the  magnetic field that can be derived as follows
\cite{clogston}, \cite{chandrasekhar}. The susceptibility of an
electron gas in the normal phase at $T=0$ is \be \chi_n=\mu_B^2
\rho\,,\ee where $\rho={gp_F^2}/(2\pi^2 v_F)$ is the density of
states at the Fermi surface, $\mu_B$ is the Bohr magneton and
$g=2$ is the electron degeneracy factor. On the other hand the
susceptibility in the superconducting phase at $T=0$ vanishes:
$\chi_s=0$, because to polarize the superconductor one has to
break the Cooper pair, which costs energy. The free energy per
unit volume $f_s$ of the superconductor, in absence of
paramagnetic effects, is:\be\label{fs}
f_s=f_n-\frac{H_c^2(T)}{8\pi}\ ,\ee where $f_n$ is the free
energy of the normal phase and $H_c(T)$ is the critical field.
Including the Pauli paramagnetism implies adding the hamiltonian
a term \be
 -\mu_B\,\psi^\dagger\bm{\sigma} \cdot{\bf H}\psi \ ;\label{ham}\ee while
  (\ref{fs}) becomes\be\label{fsh}
f_s=f_n+\frac{(\chi_n-\chi_s)H^2}2-\frac{H_c^2(T)}{8\pi}\ .\ee
Therefore the BCS superconductivity will survive at $T=0$ for
magnetic fields satisfying\be
H\le\,\sqrt{\frac{H_c^2(0)}{4\pi\chi_n}}\,\equiv \,H_P(0)\ .\ee
Now $H^2_c(0)/8\pi=\rho\Delta_0^2/4$ and therefore the Pauli
limiting field  at $T=0$ is \be H_P(0)=\frac{\Delta_0}{\sqrt
2}\,\frac 1{ \mu_B}\ .\label{pauli}\ee The identification
$\delta\mu_1=\mu_B H_P(0)$ arises by the comparison between
(\ref{ham}), (\ref{pauli}) and (\ref{exch}).

For  a type I superconductor it is difficult to reach the Pauli
limit (\ref{pauli}) because, while $H_P(0)$ is typically of the
order of $300$ K Oe, $H_c(0)$ is of the order $1$ K Oe. Therefore
superconductivity will be broken by the magnetic field well
before the Clogston-Chandrasekhar limit is achieved. This implies
the LOFF phase is unlikely to be produced by these materials  and
one has to turn to type II superconductors \cite{SjSarma} because
for some of these superconductors the upper critical field
$H_{c2}$ can be very high.
\subsection{"Clean" and strongly type II superconductors}
\label{VIB} To evaluate the possibility of the LOFF state one has
to take into account not only Pauli paramagnetism of the
electrons but also the orbital effects \cite{Ginzburg:1957ab}.
Before doing that, let us
 first distinguish between "clean" and "dirty" superconductors
 \cite{Anderson:1959ab}. One calls "clean" the
superconductors in which the electron
  mean free path $l$ is much larger than the superconducting coherence
length $\xi_0$: \be l\gg\xi_0 \ , \label{heavy1}\ee while "dirty"
superconductors are characterized by the opposite condition
 $l\ll\xi_0$. In "clean" superconductors electrons at the Fermi surface move
 with velocity ${\bf v}_F$, while in "dirty" superconductors
the electron motion is described by a diffusion equation. "Dirty"
superconductors are characterized by the presence of impurities,
which can narrow and even destroy the LOFF state
\cite{Aslamazov:1969ab,Takada:1970ab}. Therefore materials with
small $l$, e.g. PbMo$_6$S$_8$, should not  display the LOFF
phase, see e.g. \cite{Decroux:1982ab}. On the other hand the so
called heavy-fermion superconductors are favored: these materials
are indeed characterized by the small Fermi velocity of their
quasi particles \cite{Rauchschwalbe}; since \be \xi_0=
\frac{\hbar v_F}{\pi\Delta}\ ,\ee for small enough $v_F$ the
condition (\ref{heavy1}) is satisfied. For example in the heavy
fermion compound UPd$_2$Al$_3$ the superconducting coherence
length is $\xi_0\approx 85$\,\AA, much smaller than the
electronic free mean path $l=700$\,\AA; therefore it can be
considered as a "clean" superconductor.

To evaluate the possibility of the LOFF state in type II
superconductors one has to take into account not only Pauli
paramagnetism of the electrons, but also the orbital effects.
This analysis was first performed by \cite{Gruenberg:1966ab}.
These authors followed the variational method of
\cite{Werthamer}, by making an ansatz for the condensate. In
general there is a competition between the orbital and the
paramagnetic effect, the former trying to organize a structure of
Abrikosov vortices and the latter a periodic LOFF structure;
therefore the orbital effect reduces the possibility of the LOFF
state that can exist only for sufficiently high $H_{c2}$. The
quantitative criterion at $T=0$ for clean superconductors with
isotropic dispersion law is as follows. The LOFF state can
persist in a type II superconductor provided \be \label{beta2}
\alpha={\sqrt 2} \ \frac{H_{c2}(0)}{H_P(0)}>1.8 \ .\ee Here
$\alpha$ is the parameter first introduced in \cite{Maki:1964ab},
$H_{c2}(0)$ is the
 Gor'kov upper critical field  at $T=0$ in
absence of paramagnetic effects \cite{Gor'kov:1960ab} and
$H_P(0)$ is the Pauli limiting field defined in (\ref{pauli}).

In conclusion good experimental conditions to observe the LOFF
state should be provided by a clean superconductor with a large
$\alpha$ value. These features are not easily found in the most
common superconductors and therefore experimental investigations
consider unconventional superconductors, e.g. heavy-fermion,
organic or high $T_c$ superconductors. As a matter of fact many
of these materials are strongly type II superconductors, which
means that the condition (\ref{beta2}) can be satisfied. Moreover
they have often a layered structure, which
 implies that,
applying the magnetic field parallel to the layers, the orbital
effect is minimum and the Zeeman effect, on which the LOFF phase
is based, is dominant.

The condition of being very clean and simultaneously strong type
II superconductors should be more easily realized in $d-$wave
superconductors like high $T_c$ cuprate superconductors and
organic superconductors like $\kappa$-(ET)$_2$ or
$\lambda$-(ET)$_2$ salts. They will be discussed in more detail
below; suffice it here to say that in $d-$wave superconductors
the region of the LOFF phase is much more extended than in
$s-$wave superconductors. The analysis of \cite{Maki:1996ab},
where this conclusion was drawn, has been extended in
\cite{Won:2002ab} to the calculation of the LOFF free energy,
specific heat and magnetic susceptibility; in particular for
these layered $d-$wave superconductors  the energetically favored
structure at $T=0$ is found to be \be\Delta(x,y)\propto\cos q
x\,+\,\cos qy\ .\label{eq:483}\ee

Other materials where the possible existence of LOFF phase has
been investigated are ferromagnetic metals or alloys
\cite{Pickett:1999ab}, \cite{Dyugaev:2001ab}. The study of the
possible coexistence of ferromagnetism and superconductivity was
initiated by \cite{Ginzburg:1957ab} who noted that, though  the
two orderings can in principle coexist, their simultaneous
presence is practically impossible under ordinary conditions. As
a matter of fact the presence in ferromagnets of a spontaneous
magnetization $M_0$ produces at $T=0$ an internal magnetic
induction $B_0=4\pi M_0$ even in absence of external magnetic
field. For superconductivity to exist, $B_0$ should be smaller
than the lower critical field at $T=0$, in absence of
ferromagnetism:\be B_0\le H^0_{c1}(0)\label{bh}\ .\ee However,
the induction $B_0$ at $T=0$ is of the order of $10$ K Oe (e.g.
22, 18.5, 6.4, 24.8 KOe respectively for Fe, Co, Ni, Gd), while
the critical field  is in general much smaller, of the order of a
few KOe or less. Superconductivity of ferromagnets is therefore
difficult unless special conditions render the condition
(\ref{bh}) possible. They might be, for example, a reduced size
of the sample, with dimensions of the order of the penetration
depth. The formation of the vortex phase in type II
superconductors, however, screens locally the internal magnetic
induction, and allows to avoid Ginzburg's negative conclusion
\cite{Krey:1973ab}. As a matter of fact superconductivity was
recently reported in the ferromagnetic alloy
RuSr$_2$GdCu$_2$O$_8$
\cite{Tallon:1999ab,Bernhard:1999ab,Pringle:1999ab,Hadjiev:1999ab}.
This layered material becomes first ferromagnet at $T=132 $ K;
superconductivity appears at $T=35-40$ K and finally, at
$T=2.6$~K Gd ions acquire an antiferromagnetic order. The
theoretical study of \cite{Pickett:1999ab} confirms these
reports, but suggests that the superconducting phase is of the
LOFF type, because the coupling between ferromagnetism and
superconducting layers appears to be sufficiently weak to permit
superconductivity, but strong enough to require the inhomogeneous
phase. In a similar context \cite{Dyugaev:2001ab} consider the
possibility of creating the LOFF phase using ferromagnetic
materials instead of nonmagnetic bulks with paramagnetic
impurities as in the original LOFF papers. Since the impurities
create not only an exchange interaction, but also an
electromagnetic interaction, using nuclear ferromagnetism, as
they propose, would reduce the latter, since the effective field
would be proportional to the nuclear magneton and not to the Bohr
magneton. They show that in some metals, e.g. Rh, W, the BCS
condensate imbedded in a matrix of ferromagnetically ordered
nuclear spins should manifest the LOFF phase.

All the proposals we have discussed so far are rather different
from the  one discussed in the original LOFF papers. An
 extension of the LOFF analysis to these materials and unconventional
 superconductors is beyond the scope of the present
review. We will therefore limit our presentation
 to a brief survey of the
experimental results, referring the interested reader to the
specialized literature
\cite{Murthy:1995ab,Gegebwart:1996ab,Shimahara:1996ab,Samohkin:1997ab,
Shimahara:1998fb,Symington:1999ab,Agterberg:2000ab,Yang:2001ab}.

 \subsection{Heavy fermion superconductors}
 \label{VIC}
 The first
experimental investigations on the LOFF phase used heavy-fermion
compounds such as CeRu$_2$ \cite{Huxley:1993ab}, UPd$_2$Al$_3$
\cite{Gloos:1993ab} and UBe$_{13}$ \cite{Thomas:1996ab}.  For all
these materials the conditions for the formation of the LOFF
state are met.  For example CeRu$_2$ is in a metallurgically
clean state; moreover it exhibits extreme type II behavior,
because the Ginzburg-Landau parameter, which discriminates
between the two type of superconductors \cite{SjSarma}, has the
value $\kappa=16$. As another example, the compound UPd$_2$Al$_3$
used by \cite{Gloos:1993ab} is characterized by a rather large
value of the parameter $\alpha$ in (\ref{beta2}) i.e.
$\alpha=2.4$, while also being a very clean superconductor. To
quote another result, in the analysis of a high quality single
crystal of UBe$_{13}$ \cite{Thomas:1996ab}, the very high value
$H_{c2}(0)\simeq 140$ KOe was reached. All these experimental
results are however inconclusive. In a critical analysis of the
experiment of \cite{Gloos:1993ab}, \cite{Norman:1993ab} shows
that the computed Gor'kov upper critical field  does not
correspond to the experimental results reported there; for
further analysis of the compound UPd$_2$Al$_3$ see
\cite{Yin:1993ab} and \cite{Schimanski:1994ab}. In the case of
CeRu$_2$, \cite{Tenya:1999ab}
 shows that the observed effects can be explained by flux pinning mechanisms involving
disorder.  \cite{Modler:1996ab} makes a comparative study of high
quality single crystals of UPd$_2$Al$_3$ and CeRu$_2$ in the
mixed state. The order parameter exhibits a periodic array of
nodal planes perpendicular to the Abrikosov vortex lines. In the
mixed state the pinning force is very weak; however the authors
find, for $H>10$ KOe and $T<0.9\,T_c$, a first order transition
to a state characterized by strong pinning, which might be
interpreted as the formation of a LOFF state. The mechanism by
which Abrikosov vortex lines in type II superconductors are
pinned to the vortex cores is similar to the one that pins vortex
lines to non-superfluid neutrons in a rotating superfluid within
neutron stars. It will be explained in more detail in Subsection
\ref{astro} in connection to a possible role of the QCD LOFF
state in the physics of pulsars.

\subsection{Two-dimensional,
quasi-two-dimensional and organic superconductors} \label{VID} As
we already mentioned, the paramagnetic effect can dominate if the
superconducting bulk has a layered structure and the magnetic
field acts parallel to it, because
 in this case the orbital upper critical field can be extremely high and
the breaking due to the spin interaction is most significant. The
importance for two-dimensionality (2D) to favor the LOFF state
was first observed in \cite{Bulaevskii:1973ab} and
\cite{Bulaevskii:1974ab} where
 both the orbital and the spin effect were taken into account
 and the upper critical field $H_{c2}(T)$ was calculated; in
  \cite{Buzdin:1984ab}
 the analysis was carried out near
the tricritical point. For the same reason also
quasi-one-dimensional (Q1D) compounds were discussed
\cite{buzdin:1983cd,buzdin:1987ef,Dupuis:1993ab,Dupuis:1995ab},
even though the results of \cite{Shimahara:1998ab} indicate that
the 2D structures are favored in comparison the 1D ones.

These results were generalized  to arbitrary temperature and
$d-$wave superconductivity in \cite{Shimahara1997ac}. The main
result of this paper is that the critical field curve $H_{c2}(T)$
is non monotonic and consists of different pieces corresponding to
different Landau levels, characterized by $n>0$. On the contrary,
the Ginzburg-Landau theory  would predict the pair wave function
to be in the lowest energy Landau level, with $n=0$ at $H_{c2}$.
The paper \cite{Shimahara:1998ab} studies the most favored
structure for a 2D LOFF crystal with a cylindrical Fermi surface.
First, the author finds that in general the 2D structures are
favored over the 1D ones; second, it finds that the favored
crystalline structure changes with $T$. For $s-$wave the results
are as follows: at high temperature the antipodal pair
condensate\be \Delta({\bf r})\propto 2\cos{\bf q\cdot
r}\label{LOCos}\ee is favored. This was the result found by
\cite{LO} in 3D at $T=0$. Decreasing the temperature other
structures become favored: first the triangle, then the square and
finally, at low temperatures, the hexagon. For $d-$wave pairing at
high temperature again (\ref{LOCos}) is favored, while at small
temperature the square dominates; on the other hand at
intermediate temperatures the phase transition should be first
order. The result at $T=0$ has been confirmed by
\cite{Won:2002ab}, see Eq. (\ref{eq:483}). As shown in
\cite{Lebed:1986ab}, the quasi-2D superconductors can be treated
as essentially 2D and therefore the results of
\cite{Shimahara:1998ab} should hold also for quasi-2D compounds
provided the external field is sufficiently strong and is kept
parallel to the superconducting layer.

 \cite{Klein:1999ab} consider a layered
superconductor in a magnetic field of arbitrary orientation with
respect to the conducting plane. The calculation is based on the
quasi-classical Eilenberger equations \cite{Eilenberger:1968ab},
\cite{Alexander:1985ab} and allows to elucidate the structure of
the stable states below $H_{2c}$ minimizing the free energy. The
stable states are neither pure LOFF states nor pure Abrikosov
vortex states, but are two-dimensional periodic structures or
quasi-one-dimensional structures where LOFF domains are separated
by vortex chains. \cite{Barzykin:2002ab} address 2D surface
superconductivity in presence of intense magnetic fields parallel
to the surface. The spin-orbit interaction at the surface changes
the properties of the LOFF state; the authors find that strong
spin-orbit interactions significantly broadens the range of
parameters where the LOFF phase can exist and produces   periodic
superconducting stripes running along the field direction on the
surface.

 Organic superconductors are good candidates for the formation of the
 LOFF state for the reasons mentioned above: i) They have narrow electron bands and
therefore they are in principle clean type II superconductors;
ii) due to their low dimensionality the orbital pair-breaking
effect is suppressed for magnetic fields parallel to the layers
they form. For these reasons they have been discussed by several
authors, e.g.
\cite{Lebed:1986ab,Gor'kov:1987ab,Dupuis:1993ab,Shimahara:1994ab,
Dupuis:1995ab,Burkhardt:1994ab,Shimahara:1997cd}.  It is
interesting to note that in general to detect the transition from
the homogeneous BCS phase to the LOFF phase thermodynamic
signatures are chosen. This however can give ambiguous results
since the signatures can be produced by phase transitions of
different nature. Therefore Ref.
 \cite{Yang:2000cd} proposes the use of the  Josephson effect.
  According
 to this analysis, at the Josephson junction
between two superconductors, one in the BCS and the other in the
LOFF phase, the Josephson current is suppressed.

 As discussed in
 \cite{Shimahara1997ac}, the upturn of the upper critical field
$d^2H_c/dT^2>0$ is a common feature in the LOFF state in quasi-2D
systems
 and is due a Fermi surface effect. Investigations
  on the sensitivity of the LOFF state
 to the shape of the Fermi surface
 are in \cite{Aoi:1973ab,Shimahara:1994ab,Shimahara:1997cd}.
This upturn and a first order transition below the critical field
have been observed in the organic compound
$\kappa$-(BEDT-TTF)$_2$Cu(NCS)$_2$.  This quasi-two-dimensional
(Q2D) organic superconductor  is examined by a number of  authors
 \cite{Nam:1999ab,Singleton:2000ab,Ishiguro:2000ab,Houzet:2000ab,
Symington:2001wz,Singleton:2001ab} and, for a similar compound,
\cite{Goddard:2002ab}. These studies indicate that some evidence
of the formation of the LOFF state has been reached. For example
\cite{Singleton:2000ab} studied resistance and magnetic behavior
of single crystals  of this superconductor in magnetic fields  up
to 33 T and at temperatures between 0.5 K and 11 K. When the
magnetic field lies precisely in the Q2D planes of the material,
they find  evidence for a phase transition from the
superconducting mixed state into a LOFF state, manifested as a
change in the rigidity of the vortex system. \cite{Manalo:2000ab}
compare the theoretical anisotropic upper critical field $H_{c}$
of a quasi-two-dimensional d-wave superconductor with recent
$H_{c2}$ data for $\kappa$-(BEDT-TTF)$_2$Cu(NCS)$_2$ and find
agreement both with regard to the angular and the temperature
dependence of $H_{c}$. According to these authors this supports
the suggestion that the LOFF phase exists in this material for
exactly plane-parallel orientation of the magnetic field.

In \cite{Uji:2001ab} field
 induced superconductivity  was reported in an organic
 superconductor $\lambda-$(BETS)$_2$FeCl$_4$
 (BETS=bis(ethylenedithio)tetraselenafulvalene).
 A possible mechanism to create
 field
 induced superconductivity is  the Peter-Jaccarino effect
 \cite{Jaccarino:1962ab}. However the upwards convex
 nature  of the lower critical field as
  a function of the temperature casts doubts on this interpretation.
Therefore some authors, e.g. \cite{balicas:2001ab}, have proposed
that these results can be interpreted as evidence of the formation
of the LOFF state. These results were reviewed in
\cite{houzet:2002ef}  and \cite{Shimahara:2002ab}. In particular
in the latter paper, an experimental phase diagram of the
field-induced superconductivity in this organic compound was
theoretically reproduced by a combination of the LOFF state and
the Jaccarino-Peter mechanism. \cite{Tanatar:2002ab} discusses
wether  LOFF state has been observed via thermal conductivity
$\kappa(H)$ in quasi-two-dimensional organic superconductor
$\lambda$-(BETS)$_2$GaCl$_4$. For clean samples the behavior of
$\kappa(H)$ is similar to the one expected by a second order phase
transition and is consistent with the formation of a LOFF phase.

\subsection{Future developments}
\label{VIE} The superconducting LOFF phase might be realized even
if the difference in chemical potentials of two species were not
generated by a magnetic field acting on electron spins. Apart
from nuclear physics and pulsars, to be discussed below, another
context might be offered by ultracold quantum degenerate Fermi
gas of atoms comprising two hyperfine states.  The experimental
investigations of ultracold gases were first dedicated to the
study of the Bose-Einstein condensation
\cite{Anderson:1995ab,Davis:1995ab,Bradley:1997ab,Fried:1998ab};
subsequently these techniques have been extended also to
magnetically trapped ultracold alkali Fermi gases or to gases
with coexisting Bose-Einstein condensate and Fermi gas
\cite{Schreck:2001ab,Roati:2002ab,Modugno:2002ab}. In particular
two state mixtures of ultracold gases have been employed, with
$^{40}$K vapors \cite{DeMarco:1999ab,DeMarco:2001ab}, or $^{6}$Li
\cite{Ohara:2001ab,Granade:2002ab}, or a mixture of $^{6}$Li and
$^{7}$Li \cite{Mewes:1999ab}. Future developments could lead to
the observation of superconductivity and Cooper fermion pairs
condensation in these systems. As discussed in
\cite{Combescot:2000ab} it is quite likely that the two hyperfine
states would have different atomic populations, since at the
moment there are no known fast relaxation mechanisms to equalize
the two atomic populations. Therefore superconductivity for
two-state ultracold Fermi gases is likely to be of the LOFF type.
The author of Ref. \cite{Combescot:2000ab} has performed a
theoretical study of $^{6}$Li under the above mentioned
conditions; he considers not only $s-$wave interactions, but also
an anisotropic term induced by density fluctuation exchange and
shows that the range where the LOFF phase is realized increases
with the increasing role of the anisotropic term. This is an
interesting theoretical development, which adds new interest to
the experimental investigations of ultracold atomic Fermi gases.
It remains to be seen, however, if such possibility is actually
realized in Nature.
\subsection{LOFF phase in nuclear physics}
\label{VIF}Neutron-proton pair correlations and the possibility
of $n-p$ Cooper pair condensation are presently studied in
several different contexts, from heavy ion collisions to
astrophysics. They have been investigated, using the BCS theory,
in infinite nuclear matter
\cite{Alm:1990ab,Vonderfecht:1991ab,Baldo:1992ab,Alm:1993ab,Alm:1996vb,
Sedrakian:1997ab},
 and by mean-field effective interactions
in finite nuclei.  In several cases nuclear matter is highly
asymmetric, with proton concentration at most  30-40\% in
supernova matter and 10\% in neutron stars. These asymmetries are
detrimental to nucleon superconductivity; on the other hand, for
weakly asymmetric states, fermion condensation is indeed
possible. For example, weakly isospin asymmetric nuclear matter
favors the formation of Cooper pairs in the $^3S_1$-$^3D_1$
channel, due to the presence of a tensor force; gaps are of the
order of 10 MeV. Condensation in this channel might be relevant
for low density bulk matter such as dilute nuclear matter in
supernovas. On the other hand there is no evidence of large gap
isospin singlet pairing in ordinary nuclei, which might be
explained by the presence of spin orbit interaction
\cite{goodman,martinez}. The authors \cite{Sedrakian:1999cu}
study the dependence of the gap as a function of both the isospin
asymmetry $\alpha_{np}=(\rho_n-\rho_p)/\rho$
 and the temperature, using realistic nuclear interactions.
 For small asymmetries the gap develops a
 maximum at a certain intermediate temperature; for large
 asymmetries the superconducting phase exists only at finite
 temperature, because the smearing effect of the temperature on
 the Fermi surfaces favors condensation. At higher values of
$\alpha_{np}$ ($\simeq 0.11$ in their model) pairing is no longer
possible.

Also in the context of isospin asymmetric nuclear matter it is
possible to have a transition from the BCS state to a LOFF phase
instead of the normal state
\cite{Sedrakian:1997ab,Sedrakian:2000an,Isayev:2001bb}. In Ref.
\cite{Sedrakian:2000an}  the possibility of spatially
inhomogeneous condensate in asymmetric nuclear matter is studied.
Condensation is possible in different channels. The isospin
triplet channels are favored for large enough asymmetries; more
exactly the channel $^1S_0$ dominates at low densities and the
channel $^3P_2$-$^3F_2$ (or $^1P_2$) at high densities. For weak
asymmetries the dominant channels are the isospin singlets
$^3S_1$-$^3D_1$ (low densities) and $^3D_2$ (high density). The
author considers the case of low density; as the isospin singlet
$^3S_1$-$^3D_1$ has a strength much larger than the isospin
triplet $^1S_0$, he neglects the latter. The interaction is
modelled by the Paris nucleon-nucleon potential. The gap
equations are solved numerically and have non trivial solutions
for non vanishing total momentum of the pair $P$. The LOFF phase
is favored for $\alpha_{np}>0.25$ and $P=0.3p_F$, independent of
$\alpha_{np}$. For  $\alpha_{np}>0.37$ pairing exists only in
presence of non-vanishing $P$. The maximal values of
$\alpha_{np}$ and $P$ compatible with the LOFF state are 0.41 and
$0.3\,p_F$ respectively (the actual values are indicative, as a
refinement of the treatment of the nuclear interaction  may
change them by a factor as large as 3). The results are obtained
at $T=3$ MeV. From the BCS to the LOFF phase the phase transition
is first order, while one passes from the LOFF to the normal
state by a second order phase transition. No attempt is made to
determine the most favorite crystalline structure.

Under hypotheses similar to those of the previous paper
\cite{Isayev:2001bb}
 studies the effect of coupling between the isospin singlet and
 isospin triplet, since at low densities pairing between these two channels may be important
 \cite{Akhiezer:1999ab}. Besides, the author goes beyond the
 approximation of "bare" nucleon interaction, by using the Fermi-liquid
 phenomenological approach \cite{Akhiezer:1994ab}. By these
 changes one finds interesting  peculiarities at $T=0$. First, the triplet-singlet channel turns
out to be energetically favored; second,  the phase transition
from the LOFF to the normal state can be of first order, depending
on the nature of the nucleon interaction. While still model
dependent, these investigations of the LOFF phase in nuclear
interactions are interesting as they offer, in principle, a
different way to  the LOFF phase. To be closer to phenomenology
one should consider however more complicated structures such as,
for example, hyperon rich matter. Alternatively the modulation of
the order parameter might be caused by Pauli paramagnetism due to
strong magnetic fields in highly magnetized neutron stars
(magnetars). In this case one could have a splitting in the Fermi
surfaces of a nucleon pair in the $I=1$, $L=0$ channel
\cite{Sedrakian:2000an}.

\subsection{Why color LOFF superconductivity could exist  in pulsars}
\label{stars}

In this Subsection and in the next one we will be interested in
some numerical estimates of the values of the parameters needed
for the LOFF phase in pulsars to occur. In general, color
superconductivity in quark matter might be realized in compact
stars. This expectation follows from the following two facts.

First of all the BCS critical temperature is given by \be
T_c=0.57\Delta_{BCS}\ee and in QCD $\Delta_{BCS}$ is expected to
range between 20 to 100 MeV. This estimate arises from weak
coupling calculations
\cite{Son:1998uk,Schafer:1999pb,Pisarski:1998nh,
Pisarski:1999av,Pisarski:1999bf,Pisarski:1999tv,Hong:1999fh,Brown:1999aq,
Brown:1999yd,Brown:2000eh,Hsu:1999mp,Schafer:1999fe,
Beane:2000hu,Beane:2000ji,Shovkovy:1999mr,Evans:1999at,Rajagopal:2000rs}
which are valid only at asymptotically high chemical
 potentials $\mu\gg 10^8$ MeV \cite{Rajagopal:2000rs},
and from models with parameters adjusted to reproduce the physics
at zero densities
\cite{Alford:1998zt,Rapp:1998zu,Alford:1998mk,Berges:1998rc,
Rajagopal:1999cp,Wilczek:1999fr,Schafer:1999hp,Alford:2000yp,Carter:1998ji,
Evans:1998ek,Evans:1998nf,Schafer:1998na,Rapp:1999qa}. None of
these calculations is valid at chemical potentials around $400$
MeV, which correspond roughly to the density of the inner core of
a neutron star, as we shall see below. However in all these cases
one gets values of the gaps of the order quoted before.

The second fact has to do with the thermal history of a  pulsar.
The general belief is that compact stars such as pulsars are
formed in the core of a supernova explosion. The temperature at
the interior of the supernova is about $10^{11}$ K, corresponding
to $10$ MeV (1 MeV = 1.1065$\times 10^{10}$ K). The star cools
very rapidly by neutrino emission with the temperature going down
to $10^9$-$10^{10}$ K in about one day. The neutrino emission is
then dominating the cooling for one thousand years. In this
period the temperature reaches about 10$^6$ K. After this period
the star cools down due to X-ray and photon emission and in a few
million years reaches a surface temperature around $10^5$ K.
Therefore, for the greatest part of its existence a neutron star
has a temperature lower than the critical temperature, with the
possibility of forming color superconducting condensates. It
follows that also in this context a compact star can be
considered at zero temperature because its temperature is much
smaller than the typical BCS energy gap, $T_{\rm
n.s.}/\Delta_{BCS}\approx 10^{-6}-10^{-7}$.

We have seen previously that QCD favors the formation of BCS
condensates in idealized cases, e.g. two or  three massless
flavors of quarks. However in realistic cases the three quarks
have different Fermi momenta due to the mass difference. It is
interesting to have an idea of the order of magnitude of the
scales involved in the description of a neutron star with a quark
core. We begin with a very crude example of a free gas of three
flavor quarks, taking up and down massless and the strange one
with mass $M_s$ \cite{Alford:1999pb}. Requiring that the weak
interactions are in equilibrium it is easy to determine the
chemical potentials and the Fermi momenta for the quarks. We find
\bea\mu_u&=&\mu-\frac 2
3\mu_e,~~~p_F^u=\mu_u\,,\nn\\
\mu_d&=&\mu+\frac 1 3\mu_e,~~~p_F^d=\mu_d\,,\nn\\
\mu_s&=&\mu+\frac 1 3\mu_e,~~~p_F^s=\sqrt{\mu_s^2-M_s^2}\,,\eea
where $\mu$ is average chemical potential \be\mu=\frac 1
3\,(\mu_u+\mu_d+\mu_s)\,\ee and $\mu_e$ the chemical potential of
the electrons. Notice that \be \sum_{i=u,d,s}\mu_i N_i+\mu_e
N_e=\mu N_q -\mu_e Q\,,\ee where \be
N_q=\sum_{i=u,d,s}N_i,~~~Q=\frac 2 3 N_u-\frac 1
3(N_d+N_s)-N_e\,.\ee The chemical potential for the electrons is
fixed by requiring electrical neutrality, corresponding to the
following condition for the grand potential $\Omega$  at zero
temperature \be Q=\frac{\de\Omega}{\de\mu_e}=0\,.\ee $\Omega$ is
obtained from Eq. (\ref{eq:61}) (omitting the volume factor)\be
\Omega= \frac 1{\pi^2}\int_0^{p_F} p^2(E(p)-\mu)dp\,.\ee In our
case we get \be\Omega=\frac 3{\pi^2}\sum_{i=u,d,s}\int_0^{p_F^i}
p^2(E_i(p)-\mu_i)dp+\frac 1{\pi^2}\int_0^{\mu_e}
p^2(p-\mu_e)dp\,,\ee where \be
E_{u,d}(p)=p,~~~E_s(p)=\sqrt{p^2+M_s^2}\,.\ee Although the
integral is feasible its expression is algebraically involved and
it is easier to do all calculations numerically. In particular
the result for the chemical potential of the electrons for
different values of $\mu$ as a function of $M_s$ is given in Fig.
\ref{fig:14}.
\begin{center}
\begin{figure}[htb]
\epsfxsize=10truecm \centerline{\epsffile{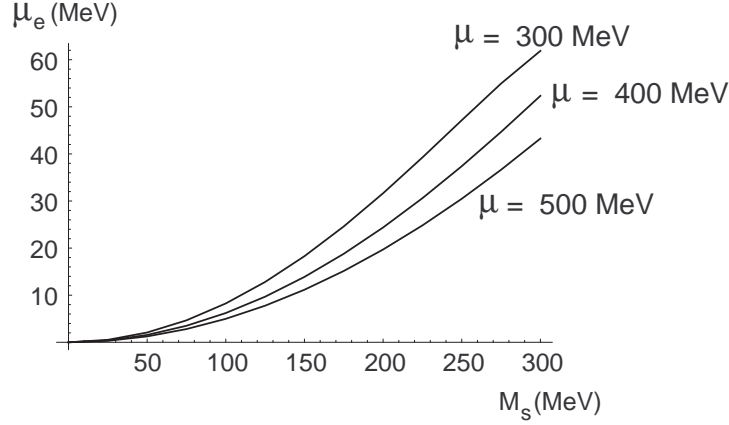}}
\caption{{\it The chemical potential of the electrons vs. $M_s$
for three values of the average chemical
potential.\label{fig:14}}}
\end{figure}
\end{center}
We can get an analytical expression  by performing an expansion
up to the order $M_s^4/\mu^4$. One gets \be \mu_e\approx
\frac{M_s^2}{4\mu}\ee and \be \Omega\approx -\frac
3{4\pi^2}\mu^4+\frac
3{4\pi^2}M_s^2\mu^2-\frac{7-12\log(M_s/2\mu)}{32\pi^2}M_s^4\,.\ee
The baryon density is obtained as \be \rho_B=-\frac 1
3\frac{\de\Omega}{\de\mu}=\frac{1}{3\pi^2}\sum_{i=u,d,s}
(p_F^i)^3\,.\ee The plot of the  ratio of the baryon density to
the nuclear baryon density is given in Fig. \ref{fig:15}. The
nuclear baryon density has been assumed  as the inverse of the
volume of a sphere of radius about 1.2 fm. Within the same
approximation as before one finds \be \rho_B\approx
\frac{\mu^3}{\pi^2}\left[1-\frac 1
2\left(\frac{M_s}\mu\right)^2\right]\,.\ee We note that densities
in the core are of the order of $10^{15}\,g/cm^3$, corresponding
to a chemical potential of the order of 400 MeV, as shown in Fig.
\ref{fig:15}.
\begin{center}
\begin{figure}[thb]
\epsfxsize=10truecm \centerline{\epsffile{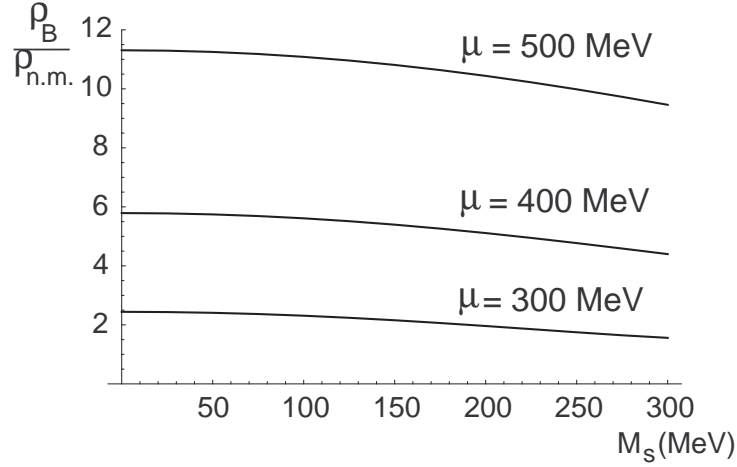}}
\caption{{\it The ratio of the baryon density of the free quark
gas to the nuclear baryon density  vs. $M_s$, for three values of
the average chemical potential.\label{fig:15}}}
\end{figure}
\end{center}
In particular let us  discuss  the range of values around 400 MeV
of the average chemical potential, with a strange mass of the
order 200-300 MeV (the strange mass here is not the current mass
but an effective density dependent mass). With $M_s=300$ MeV one
finds $\mu_e=53$ MeV (56 MeV from the approximate equation) with
Fermi momenta \be p_F^u=365\,{\rm MeV},~~~p_F^d=418\,{\rm
MeV},~~~p_F^s=290\,{\rm MeV}\,, \ee and a baryon density about
4.4 times the nuclear matter density. With  $M_s=200$ MeV the
result is $\mu_e=24$ MeV (25 MeV from the approximate equation)
and \be p_F^u=384\,{\rm MeV},~~~p_F^d=408\,{\rm
MeV},~~~p_F^s=357\,{\rm MeV}\,,\ee and a baryon density about 5.1
times the nuclear matter density. To go to baryon densities
relevant to the central core of the star, i.e. densities from 6
to 8 times the nuclear matter density, one needs to go to higher
values of $\mu$ and lower values of $M_s$ where the difference
among of the Fermi momenta is lower. This can be seen from Fig.
\ref{fig:16}, or using our approximate expression for $\mu_e$:\be
p_F^u\approx \mu-\frac{M_s^2}{6\mu},~~~ p_F^d\approx
\mu+\frac{M_s^2}{12\mu},~~~p_F^s\approx
\mu-\frac{5M_s^2}{12\mu}\,,\ee with \be p_F^d-p_F^u\approx
p_F^u-p_F^s\approx \frac{M_s^2}{4\mu}\,.\ee
\begin{center}
\begin{figure}[htb]
\epsfxsize=9truecm \centerline{\epsffile{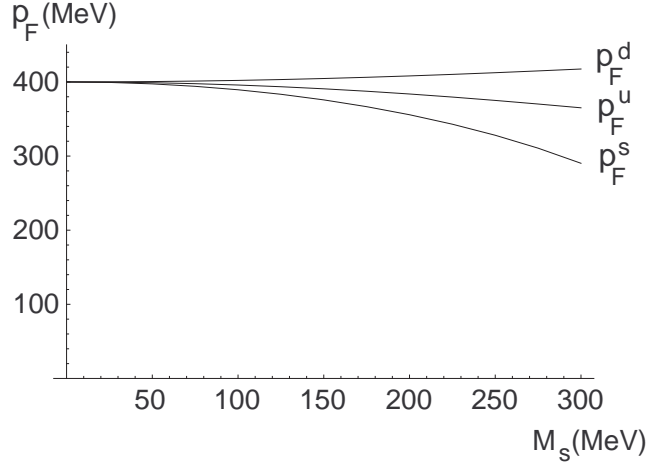}}
\caption{{\it The Fermi momenta of the three quarks vs.
$M_s$.\label{fig:16}}}
\end{figure}
\end{center}
The previous results are rather general, but in order to discuss
the possible astrophysical applications  we need to fix a value
for $\Delta_0$. Notice that we can trade the coupling constant
$G$ (see Eq. (\ref{eq:291})) for $\Delta_0$ since $G$ is fixed
once we give the cutoff $\delta$. On the other hand, the equation
for the chiral gap \cite{Rajagopal:2000wf} gives a relatione
between the NJL cutoff $\Lambda=\delta+\mu$, the coupling $G$ and
the constituent quark mass. By taking the constituent mass around
$300-400$ MeV and fixing $\Lambda$, one has still a parameter to
play around and it is possible to get values of $\Delta_0$ from
about 20 MeV up to about 100 MeV. In the present case, since the
typical value of $\delta\mu$ inside the LOFF window is
$0.7\Delta_0$ and \be \delta\mu=\frac 1 2(\mu_d-\mu_u)=\frac 1
2\mu_e\,,\ee we can reproduce approximately the situation
illustrated at the beginning of this Subsection with $M_s=300$
MeV by choosing $\Delta_0=40$ MeV. With this choice the LOFF
grand potential at $\delta\mu=\delta\mu_1$ is of the order
$10^{-7}$ GeV$^4$ which, as we shall see  in Section \ref{astro},
is of the right order of magnitude to give rise to the glitch
phenomena \cite{Alford:1999pb}. Notice also that the LOFF
condensate evaluated at $\delta\mu_1$ \be
\Delta_{LOFF}(\delta\mu_1)\approx 0.25 \Delta_0=10\,{\rm MeV}\ee
is much larger that a typical temperature of neutron stars, of
the order of  keVs.

\subsection{Astrophysical implications of the QCD LOFF
phase\label{astro}}

While a great experimental effort is devoted to the search of the
LOFF phase in condensed matter, so far nothing similar happens
for the crystalline phase of QCD. The reason is that it is
difficult to create the experimental conditions of high density
and low temperature for hadronic matter. The crystalline
superconducting phase of quarks may however result relevant for
astrophysical dense systems, in particular in the explanation of
pulsar glitches. Pulsars are rapidly rotating stellar objects,
characterized by the presence of strong magnetic fields and by an
almost continuous conversion of rotational energy into
electromagnetic radiation. The rotation periods can vary in the
range $10^{-3}$ sec up to a few seconds; these periods increase
slowly and never decrease except for occasional glitches, i.e.
sudden increases of the rotational frequencies, when the pulsar
spins up with a variation in frequency of the order of  $
\delta\Omega/\Omega\approx 10^{-6}$ or smaller. Glitches are a
typical phenomenon of the pulsars, since probably all the pulsars
experience them.

Pulsars are commonly identified with neutron stars; these compact
stars are characterized by a complex structure comprising a core,
an intermediate region with superfluid neutrons and a metallic
crust.  With a chemical potential of the order of 400 MeV, as we
have seen, the conditions for color superconductivity in the CFL
or the LOFF versions might be reached in the core. Before
examining this possibility, let us however describe the standard
explanation of glitches, in the form originated by the papers
\cite{Anderson:1973ab,Alpar:1984ab}. This model is based on the
idea that the sudden jumps of $\Omega$
 are due to the movements outwards of rotational vortices in the neutron superfluid and their
interaction with the crust. Crucial ingredients of the model are
therefore the existence of a superfluid and a crystal (the
metallic crust). This is one of the main reasons that allows the
identification of pulsars with neutron stars, as only neutron
stars are supposed to have a metallic crust. The LOFF state can
be relevant in this context because, if there is a LOFF phase
inside the pulsar, the superfluid might interact with the LOFF
crystal instead of the crust, thus providing an alternative or
complementary mechanism for the glitches. Thus far, there is no
developed model for the pinning of the superfluid vortices to the
QCD  LOFF crystals within compact stars. Therefore we limit our
survey to an introduction to the subject, along the lines of
\cite{Alford:2000ze,Alford:2000yp,Nardulliconf}.

Let us consider a compact star whose metallic crust rotates
 with angular
velocity $\bf\Omega$. The superfluid inside the star
 should not rotate because, in absence
 of friction, the crust cannot communicate
its rotation to the superfluid component.  The velocity of the
superfluid is ${\bf v}_s=\hbar\,{\bm{ \nabla}}\Phi/m\label{vs}$
where $\Phi$ is the phase of the superfluid condensate wave
function. The consequence of this formula would be
$\oint_\gamma{\bf v}_s\cdot d {\bm{\ell}}=0$. This would imply
the absence of rotation in the superfluid, which however does not
correspond to the state of minimal energy (for a discussion see
\cite{Landau2}). The correct condition is \be\oint_\gamma{\bf
v}_s\cdot d{\bm{\ell}}\,=\,2\pi n\kappa\ , \label{g5}\ee where
 $\kappa$ is the quantum of vorticity: $
\kappa={\hbar}/{m}$. For Eq. (\ref{g5}) to hold the curve
$\gamma$ must wind a singular point; the integer $n$ is the
winding number which counts the number of times the curve goes
around the singular point; the most energetically favorable
condition is realized by $n=1$. If $\gamma$
 is in a plane the condition (\ref{g5}) holds for any plane and the
 {\it locus} of the singular points is a vortex line (v.l.).
 In absence of rotation there are no v.l.'s; the minimal angular
velocity $\Omega_{min}$ for the formation of the first vortex
line is\be \Omega_{crit}=\frac\hbar {mR^2}\ln\frac R a \ .\ee
Here we are assuming
 a
cylindrical configuration with radius $R$;  $a$ is a cutoff of
the order of the interatomic distances. By increasing $\Omega$
also the number $N$ of vortex lines per unit area in the
superfluid increases according to the formula:\be
N=\frac{m\Omega}{\pi\hbar} \label{gn}\ee and one gets, instead of
(\ref{g5}), \be\displaystyle \oint_\gamma d{\bm{\ell}}\cdot{\bf
v}_s= NA2\pi\kappa\label{n} \ ,\ee where $A$ is the area
encircled by $\gamma$. Eventually the v.l.'s tend to fill in all
the space. As a numerical example one can estimate $N$ for the
pulsar in the Crab nebula. Here $m=2m_N$ (the condensate is
formed by neutral bosons: pairs of neutrons) and
$\Omega=\Omega_{pulsar}$ gives $ N\simeq 1.9\times 10^{5}cm^{-2}$
with an average distance between vortex lines $d\sim N^{-1/2}\sim
10^{-2}cm$. If the vortex line is a straight line,
 $\bf v_s$ is perpendicular both to the
vortex line and to the radius joining the singular point and the
point at which we compute $v_s$. At a distance $r$ from the
singular point one has\be v_s=\frac{n \kappa}{r}\ ,\label{g7}\ee
as can be immediately seen from (\ref{g5}). More generally: \be
{\bf v}_s=\frac{\kappa}{2}\int_{v.l.}
\frac{d{\bm{\ell}}\wedge{\bf R}}{R^3}\ ,\ee where $\bf R$ is the
distance vector from the vortex line to the point at which we
compute the superfluid velocity.

During the rotation the vortex lines follow the rotational motion
of the vessel, which is clear because they are pinned at the
boundary of the superfluid; in particular, for rotations around
an axis, the vortex lines are, by symmetry, straight lines
parallel to the rotation axis. Their motion imitate the motion of
the liquid as a whole and, as  a consequence, also for the
superfluid one can use the  hydrodynamical formula
\be\displaystyle {\bf\Omega}=\frac 1 2{\bm{\nabla}}\wedge{\bf
v}_s\ ,\ee which in principle would be  valid only for the fluid
normal component.

 Let now
$\nu(r) $ be the number of vortices per unit area
 at a distance $r$ from the
rotation axis; if ${\bf v}={\bf v}_s$ is the superfluid velocity,
one gets \be \oint_\gamma d{\bm{ \ell}}\cdot{\bf v}=\int_0^r
d{\bf S}\cdot {\bm{ \nabla}}\wedge{\bf v}=2\pi\kappa\int_0^r 2\pi
r^\prime\nu(r^\prime)dr^\prime\ .\label{alp1}
 \ee
 We put $
 k=2\pi\kappa= h/{2m_n}$
 and write (\ref{alp1}) as follows:
 \be
2\pi\,r^2\,\Omega(r)=k\,\int_0^r 2 \pi
r^\prime\nu(r^\prime)dr^\prime\ ,\label{alp2}
 \ee
 which implies
 \be
k\nu(r)=2\Omega(r)\,+\,r\,\frac{\partial\Omega}{\partial r}\ .
 \label{alp4}\ee
 Since the  total number
 of v.l.'s is conserved, one has
 \be{\partial
 \nu}/{\partial t}+{\bm{\nabla}}\cdot(\nu{\bf v}_r)=0\,,\label{eq:519}\ee
where ${\bf v}_r$ is the radial component of the superfluid
velocity. We write (\ref{alp2}) as\be
2\pi\,r^2\,\Omega(r)=k\,\int_0^r\, \nu\,dS \label{alp3}
 \ee and take the time derivative, using
(\ref{eq:519}) to get\be
2\pi\,r^2\,\frac{\partial\Omega}{\partial t }=-k\,\int_0^r dS\,
\bm{\nabla}\cdot(\nu{\bf v}_r)\ .\label{alp5}
 \ee Using the Gauss theorem one gets
 $\displaystyle 2\pi\,r^2\,\frac{\partial\Omega}{\partial t
}=-k\,2\pi r \nu v_r$,
 i.e.
\be \frac{\partial\Omega}{\partial t }=-\frac{k \nu
v_r}r=-\left(2\Omega(r)\,+\,r\,\frac{\partial\Omega}{\partial
r}\right)\,\frac{v_r}r \ .\label{alp7}
 \ee
 Eq. (\ref{alp7}) shows that {\it the only possibility for the
 superfluid to change its angular velocity} ($\dot\Omega\neq 0$) {\it
 is by means of a
 radial motion, i.e. $v_r\neq 0$}.

 Let us now consider a rotating superfluid in contact with rotating normal matter
 on which an external torque is acting \cite{Alpar:1984ab}. We denote by
 $I_c$ and $\Omega_c$ the moment of inertia and
 angular velocity of the
 normal components  that, in a neutron star, includes
  the crust and possibly other normal components.
  The equation of motion of the normal component is
\be I_c\dot\Omega_c(t)=M_{ext}+M_{int}\ .\label{alp8}\ee Besides
the external torque $M_{ext}$, basically related to the spin down
of the pulsar
 (or the steady accretion in binary pulsars), we have included the internal torque
 $M_{int}$ due to the interaction with the superfluid:
 \be
 M_{int}=-\int dI_p\,\dot\Omega (r,t)\label{alp9}
 \ee where $dI_p$ is the infinitesimal moment of inertia of the
 superfluid component.
 Eqs. (\ref{alp7}-\ref{alp9}) are the equations of
 motion for the
 angular velocities $\Omega$ and $\Omega_c$ (superfluid and crust).
  The two velocities  are coupled not
 only through $M_{int}$, but also by $v_r$, because we will show
 below that $v_r$ depends on the difference $\Omega-\Omega_c$.
 We note again that
 fundamental for this model is the existence of radial motion,
 for, if $v_r=0$, then $\Omega= const.$ and only $\Omega_c$
 changes, due to the external torque alone.

 In the neutron star, superfluid neutrons (in Cooper pairs)
 coexist with nuclei of the crust. Also in the crust there are
superfluid neutrons, but they are characterized by a different
(and smaller)
 $\Delta$. Computing the difference in the free energies
 between the two phases one obtains the difference of pressures
 and, consequently, the force per unit length of vortex line.
Let  $b$ be the average distance
 between the nuclei; $b$ is also the average
 distance between two consecutive pinning centers. Let us assume \be
 2\xi_0<b \  ,\label{pin1}\ee
 where $\xi_0$ is the superconducting  coherence length, which also
gives the dimension of the vortex core, since $\xi_0$ is of the
order
 of the spatial  extension of the Cooper pair.
 The maximum pinning  force is obtained when the vortex passes through one layer of
 the lattice; therefore  the maximum force per unit
 length  of vortex line is
 \be
 f_p\simeq \frac{\delta E_p}{b\xi}\ ,\label{force1}\ee
where \be\delta E_p=F_s-F_c\propto\rho\Delta_s^2\,,\label{17}\ee
where $F_s$ and $F_c$ are the free
 energies of the superfluid neutrons and the nucleons in the crust;
  $\Delta_s$ is
  the gap for superfluid neutrons and one can neglect $\Delta_c$,
   the gap of superfluid neutrons in the crust since $\Delta_c\ll\Delta_s$.  Eq. (\ref{17})
implies that neutrons tend to remain in the volume $V$ of the
vortex core because they experience a force repelling them from
the superconducting phase (if neutron rich nuclei are present,
the
 repulsion will be less important). Typical values for the pinning
energy per nucleus  $\delta E_p$ at densities $3\times10^{13}\, -
\,1.2\times 10^{14}$
 g/cm$^3$ are\be\delta
E_p=1\,-\,3\,{\rm MeV}~,\label{TYPICAL}\ee while $b=
25\,-\,50\,{\rm fm}$ and $\xi_0= 4\,-\,20\,{\rm fm}$ give \be
f_p=40\,-\,1200\,{\rm MeV}^3~.\label{TYPICAL2}\ee

On the basis of these considerations let us now sketch a possible
mechanism for the formation
  of glitches \cite{Anderson:1973ab,Alpar:1977ab,Alpar:1984ab,
  Alpar:1984cd} (for further references see below). We
  consider the rotating neutron star with superfluid
 neutrons in its interior and a metallic crust, which is a
simplified model, but adequate to our purposes. As stressed
already, we
 distinguish between  the superfluid
 velocity  $\Omega$ and the crust velocity $\Omega_c$. Let us
 suppose that they are initially equal, which is a consequence of
 the pinning.
  Due to the spinning down
 of the star, $\Omega_c$ decreases; as long the vortex cores are
 pinned to the crust lattice, the neutron superfluid cannot spin down,
  because the radial motion is forbidden. There is therefore a
  relative velocity of the superfluid with respect
  to the  pinned vortex core because $\Omega>\Omega_c$:
  \be \delta{\bf v}=({\bf  \Omega}-{\bf \Omega}_c)\wedge {\bf  r}\ .\ee The interaction
  between the normal matter in the core of
  the v.l. and the rest of normal matter (nuclei in the lattice,
  electrons, etc.) produces a Magnus force per unit length
  given by
 \be{\bf f}=\rho {\bf  k}\wedge\delta{\bf  v}\ ,\ee
where the direction of $\bf k$ coincides with the rotation axis
and its modulus is equal to the quantum of vorticity. $f$ is the
force exerted on the vortex line; as it cannot be larger than
$f_p$ there is a maximum difference of angular velocity that the
system can maintain: \be
\omega_{cr}=\left(\Omega-\Omega_c\right)_{max}= \frac{f_p}{\rho k
r}=\frac{E_p}{\rho k\xi b}\ .\ee If $\omega <\omega_{cr}$ the
vortices remain pinned at the lattice sites instead of flowing
with the superfluid as they generally do inside it (see
discussion above). On the contrary, if $\omega
>\omega_{cr}$, the hydrodynamical forces arising from the mismatch
between the two angular velocities ultimately break the crust and
produce the conditions for the glitch. A possible way to get it
is by the observation following Eq. (\ref{alp7}). If a bunch of
vortex lines are unpinned and move outwards then eq. (\ref{alp7})
implies that the angular velocity (and  the angular momentum) of
the superfluid decreases, and, therefore, the angular momentum of
the crust increases, which is revealed from outside as a spin up
of the star, i.e. a glitch. A numerical analysis would imply
solving the set of Eqs. (\ref{alp7}-\ref{alp9}), but this is
outside the scope of the present review \footnote{Models differ
in the mechanism by which angular momentum is released; instead
of performing outward movements for example v.l.'s might break
the crust or rearrange it. For reviews see
\cite{Pines:1985ab,Alpar:1995ab} and, more recently,
\cite{Ruderman:1991ab,Epstein:1992ab,Alpar:1993ab,Link:1996ab,Ruderman:1998ab}.}.
Let us instead discuss the possible role of the LOFF phase in
this context.  The QCD LOFF phase provides a lattice structure
independently of the crust. Therefore it meets one of the two
requirements of the model for glitches in pulsars we have
outlined above, the other being the presence of a superfluid. On
the other hand the only existing calculations  for the
inhomogeneous phase in color superconductivity have been
performed for the case of two flavors, which however, in the
homogeneous case, does not present superfluidity, since there are
no broken global symmetries. Superfluidity is  on the other hand
manifested by the CFL phase of QCD. Therefore a realistic
application to QCD superfluid has to wait until a calculation of
the LOFF phase with three flavors will be completed. For the time
being one can give some order of magnitude estimates
\cite{Alford:2000ze}. Let us assume the following choice  of the
parameters: $\Delta_{2SC}=40 $ MeV, $\Delta_{LOFF}\approx 8$ MeV,
corresponding to the Fulde-Ferrel state; since $q\approx
1.2\delta\mu\approx\,0.7\Delta_{2SC}$, one would get for the
average distance between nodal planes $b=\pi/(2|{\bf q}|)\approx
9$ fm and  for the superconducting coherence length $\xi_0=6$ fm.
From (\ref{044}), with $\delta\mu=\delta\mu_1$ and an extra
factor of 4 to take into account the two flavors and the two
colors, we get the free energy per volume unit as
follows\footnote{Using the exact expression instead of
(\ref{044}), that is  valid only in the weak coupling limit, one
would get $|F_{LOFF}|=5\times(10\,MeV)^4$.}:
$|F_{LOFF}|=8\times(10\,MeV)^4$ and therefore, from (\ref{17}),
the pinning energy of the vortex line is \be \delta
E_p=|F_{LOFF}|\times b^3=6\ {\rm MeV}\ .\ee To get the pinning
force we cannot use (\ref{force1}) since (\ref{pin1}) does not
hold in this case. For an order of magnitude estimate one can use
\be
 f_p\simeq \frac{\delta E_p}{b^2}\ ,\label{force2}\ee giving a
 pinning  force per unit length of the vortex of the order of \be\label{order}
 f_p\approx 3\times 10^3\ {\rm MeV}^3\,.\ee
Comparing these numerical values with Eqs. (\ref{TYPICAL}) and
(\ref{TYPICAL2}) one can see that these  order of magnitude
estimates give figures similar and therefore some of the glitches
in neutron stars may be generated well inside the star by
vortices related to the LOFF phase of QCD.

As we already stated these conclusions are tentative because the
analysis of the QCD LOFF phase needs extension to the three
flavor case; moreover the true ground state is likely to be
different from the Fulde-Ferrel one plane wave structure.
Nevertheless they are encouraging and leave open the possibility
that neutron stars might give another laboratory where to study
the inhomogeneous superconducting phase. It can be useful to
stress that even in quark stars, in the QCD superconducting LOFF
phase, one would get a crystal structure given by a lattice
characterized by a geometric array where the gap parameter varies
periodically. This would overcome the objection that pulsars
cannot be strange stars. This objection is based on the following
observation:  If strange matter there exists, quark stars should
be rather common; however, in absence of metallic crusts, strange
stars can hardly develop vortices, at least by the model we have
described here. On the contrary, if the color superconductivity
is able to produce a crystalline structure it could also give
rise to glitches and the argument in favor of the existence of
strange stars would be reinforced.


\section{Conclusions}
Inhomogeneous crystalline superconductivity was predicted forty
years ago by Larkin, Ovchinnikov, Fulde and Ferrell, but
realistic conditions for its experimental investigations became
available only a few years ago. In condensed matter the existence
of the LOFF phase, with its characteristic space modulation of
the energy gap, still awaits complete confirmation. This is due
to the fact that it is indeed a subtle effect. It arises when the
Fermi surfaces of the two species participating in the Cooper
pairing are different. However for large separation pairing is
not possible at all and superconductivity disappears altogether.
In condensed matter the separation of the Fermi surfaces is
obtained by a Zeeman splitting due to an exchange interaction due
to a magnetic field. However the needed field strength is such to
destroy superconductivity due to diamagnetic effects. As we
discussed in the paper, the way to avoid the problem was to use
unconventional superconductors such as organic compounds. These
materials have in fact a layered structure and therefore, if the
magnetic field is parallel to the layers, the orbital effects can
be controlled.

New opportunities have recently arisen to detect the LOFF phase
in atomic physics (ultracold atomic gases), nuclear physics and
especially quark matter. This last development is a consequence
of the recent excitement generated by the study of QCD at high
density and small temperature. Inhomogeneous crystalline
superconductivity in this context could be generated by the
difference in quark chemical potentials induced by weak
interactions in the inner core of pulsars. Their main
phenomenological effect might therefore be to provide a mechanism
for the explanation of glitches in pulsars. If pulsars are
neutron stars with a core made up by color superconducting
matter, this mechanism would be complementary to the standard
models of glitches. If pulsars are strange stars, then the
crystalline structure of the condensate would provide the
possibility for pinning the superfluid vortices and eventually
creating the glitches.

This paper was mainly aimed to the presentation of a unified
formalism to describe the LOFF phase both in condensed and
hadronic matter. The simplest way in our opinion to describe
superconductivity effects, including the LOFF state, is by the
effective lagrangian approach. Since they are based on the
general mathematical ground of the Renormalization Group,
effective lagrangians allow the conditions for this unification.
The existence of a common mathematical basis should allow experts
of one side to fully appreciate and take advantage of the
progresses made in the other. We would be gratified if this paper
turned out to be useful to this end.
\begin{acknowledgments}
One of us (G.N.) would like to thank CERN Theory Division for the
kind hospitality offered during the completion of this paper. We
would like to thank R. Gatto for his invaluable help in common
work and for reading the manuscript and  E. Fabiano, and M.
Mannarelli for the very pleasant scientific collaboration. We
also wish to thank K. Rajagopal for a number of useful
discussions on color superconductivity and the LOFF phase. Our
thanks are finally due to M. Alford, J. Bowers, M. Ciminale, R.
Combescot and  H. Shimahara for useful correspondence on the
present review

\end{acknowledgments}
\appendix
\section{Calculation of $J$  and $K$}\label{app:A}
 We give here an outline of the calculation of
 the integrals $J$ and $K$ appearing in the GL expansion at $T=0$.
Using the definition of $J$, Eq. (\ref{gei}), and $K$, eq.
(\ref{kappa}), we have \bea J&\equiv&J({\bf q_1},{\bf q_2},{\bf
q_3},{\bf q_4})=+\frac{ig\rho}2 \int \frac{d{ \bf \hat w}}{4\pi}
\int_{-\delta}^{+\delta}
d\xi\int_{-\infty}^{+\infty}\frac{dE}{2\pi} \prod_{i=1}^4\,
f_i(E,\delta\mu,\{{\bf q}\})\cr &=& +\frac{ig\rho}2 \int
\frac{d{\bf \hat w}}{4\pi} \int_{-\delta}^{+\delta}
d\xi\int_{-\infty}^{+\infty}\frac{dE}{2\pi} \prod_{i=1}^2\Big\{
 \frac{1}{E+i\epsilon sign E+\xi-\delta\mu+2\,{\bf w\cdot k_i}}
 \cr&\times&
 \frac{1}{E+i\epsilon sign E-\xi-\delta\mu-2\,{\bf w} {\bf\cdot\bm {\ell}_i}}
\Big\} \ , \label{gei1}\eea \bea && K=K({\bf q_1},{\bf q_2},{\bf
q_3},{\bf q_4},{\bf q_5},{\bf q_6})=\frac{ig\rho}2 \int
\frac{d{\bf \hat w}}{4\pi}\int_{-\delta}^{+\delta}
d\xi\int_{-\infty}^{+\infty}\frac{dE}{2\pi}\prod_{i=1}^6
f_i(E,\delta\mu,\{{\bf q}\})\cr &&=\frac{ig\rho}2 \int
\frac{d{\bf \hat w}}{4\pi}\int_{-\delta}^{+\delta}
d\xi\int_{-\infty}^{+\infty}\frac{dE}{2\pi}\prod_{i=1}^3\Big\{
 \frac{1}{E+i\epsilon sign E+\xi-\delta\mu+2\,{\bf w\cdot k_i}}
 \cr&\times&
 \frac{1}{E+i\epsilon sign E-\xi-\delta\mu-2\,{\bf w}
 {\bf\cdot\bm {\ell}_i}}
\Big\} \,,\label{kappabis}\eea where we have defined\bea{\bf
k_1}&=&\bf 0, ~~~{\bf k}_2={\bf q_1}-{\bf q_2},~~~~~~~~~{\bf
k}_3={\bf q_1}-{\bf q_2}+{\bf q_3}-{\bf q_4}\,,\cr {\bm
{\ell}_1}&=&{\bf q_1},~~{\bm {\ell}_2}\,={\bf q_1}-{\bf q_2}+{\bf
q_3},~~{\bf \bm {\ell}_3}= {\bf q_1}-{\bf q_2}+{\bf q_3}-{\bf
q_4}+{\bf q_5}\,, \eea with the conditions ${\bf q_1}-{\bf
q_2}+{\bf q_3}-{\bf q_4}=0$ and
 ${\bf q_1}-{\bf q_2}+{\bf q_3}-{\bf q_4}+{\bf q_5}-{\bf q_6}=0$
  for $J$ and $K$ respectively.
We introduce the Feynman variables $x_j,\,y_j$ ($j=1,2$ for $J$
and $j=1,2,3$ for $K$) to form the vectors ${\bf k}=\sum_i
x_i{\bf k_i}$ and ${\bm {\ell}}=\sum_i y_i{\bm {\ell}_i}$; after
rotation of the energy integration contour $E\to \,i\,p_4$ we
get: \bea\int\frac{dE}{2\pi}\prod_{i}\frac 1{E+\xi-\delta\mu+2\,
{\bf w\cdot k_i}}&=&
\int\frac{idp_4}{2\pi}\frac{\delta\left(1-\sum x_k\right)}
{[ip_4+\xi-\delta\mu+2\,{\bf w\cdot k}]^2}\prod_{n} dx_n\, ,~~~~
\cr \int\frac{dE}{2\pi}\prod_{i} \frac{1}{E-\xi-\delta\mu-2\,{\bf
w} {\bf\cdot\bm {\ell}_i}}
&=&\int\frac{idp_4}{2\pi}\frac{\delta\left(1-\sum y_k\right)}
{[ip_4-\xi-\delta\mu-2\,{\bf w\cdot \bm
{\ell}}]^2}\prod_{n}dy_n~;~~~ \eea next we perform the $\xi$
integration by the residues method and the angular integration;
for $J$ the result is\be J=\,-\,\frac{ig\rho}{8}\int
dp_4\epsilon(p_4)\left(\delta\mu-ip_4\right)
\int\left[\prod_{n=1}^2dx_ndy_n\right] \frac{\delta\left(1-\sum
x_k\right)\delta\left(1-\sum y_k\right)}
{[(\delta\mu-ip_4)^2-v_F^2|{\bf k}-{\bm {\ell}}|^2]^2}\,.
\label{138}\ee After the energy  integration we remain with \be
J=\,-\,\frac{g\rho}{8}\int\left[\prod_{n=1}^2dx_ndy_n\right]
\delta\left(1-\sum x_k\right)\delta\left(1-\sum y_k\right)
\frac{1} {v_F^2|{\bf k}-{\bm {\ell}}|^2-\delta\mu^2}\,. \ee This
expression is general; we can specialize it to the various
crystal structures, as explained in the text.

For $K$ we get, instead of (\ref{138}), the result \bea
K&=&-\frac{3ig\rho}{8}\int dp_4\epsilon(p_4)
\int\left[\prod_{n=1}^3dx_ndy_n\right] \delta\left(1-\sum
x_k\right)\delta\left(1-\sum
y_k\right)\cr&\times&\left(\delta\mu-ip_4\right)\,
\frac{(\delta\mu-ip_4)^2+v_F^2|{\bf k}-{\bm
{\ell}}|^2}{[(\delta\mu-ip_4)^2-v_F^2|{\bf k}-{\bm {\ell}}|^2]^4}
\label{138K}\ ,\eea which, after energy integration, becomes \be
K=-\frac{g\rho}{16}\int\left[\prod_{n=1}^3dx_ndy_n\right]\delta\left(1-\sum
x_k\right)\delta\left(1-\sum y_k\right)
\frac{\delta\mu^2\,+\,3\,v_F^2|{\bf k}-{\bm {\ell}}|^2}
{\left[v_F^2|{\bf k}-{\bm
{\ell}}|^2-\delta\mu^2\right]^3}\,.\label{138Kb} \ee
\section{Expansion of $\Pi$ around the tricritical point}
\label{appendixb} Let us consider the expansion of $\Pi(q)$ in
${\bf Q}={\bf q} v_F$, at finite $T$ and $\mu$, which can be
obtained from Eq. (\ref{eq:143}) after introducing the Matsubara
frequencies:\be \Pi(q)=-\frac 1 2 g\rho T\int\frac{d{\hat{\bf
w}}}{4\pi}\int_{-\delta}^{+\delta}d\xi\sum_{n=-\infty}^{+\infty}
\frac{1} {(i\omega_n-\delta\mu-\xi-2{\hat{\bf{w}}}\cdot {\bf q}
v_F)(i\omega_n-\delta\mu+\xi)}\,.\ee Expanding the first
denominator in the momentum ${\bf q}$ we find \be \Pi(q)=\frac 12
g\rho T \sum_{n=-\infty}^{+\infty}\int\frac{d{\hat{\bf
w}}}{4\pi}\int_{-\infty}^{+\infty}d\xi\sum_{m=0}^{\infty}\frac{1}
{\bar\omega_n^2+ \xi^2}\frac{(2{\hat{\bf{w}}}\cdot {\bf q}
v_F)^{2m}}{(i\bar\omega_n-\xi)^{2m}}\,,\label{eq:191}\ee where,
as in Eq. (\ref{eq:70}),
\be\bar\omega_n=\omega_n+i\delta\mu\,.\ee Notice that we have
inverted the sum over the Matsubara frequencies with the
integration over $\xi$. In this way, as we did for the
homogeneous case, we are converting the divergence in $\xi$ into
a divergence in the series, which can be treated as before by
introducing a cutoff in the sum. Performing the angular
integration and the integration over $\xi$ with the help of the
following integral \be
\int_{-\infty}^{+\infty}d\xi\frac{1}{\bar\omega_n^2+ \xi^2}\frac
1{(i\bar\omega_n-\xi)^{2m}}=(-1)^m\frac{\pi}{2^{2m}\bar\omega_n^{2m+1}}\,,\ee
we get eventually \be \Pi(q)=\frac 12 g\pi\rho T
\sum_{n=-\infty}^{+\infty}\sum_{m=0}^{\infty}\frac
{(-1)^m}{2m+1}\frac {Q^{2m}}{\bar\omega_n^{2m+1}}\,.\ee By using
the definition of the first term in the grand potential as
$1-\Pi(q)$ multiplied by $2/g$ we recover easily the expression
(\ref{eq:162}) for $\tilde\alpha$. In analogous way, to get
$\tilde\beta$ and $\tilde\gamma$  one proceeds expanding $J$ (see
Eq. (\ref{gei}) and $K$ (see Eq. (\ref{kappa})).


\end{document}